\documentclass[aps,pra,reprint,amsmath,amssymb,superscriptaddress,longbibliography]{revtex4-2}

\usepackage[utf8]{inputenc}
\usepackage{graphicx}
\usepackage{booktabs}
\usepackage{hyperref}

\begin{document}

\title{Optimization Landscape Geometry in VQE for Frustrated Quantum Spin Models}

\author{Vojt\v{e}ch Nov\'{a}k}
\email{vojtech.novak.st1@vsb.cz}
\affiliation{Department of Computer Science, Faculty of Electrical Engineering and Computer Science, VSB--Technical University of Ostrava, Ostrava, Czech Republic}
\affiliation{IT4Innovations National Supercomputing Center, VSB--Technical University of Ostrava, 708 00 Ostrava, Czech Republic}
\affiliation{Department of Informatics and Statistics, Marine Research Institute, Klaipeda University, Lithuania}

\author{Ivan Zelinka}
\affiliation{Department of Computer Science, Faculty of Electrical Engineering and Computer Science, VSB--Technical University of Ostrava, Ostrava, Czech Republic}
\affiliation{IT4Innovations National Supercomputing Center, VSB--Technical University of Ostrava, 708 00 Ostrava, Czech Republic}
\affiliation{Department of Informatics and Statistics, Marine Research Institute, Klaipeda University, Lithuania}

\author{Swagatam Das}
\affiliation{Electronics and Communication Sciences Unit, Indian Statistical Institute, Kolkata 700108, India}

\author{Martin Beseda}
\affiliation{Dipartimento di Ingegneria e Scienze dell'Informazione e Matematica, Universit\`{a} dell'Aquila, Via Vetoio, I-67010 Coppito, L'Aquila, Italy}

\begin{abstract}
We benchmark eight classical optimizers for exact-statevector VQE calculations on a controlled hierarchy of frustrated spin models, ranging from a diagonal Ising glass to transverse-field Ising and anisotropic Heisenberg models. The benchmark includes local, stochastic-gradient, evolutionary, covariance-adaptation, and swarm-based optimization methods under matched function-evaluation budgets. To understand their performance beyond final energies, we characterize the underlying Hamiltonian--ansatz landscapes in terms of local minima, gradients, curvature, and ground-state reachability. We use simple variational circuits, from an $R_y$ product-state ansatz for the diagonal model to shallow $R_y$--CNOT hardware-efficient circuits for the noncommuting models, and study how increasing circuit depth changes their expressivity, reachability, and optimization geometry. We find that optimizer performance changes substantially across the model hierarchy and is closely connected to landscape structure, while the variational gap represents a separate source of error. These results show how classical optimization, variational expressivity, and landscape geometry jointly determine VQE performance for frustrated spin models.

\end{abstract}


\maketitle

\section{Introduction}

Variational quantum algorithms (VQAs) \cite{Cerezo2021} formulate quantum
problems as hybrid optimization tasks in which a classical optimizer updates
the parameters of a quantum circuit. In VQE \cite{fedorov2022vqe}, the circuit fixes the variational manifold; the classical optimizer determines
how that manifold is searched. Local minima, ill-conditioning, and barren
plateaus can therefore limit performance even for expressive ansatze
\cite{McClean2018,Cerezo2021}. These mechanisms are distinct: shallow local
VQAs can retain sizeable gradients while still containing many poor local
minima \cite{AnschuetzKiani2022}. Recent work further shows that low-quality
traps can also proliferate within barren-plateau regimes~\cite{Nemkov2025Traps}.

Optimizer performance in VQAs is strongly problem dependent. Broad comparisons
show changes in optimizer ranking with Hamiltonian, ansatz, initialization,
evaluation budget, and gradient-estimation strategy \cite{Jones2025}. Our
related studies show the same dependence in metaheuristic screening, finite-shot
chemistry, decoherence-noise benchmarks, QAOA parameter-activity analysis, and
noise-aware VQE \cite{Novak2025NoisyLandscapes,Illesova2025VHA,
Illesova2025Statistical,Bezdek2025ClassicalOptimization,Novak2025Reliable}.
Optimizer choice is therefore part of the VQA specification.

Population-based optimization has several VQE precedents. Fa\'ilde et al.\
showed that Differential Evolution (DE) can escape local minima that trap
standard local optimizers on Ising and Hubbard benchmarks \cite{Failde2023}.
Swarm methods have also been studied for molecular VQE \cite{Mei2024}, and a
30-optimizer comparison on 372 Fermi--Hubbard instances found SPSA and CMA-ES
to be competitive when function calls are counted explicitly \cite{Jones2025}.
These results establish population methods as viable VQA optimizers without
implying a universal ranking.

Frustrated and disordered spin systems provide a useful test regime. Competing
interactions generate many low-energy configurations, while noncommuting terms
introduce quantum correlations. Frustrated Ising models have been studied with
QAOA \cite{Lotshaw2023}, while VQE has been analyzed directly for frustrated
quantum systems~\cite{Uvarov2020Frustrated}; VQE studies of Heisenberg systems
also report strong optimizer and initialization dependence \cite{Jattana2022, illesova2025qmetric}. Kirmani et al.\
used independent restarts to mitigate SPSA trapping in a frustrated
transverse-field Ising VQE \cite{Kirmani2025}, and Pelofske and Eidenbenz
reported substantial convergence difficulty for anisotropic Heisenberg
spin-glass VQE \cite{Pelofske2026}. Here the focus is the continuous
variational landscape induced by the circuit, not classical spin-glass search
itself \cite{Bautu2008}.

Modern adaptive DE methods provide a useful comparison with established VQE optimizers because they were developed for difficult continuous optimization problems, including complex, rotated, and multimodal objective functions. We therefore include iL-SHADE, jSO, and L-SRTDE, which combine success-history adaptation and population-size reduction and have achieved strong results on CEC single-objective benchmarks \cite{Brest2016iLSHADE,Brest2017jSO,Stanovov2024LSRTDE,Novak2026CEC}. We compare these methods with classical DE as a canonical evolutionary baseline, together with BFGS, SPSA \cite{spall1998implementation}, CMA-ES \cite{cma}, and iSOMA \cite{zelinka2023isoma}, thereby spanning local, stochastic-gradient, evolutionary, covariance-adaptation, and swarm-based optimization under common function-evaluation (FE) budgets.

The analysis combines optimizer benchmarking with direct characterization of
the same Hamiltonian--ansatz landscapes. We measure basin multiplicity,
curvature, random-gradient scale, physical reachability, and controlled
one-dimensional Hamiltonian interpolations while keeping optimization error
separate from the variational gap. For the anisotropic Heisenberg model, we additionally vary the entangling-block
depth and combine Schmidt-rank \cite{van2024schmidt} bounds with fidelity, gradient, Hessian, and
quantum-geometric \cite{stokes2020quantum} diagnostics to separate expressivity from
landscape trainability. A separate FE-matched multistart BFGS control tests
restart-based basin coverage within the same evaluation budget, while an additive-noise control probes the sensitivity of the
numerical-gradient restart result. The central observation is that optimizer reversals track basin
accessibility and local geometry, whereas ansatz reachability and classical
optimization difficulty remain distinct.

\section{Methods}

The three benchmark Hamiltonians are special cases of a common two-parameter Hamiltonian. This keeps the change in Hamiltonian structure explicit
across the diagonal Ising, transverse-field, and anisotropic Heisenberg cases.
All variational energies are evaluated by exact statevector simulation, so the benchmark isolates the intrinsic Hamiltonian--ansatz landscape from finite-shot noise. Table~\ref{tab:optimizers_summary} summarizes the optimizers; full
settings are given in Appendix~\ref{app:optimizer-configurations}.

\subsection{Unified Hamiltonian family and variational ansatze}
\label{subsec:unified-model}

For a parameterized circuit $U(\boldsymbol{\theta})$, all optimizers minimize
the same variational objective \cite{Peruzzo2014,Cerezo2021},

\begin{equation}
    E(\boldsymbol{\theta};\Gamma,\lambda)
    =
    \langle\psi(\boldsymbol{\theta})|
    H(\Gamma,\lambda)
    |\psi(\boldsymbol{\theta})\rangle,
    \qquad
    |\psi(\boldsymbol{\theta})\rangle
    =
    U(\boldsymbol{\theta})|0\rangle .
\end{equation}

The three spin-glass benchmarks are embedded in the unified Hamiltonian

\begin{equation}
\label{eq:unified_hamiltonian}
    H(\Gamma,\lambda)
    =
    H_{\mathrm{SG}}
    -
    \Gamma\sum_{i=1}^{N}X_i
    +
    \lambda\sum_{(i,j)\in E}
    \left(
        J^x_{ij}X_iX_j
        +
        J^y_{ij}Y_iY_j
    \right),
\end{equation}

with the common frustrated Ising backbone

\begin{equation}
\label{eq:ising_backbone}
    H_{\mathrm{SG}}
    =
    \sum_{(i,j)\in E}J^z_{ij}Z_iZ_j
    +
    \sum_{i=1}^{N}h_i Z_i .
\end{equation}

Here the interaction graph $E$ is a nearest-neighbour ring supplemented by
long-range matching edges, yielding a sparse degree-three interaction
structure. For the Ising and transverse-field benchmarks,
$J^z_{ij}\equiv J_{ij}\in\{-1,+1\}$, and small longitudinal fields $h_i$
remove trivial degeneracies. For the Heisenberg benchmark, the additional
$J^x_{ij}$ and $J^y_{ij}$ couplings are fixed by the original Q3 instance
construction. For $N=10$ and $N=12$, the $J^z_{ij}$ and $h_i$ disorder
backbone is the same saved realization used across the corresponding model
constructions.

The original benchmark labels are therefore recovered as parameter cuts of
Eq.~\eqref{eq:unified_hamiltonian}:

\begin{equation}
\begin{aligned}
    \mathrm{Q1}:\quad & H(0,0)=H_{\mathrm{SG}},\\
    \mathrm{Q2}:\quad & H(\Gamma,0)
        =H_{\mathrm{SG}}-\Gamma\sum_i X_i,\\
    \mathrm{Q3}:\quad & H(0,\lambda)
        =H_{\mathrm{SG}}
        +\lambda\sum_{(i,j)\in E}
        \left(J^x_{ij}X_iX_j+J^y_{ij}Y_iY_j\right).
\end{aligned}
\end{equation}

The fixed benchmark points used in the present comparisons are
$\Gamma=0.3$ for Q2 and $\lambda=0.5$ for Q3. This formulation also makes the two coordinate scans
$H(\Gamma,0)$ and $H(0,\lambda)$ directly comparable and permits combined
$(\Gamma,\lambda)$ points without redefining the Hamiltonian.

The variational family is chosen according to the corresponding cut through
Hamiltonian space. At the diagonal point Q1, we use the product-state ansatz

\begin{equation}
    U_{\mathrm{prod}}(\boldsymbol{\theta})
    =
    \bigotimes_{i=1}^{N}R_y(\theta_i),
\end{equation}

which contains $D=N$ parameters. Since
$\theta_i\in\{0,\pi\}$ generates every computational-basis configuration,
the exact ground state of the diagonal Hamiltonian is contained in the
variational family. The corresponding objective can be written analytically
as

\begin{equation}
    E(\boldsymbol{\theta};0,0)
    =
    \sum_{(i,j)\in E}J^z_{ij}\cos\theta_i\cos\theta_j
    +
    \sum_i h_i\cos\theta_i .
\end{equation}

Thus, Q1 separates classical optimization difficulty from ansatz
expressivity and circuit-depth limitations.

For the non-commuting Q2 and Q3 cuts, we use the same shallow
$2N$-parameter hardware-efficient ansatz,

\begin{equation}
\label{eq:hea_ansatz}
    U_{\mathrm{HEA}}(\boldsymbol{\theta})
    =
    \left[\bigotimes_{i=1}^{N}R_y(\theta_{N+i})\right]
    U_{\mathrm{ent}}
    \left[\bigotimes_{i=1}^{N}R_y(\theta_i)\right].
\end{equation}

The fixed entangling layer is a nearest-neighbour CNOT chain. With qubits
ordered as $1,\ldots,N$, the gates
$\mathrm{CNOT}_{1\rightarrow2},\mathrm{CNOT}_{2\rightarrow3},\ldots,
\mathrm{CNOT}_{N-1\rightarrow N}$ are applied once in this order, equivalently

\begin{equation}
\label{eq:entangling_layer}
    U_{\mathrm{ent}}
    =
    \mathrm{CNOT}_{N-1\rightarrow N}\cdots
    \mathrm{CNOT}_{2\rightarrow3}
    \mathrm{CNOT}_{1\rightarrow2}.
\end{equation}

\begin{figure}[htpb]
    \centering
    \begin{minipage}[t]{0.28\columnwidth}
        \centering
        \textbf{(a)}\\[0.3em]
        \includegraphics[width=\linewidth]{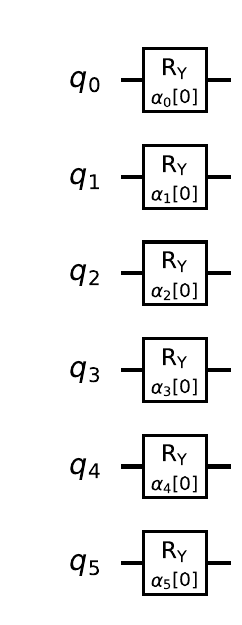}
    \end{minipage}\hfill
    \begin{minipage}[t]{0.70\columnwidth}
        \centering
        \textbf{(b)}\\[0.3em]
        \includegraphics[width=\linewidth]{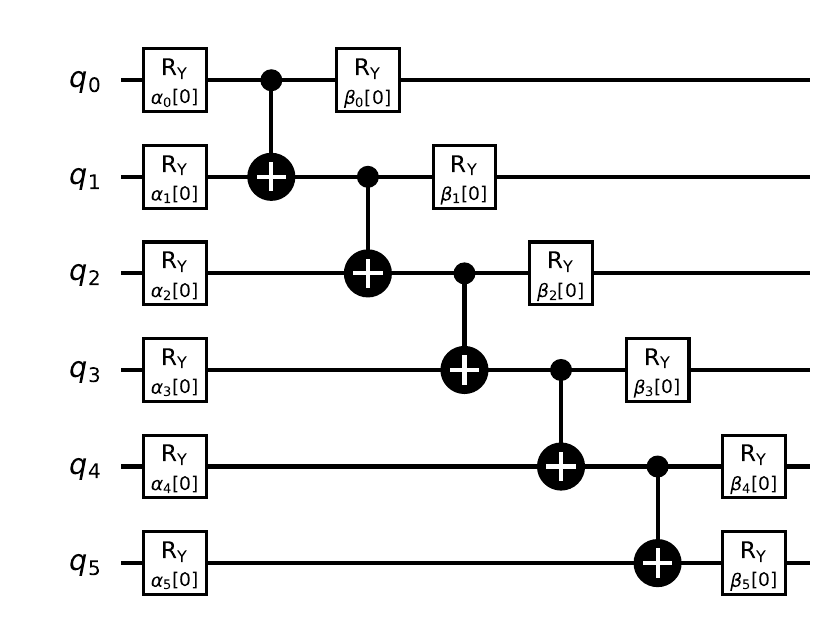}
    \end{minipage}
    \caption{(a) Product-state ansatz with parameter dimension $D=N$. (b) Hardware-efficient $R_y$--CNOT--$R_y$ ansatz with $D=2N$. Schematics illustrate $N=6$ qubits; the entangling layer in (b) follows Eq.~\eqref{eq:entangling_layer}.}
    \label{fig:ansatz-circuits}
\end{figure}

The two ansatz architectures are shown in Fig.~\ref{fig:ansatz-circuits};
$N=6$ is used only for visual clarity.

Exact ground-state energies are obtained by enumeration or exact
diagonalization whenever computationally feasible.

\subsection{Optimizers}
\label{subsec:optimizers}

Eight optimization methods are compared:

\begin{table*}[htbp]
\centering
\small
\caption{Summary of compared optimization methods, their algorithmic paradigms, and corresponding references.}
\label{tab:optimizers_summary}
\begin{tabular}{lllp{6.5cm}}
\hline
\textbf{Optimizer} & \textbf{Paradigm} & \textbf{Reference} & \textbf{Description} \\ \hline
BFGS & Quasi-Newton & \cite{nocedal2006numerical} & Local gradient-based reference method \\
SPSA & Stochastic Gradient & \cite{Spall1992} & Simultaneous perturbation stochastic approximation commonly used in VQAs \\
CMA-ES & Evolution Strategy & \cite{HansenOstermeier2001} & Covariance matrix adapting evolutionary strategy \\
SciPy-DE & Differential Evolution & \cite{StornPrice1997} & Classical Differential Evolution baseline \\
iL-SHADE & Adaptive DE & \cite{Brest2016iLSHADE} & Success-history adaptive DE with linear population size reduction \\
L-SRTDE & Adaptive DE & \cite{Stanovov2024LSRTDE} & Success-rate-adaptive DE developed for the CEC 2024 competition \\
jSO & Adaptive DE & \cite{Brest2017jSO} & Success-history-based adaptive DE variant \\
iSOMA & Swarm Intelligence & \cite{Diep2022iSOMA} & Self-organizing migrating algorithm with search space narrowing \\ \hline
\end{tabular}
\end{table*}

The panel spans standard VQA baselines and modern evolutionary or swarm
methods. The adaptive DE variants represent mechanisms successful on recent CEC
benchmarks \cite{Novak2026CEC}; SPSA, CMA-ES, DE, and related population methods
provide established VQA references \cite{Failde2023,Jones2025,
Novak2025NoisyLandscapes,Illesova2025VHA,Novak2025Reliable}.

All methods are compared using the number of objective-function evaluations
(FEs) as the primary computational budget, following the function-call-aware
perspective used in broad VQE optimizer benchmarking \cite{Jones2025}.
Except when an optimizer terminates according to its own convergence criterion,
each method receives the same maximum FE budget. Wall-clock time is recorded
as a secondary measure.

\subsection{Evaluation protocol and landscape diagnostics}
\label{sec:landscape_methods}

For each Hamiltonian instance and optimizer, independent runs are performed
from randomly generated initial populations or parameter vectors. The main
benchmark tables use 25 independent stochastic runs with a common set of
reproducibility seeds across optimizers for each fixed benchmark condition.
Hamiltonian disorder seeds and optimizer random seeds are controlled
separately. Repeated runs therefore quantify stochastic optimization
variability on a fixed disorder realization, while comparisons across model
cuts quantify changes associated with the selected Hamiltonian--ansatz
landscape rather than population-level variability over a large disorder
ensemble.

The energy reference is chosen to distinguish three quantities used
throughout: the ground-state energy error $E_{\mathrm{best}}-E_0$, the
optimization error $E_{\mathrm{best}}-E_{\mathrm{ref}}$, and the variational
gap $E_{\mathrm{ref}}-E_0$. Here $E_{\mathrm{ref}}$ denotes the saved
variational reference used for the corresponding comparison. For Q1, the
product-state ansatz contains every computational-basis state, so the exact
ground-state energy $E_0$ is reachable and

\begin{equation}
    \Delta E
    =
    E_{\mathrm{best}}-E_0
\end{equation}

is the ground-state energy error; because $E_{\mathrm{ref}}=E_0$ for Q1, it is
also the optimization error. For Q2, the exact ground-state energy remains
the primary physical reference and is complemented by the ground-state
fidelity

\begin{equation}
    F
    =
    |\langle\psi_0|\psi(\boldsymbol{\theta})\rangle|^2,
\end{equation}

because low variational energy need not imply large ground-state overlap in a
frustrated quantum problem \cite{Pelofske2026}. For Q3, the fixed shallow
ansatz does not in general contain the exact physical ground state. Optimizers
are therefore additionally compared with the best-known variational reference
$E_{\mathrm{ref}}$ through

\begin{equation}
    \Delta E_{\mathrm{opt}}
    =
    E_{\mathrm{best}}-E_{\mathrm{ref}},
\end{equation}

so $\Delta E_{\mathrm{opt}}$ is the optimization error and
$E_{\mathrm{ref}}-E_0$ is the variational gap. For Q3,
$E_{\mathrm{ref}}$ is a best-known reference rather than a certified global
optimum. For
$N=8$ the supplied reference energy is retained after consistency checks, with
an archived BFGS endpoint at the same energy providing the stored parameter
vector. For $N=10$ and $N=12$, the reference is constructed from 20 independent
Differential-Evolution searches (\texttt{best1bin}, \texttt{popsize}=30,
200 generations, tolerance $10^{-10}$), each followed by BFGS polishing
($g_{\rm tol}=10^{-10}$, at most 5000 iterations); the lowest polished energy
is retained. Any subsequent benchmark endpoint improving the stored value by
more than $10^{-12}$ replaces it, after which all Q3 optimization errors are
recomputed relative to the final reference.

In the cross-model figures below, $\Delta E_{\mathrm{opt}}$ denotes error to
the saved reference used for that model: $E_0$ for Q1, the stored
continuation reference for Q2, and $E_{\mathrm{ref}}$ for Q3.

In addition to final error, we record success probability at predefined energy
tolerances, convergence as a function of FEs, the number of FEs required to
reach a target accuracy, and wall-clock time. Because final-error
distributions are often non-Gaussian and may contain point masses close to
numerical zero, descriptive comparisons emphasize medians and interquartile
ranges. The common integer seeds are used for reproducibility but do not define
paired observations across heterogeneous optimizers. Inferential comparisons
therefore use independent-sample Kruskal--Wallis tests followed by two-sided
Mann--Whitney $U$ tests with Holm correction within each
model--size--budget condition. The complete statistical analysis, including
effect sizes, is reported in Appendix~\ref{app:statistical-comparison}.

The optimizer benchmark is complemented by diagnostics that characterize the
same fixed Hamiltonian--ansatz landscapes independently of the eight-method
comparison. For Q1 at moderate system sizes, all $2^N$ spin configurations
are enumerated. Besides the exact ground-state energy, we determine the number
and energy distribution of one-spin-flip local minima. A spin configuration
$\mathbf{s}$ is classified as locally stable when

\begin{equation}
    E(\mathbf{s}^{(i)})\geq E(\mathbf{s})
    \qquad \forall i,
\end{equation}

where $\mathbf{s}^{(i)}$ differs from $\mathbf{s}$ by one spin flip. These
minima provide a direct discrete measure of the multimodal structure
underlying the continuous Q1 variational objective.

For each original model--size condition, 24 parameter vectors are sampled
uniformly from $[-\pi,\pi)^D$ and quenched with a common L-BFGS-B local-descent
rule using exact analytic or adjoint gradients. By quench we mean local minimization or a local-descent run. L-BFGS-B is used only as a
deterministic geometry probe. Endpoint energies within $10^{-6}$ are
identified as one energy level, and the fraction of quenches within $10^{-2}$
of the best observed variational energy measures the volume of near-optimal
local-descent basins. Hessians are evaluated at the three lowest distinct
quench minima, while 64 additional uniformly sampled points per condition are
used to measure
$g_{\rm RMS}=\|\nabla E\|_2/\sqrt{D}$.

For Q2 and Q3, ansatz reachability is probed separately by maximizing exact
physical-ground-state fidelity from 32 independent starts. The lowest 32 exact
eigenpairs are used to resolve the spectral content of selected low-energy
variational states. The unified Hamiltonian in
Eq.~\eqref{eq:unified_hamiltonian} also provides a common interpolation
language: the transverse-field cut is $H(\Gamma,0)$ and the anisotropic
exchange cut is $H(0,\lambda)$. The controlled $N=12$ interpolation follows
both cuts over $\Gamma,\lambda\in\{0,0.1,\ldots,1\}$ on the same disorder
backbone and with the same hardware-efficient ansatz. BFGS and SciPy-DE are
re-evaluated along these cuts together with local-quench, Hessian, and
reachability diagnostics; these sweep runs are a diagnostic extension and are
not included in the main eight-optimizer ranking tables. Detailed clustering,
Hessian, and random-field-inspired diagnostics are retained in
Appendix~\ref{app:landscape-characterization}. Because single-start BFGS often
terminates before exhausting the nominal FE budget, Appendix~\ref{app:multistart-bfgs}
reports a separate FE-matched multistart control at $N=12$. This control isolates
restart-based basin coverage and is not included in the main eight-optimizer
ranking tables.

For the considered ansatze, $\boldsymbol{\theta}=\mathbf{0}$ prepares the computational-basis state $|0\rangle^{\otimes N}$, which is generally not a ground state of the frustrated disordered Hamiltonians; hence, the parameter-space origin has no privileged relation to the variational minimum.

\subsection{Ansatz-depth expressivity and reachability diagnostics}

Insufficient circuit expressivity creates an intrinsic variational gap that no classical optimizer can overcome~\cite{Akshay2021reachability,Bharti2022NISQ,Grimsley2019adaptvqe}. To separate this structural limitation from landscape trainability and local trapping, we evaluate depth-dependent Schmidt rank bounds alongside direct numerical diagnostics.

We evaluate the layered alternating ansatz with linear CNOT entanglement across depths $p=1,\ldots,6$:
\begin{equation}
U_p(\boldsymbol{\theta})
=
R_y^{(p)} U_{\mathrm{ent}}
R_y^{(p-1)} U_{\mathrm{ent}}
\cdots
U_{\mathrm{ent}} R_y^{(0)},
\end{equation}
which contains $D=(p+1)N$ variational parameters and $p(N-1)$ CNOT gates. The primary benchmarks use $N=8,10,12$, with auxiliary $N=4,6$ systems used to verify exact ground-state recovery.

For any contiguous bipartition $c=A\vert B$, each entangling block contains exactly one CNOT crossing the cut. The bipartite Schmidt rank is therefore bounded by $r_{p,c} \leq \min(2^p, 2^{\min(|A|,|B|)})$~\cite{van2024schmidt}. 

Let $s_{j,c}$ denote the exact Schmidt coefficients of the target ground state. The attainable state fidelity is upper-bounded by the truncated Schmidt weight $F_{\mathrm{Sch},c}^{(p)} = \sum_{j=1}^{r_{p,c}} s_{j,c}^2$. Taking the strictest cut yields $F_{\mathrm{Sch}}^{(p)} = \min_c F_{\mathrm{Sch},c}^{(p)}$. Combined with the exact spectral gap $\Delta_{\mathrm{gap}}=E_1-E_0$, this gives an optimizer-independent lower bound on the variational energy error:
\begin{equation}
\Delta E \geq (1 - F_{\mathrm{Sch}}^{(p)})\Delta_{\mathrm{gap}}.
\end{equation}
A Schmidt-rank deficiency thus strictly rules out exact state preparation.

To complement the theoretical bounds, we independently minimize the variational energy and the diagnostic state infidelity $\mathcal{L}_{\mathrm{fid}}=1-|\langle\psi_0|\psi(\boldsymbol{\theta})\rangle|^2$. Global exploration is performed using iL-SHADE, followed by local refinement with BFGS using exact parameter-shift gradients. In parallel, multi-start random BFGS runs probe the local basin structure. 

At the polished global best point, we evaluate the residual energy error $\Delta_{\mathcal{M}} = E_{\mathrm{best}} - E_0$, target fidelity, average gradient norms, Hessian eigenspectra, Quantum Fisher Information Matrix (QFIM) rank, and 1D random-line nonquadraticity. Full hyperparameters, function evaluation budgets, and numerical tolerances are detailed in Appendix~\ref{app:q3-depth-study}.

\section{Results}

In this section we present the optimizer benchmarks together with the independent landscape diagnostics defined above.

\subsection{Frustrated Ising spin-glass benchmark}
\label{sec:results_ising_spin_glass}

Q1 isolates classical optimization difficulty because the product-state ansatz
contains the exact ground state. Table~\ref{tab:ising_median_errors} summarizes
25 runs per optimizer for $N\in\{12,14,16\}$ at $10\,000$ and $30\,000$ FEs.
The main trend is a clear advantage for global population-based methods at the
smaller sizes: jSO reaches a median error of $3.61\times10^{-12}$ at
$N=12$ and $10\,000$ FEs, while iSOMA reaches a zero median at $30\,000$ FEs.
At $N=14$, jSO and iSOMA both reach zero median error by $30\,000$ FEs. At
$N=16$ the ranking becomes less stable: L-SRTDE attains the lowest
$30\,000$-FE median, $2.90\times10^{-3}$, but with a very broad
interquartile range $[1.07\times10^{-4},1.95]$, showing that exceptionally
accurate runs coexist with poor outcomes.

\begin{table*}[t]
    \centering
    \caption{Median final ground-state energy error
    $\Delta E=E_{\mathrm{best}}-E_0$ over 25 independent runs.
    Lower values are better. A value of zero denotes agreement with the
    exact ground-state energy within numerical precision. The lowest median
    in each configuration is shown in bold.}
    \label{tab:ising_median_errors}
    \begin{tabular}{lcc|cc|cc}
        \toprule
        & \multicolumn{2}{c|}{$N=12$}
        & \multicolumn{2}{c|}{$N=14$}
        & \multicolumn{2}{c}{$N=16$} \\
        \cmidrule(lr){2-3}
        \cmidrule(lr){4-5}
        \cmidrule(lr){6-7}
        Optimizer
        & $10$k & $30$k
        & $10$k & $30$k
        & $10$k & $30$k \\
        \midrule
        BFGS
        & $1.706$ & $1.706$
        & $0.822$ & $0.822$
        & $2.248$ & $2.248$ \\

        SPSA
        & $1.999$ & $1.996$
        & $0.984$ & $0.979$
        & $2.273$ & $2.268$ \\

        CMA-ES
        & $0.683$ & $0.646$
        & $1.737$ & $0.809$
        & $1.860$ & $2.388$ \\

        SciPy-DE
        & $0.0371$ & $0.0368$
        & $0.530$ & $0.518$
        & $1.452$ & $1.447$ \\

        iL-SHADE
        & $0.127$ & $0.0368$
        & $0.518$ & $0.518$
        & $1.447$ & $1.447$ \\

        L-SRTDE
        & $0.124$ & $0.0368$
        & $0.558$ & $0.518$
        & $2.127$ & $\mathbf{2.90\times10^{-3}}$ \\

        jSO
        & $\mathbf{3.61\times10^{-12}}$ & $0.0368$
        & $\mathbf{2.43\times10^{-11}}$ & $\mathbf{0}$
        & $1.402$ & $0.458$ \\

        iSOMA
        & $1.50\times10^{-5}$ & $\mathbf{0}$
        & $6.93\times10^{-8}$ & $\mathbf{0}$
        & $\mathbf{0.458}$ & $0.458$ \\
        \bottomrule
    \end{tabular}
\end{table*}

Under the single-start benchmark protocol, the diagonal model therefore favors methods that provide broader basin coverage, while the increasing run-to-run dispersion with $N$ makes a single median progressively less representative. The $N=12$ behavior is compared directly with Q2 and Q3 in
Sec.~\ref{sec:results_cross_model}.

\subsection{Transverse-field frustrated Ising spin-glass benchmark}
\label{sec:results_tf_spin_glass}

Q2 adds the non-commuting transverse field at $\Gamma=0.3$ and uses the
$2N$-parameter hardware-efficient ansatz. The benchmark covers $N\in\{10,12\}$
with the same two FE budgets and 25 runs per condition. As summarized in
Table~\ref{tab:tfsg_median_errors}, the leading method is budget dependent.
At $N=10$, BFGS gives the lowest $10\,000$-FE median,
$\Delta E=0.179$, whereas jSO improves to $0.0889$ and leads at
$30\,000$ FEs. At $N=12$, iL-SHADE leads at $10\,000$ FEs with
$\Delta E=0.0723$, while jSO reaches $0.0368$ at $30\,000$ FEs; SciPy-DE
and iL-SHADE follow at $0.0518$ and $0.0652$, respectively. BFGS changes
little with the larger budget because it usually terminates before exhausting
the FE allowance.

\begin{table}[htpb]
    \centering
    \caption{Median final ground-state energy error
    $\Delta E=E_{\mathrm{best}}-E_0$ over 25 independent runs for the
    transverse-field spin-glass benchmark at $\Gamma=0.3$.
    Lower values are better. The smallest median error in each configuration
    is shown in bold.}
    \label{tab:tfsg_median_errors}
    \begin{tabular}{lcc|cc}
        \toprule
        & \multicolumn{2}{c|}{$N=10$}
        & \multicolumn{2}{c}{$N=12$} \\
        \cmidrule(lr){2-3}
        \cmidrule(lr){4-5}
        Optimizer
        & $10$k & $30$k
        & $10$k & $30$k \\
        \midrule
        BFGS
        & $\mathbf{0.179}$ & $0.179$
        & $0.124$ & $0.124$ \\

        SPSA
        & $0.244$ & $0.216$
        & $0.541$ & $0.531$ \\

        CMA-ES
        & $0.238$ & $0.202$
        & $0.171$ & $0.189$ \\

        SciPy-DE
        & $0.262$ & $0.139$
        & $1.859$ & $0.0518$ \\

        iL-SHADE
        & $0.225$ & $0.226$
        & $\mathbf{0.0723}$ & $0.0652$ \\

        L-SRTDE
        & $0.847$ & $0.219$
        & $5.913$ & $0.102$ \\

        jSO
        & $0.418$ & $\mathbf{0.0889}$
        & $0.184$ & $\mathbf{0.0368}$ \\

        iSOMA
        & $0.275$ & $0.206$
        & $0.265$ & $0.231$ \\
        \bottomrule
    \end{tabular}
\end{table}

Energy alone does not determine the physical state reached. The fidelities in
Table~\ref{tab:tfsg_fidelity} remain low for all $N=10$ methods; even at
$30\,000$ FEs the largest median is $0.348$ for jSO. At $N=12$, iL-SHADE
already reaches a median fidelity of $0.959$ at $10\,000$ FEs, while jSO
reaches $0.983$ at $30\,000$ FEs. BFGS is notably split: its median fidelity
is only $9.23\times10^{-4}$ even though its upper quartile is approximately
$0.998$, so similar low energies can arise from qualitatively different states.

\begin{table}[htpb]
    \centering
    \caption{Median fidelity with the exact quantum ground state over
    25 independent runs. Higher values are better. The largest median
    fidelity in each configuration is shown in bold.}
    \label{tab:tfsg_fidelity}
    \begin{tabular}{lcc|cc}
        \toprule
        & \multicolumn{2}{c|}{$N=10$}
        & \multicolumn{2}{c}{$N=12$} \\
        \cmidrule(lr){2-3}
        \cmidrule(lr){4-5}
        Optimizer
        & $10$k & $30$k
        & $10$k & $30$k \\
        \midrule
        BFGS
        & $0.0665$ & $0.0665$
        & $9.23\times10^{-4}$ & $9.23\times10^{-4}$ \\

        SPSA
        & $0.0650$ & $0.0678$
        & $5.68\times10^{-4}$ & $1.72\times10^{-4}$ \\

        CMA-ES
        & $0.0703$ & $0.0808$
        & $0.0133$ & $1.64\times10^{-4}$ \\

        SciPy-DE
        & $\mathbf{0.0779}$ & $0.125$
        & $0.00134$ & $0.474$ \\

        iL-SHADE
        & $0.0275$ & $0.0577$
        & $\mathbf{0.959}$ & $0.972$ \\

        L-SRTDE
        & $0.0220$ & $0.0762$
        & $0.00857$ & $4.69\times10^{-4}$ \\

        jSO
        & $0.0602$ & $\mathbf{0.348}$
        & $0.914$ & $\mathbf{0.983}$ \\

        iSOMA
        & $0.0614$ & $0.323$
        & $0.887$ & $0.898$ \\
        \bottomrule
    \end{tabular}
\end{table}

The stored continuation solution lies approximately $2.55\times10^{-2}$ above
$E_0$ for $N=10$ and $5.11\times10^{-3}$ above $E_0$ for $N=12$; the
independent quench analysis improves the latter best-observed gap to
$4.25\times10^{-3}$. Thus Q2 combines budget-sensitive optimizer rankings with
a nontrivial distinction between low energy, ansatz reachability, and physical
fidelity.

\subsection{Anisotropic Heisenberg spin-glass benchmark}
\label{sec:results_heisenberg_glass}

Q3 considers the anisotropic Heisenberg glass at $\lambda=0.5$ for
$N\in\{8,10,12\}$. Because the shallow ansatz does not generally contain the
physical ground state, optimizer performance is measured by
$\Delta E_{\mathrm{opt}}=E_{\mathrm{best}}-E_{\mathrm{ref}}$.
The estimated variational gaps $E_{\mathrm{ref}}-E_0$ are approximately
$1.229$, $1.243$, and $0.895$ for $N=8$, $10$, and $12$, respectively. The
supplied $N=8$ variational reference is recovered to numerical precision, and
neither the $N=10$ nor $N=12$ reference is improved by the benchmark runs; the
independent local-quench probe recovers the same best-known variational levels.

Table~\ref{tab:heis_median_errors} shows an ordering unlike Q1. At $N=8$,
BFGS and SciPy-DE both reach a median
$\Delta E_{\mathrm{opt}}\simeq6.31\times10^{-2}$ at $10\,000$ FEs. At
$N=10$, SciPy-DE gives the lowest medians, $0.398$ and $0.360$ at
$10\,000$ and $30\,000$ FEs, while BFGS remains at $0.475$ under both
budgets. The $N=12$ behavior is sharper: BFGS reaches
$2.08\times10^{-3}$ already at $10\,000$ FEs, whereas SciPy-DE falls from
$1.33$ to $2.16\times10^{-3}$ only when the budget is increased to
$30\,000$ FEs.

\begin{table}[htpb]
    \centering
    \caption{Median final optimization error
    $\Delta E_{\mathrm{opt}}
    =E_{\mathrm{best}}-E_{\mathrm{ref}}$
    over 25 independent runs for the anisotropic Heisenberg spin-glass
    benchmark at $\lambda=0.5$. Lower values are better. The lowest
    median in each configuration is shown in bold; values indistinguishable
    at the displayed precision are both emphasized.}
    \label{tab:heis_median_errors}
    \small
    \begin{tabular}{lcc|cc|cc}
        \toprule
        & \multicolumn{2}{c|}{$N=8$}
        & \multicolumn{2}{c|}{$N=10$}
        & \multicolumn{2}{c}{$N=12$} \\
        \cmidrule(lr){2-3}
        \cmidrule(lr){4-5}
        \cmidrule(lr){6-7}
        Optimizer
        & $10$k & $30$k
        & $10$k & $30$k
        & $10$k & $30$k \\
        \midrule
        BFGS
        & $\mathbf{0.0631}$ & $\mathbf{0.0631}$
        & $0.475$ & $0.475$
        & $\mathbf{0.0021}$
        & $\mathbf{0.0021}$ \\

        SPSA
        & $0.230$ & $0.0944$
        & $0.540$ & $0.528$
        & $0.662$ & $0.656$ \\

        CMA-ES
        & $0.766$ & $0.465$
        & $0.581$ & $0.591$
        & $0.822$ & $0.775$ \\

        SciPy-DE
        & $0.0631$ & $\mathbf{0.0631}$
        & $\mathbf{0.398}$ & $\mathbf{0.360}$
        & $1.334$ & $0.0022$ \\

        iL-SHADE
        & $0.412$ & $0.0962$
        & $0.541$ & $0.475$
        & $0.818$ & $0.737$ \\

        L-SRTDE
        & $1.128$ & $0.0633$
        & $0.760$ & $0.475$
        & $4.757$ & $0.955$ \\

        jSO
        & $0.766$ & $0.766$
        & $1.421$ & $0.432$
        & $2.952$ & $1.073$ \\

        iSOMA
        & $0.767$ & $0.766$
        & $1.367$ & $1.367$
        & $1.902$ & $1.902$ \\
        \bottomrule
    \end{tabular}
\end{table}

At $N=12$, BFGS reaches $\Delta E_{\mathrm{opt}}\leq10^{-2}$ in $80\%$ of
runs under both budgets, while SciPy-DE increases from $0\%$ to $64\%$ when
the budget is raised from $10\,000$ to $30\,000$ FEs. This sharp reversal
relative to Q1 motivates the direct cross-model comparison below.

\subsection{Q3 ansatz-depth expressivity}
\label{sec:results_q3_depth}

The shallow Q3 ansatz leaves a substantial variational gap. We
therefore study how this gap changes as the depth of the same
hardware-efficient ansatz is increased, and how the corresponding optimization
landscape changes with circuit depth. In particular, we track ground-state
fidelity and Schmidt-rank constraints together with gradient scale, curvature,
parameter coupling, and quantum-geometric diagnostics.

We retain the same hardware-efficient architecture at every depth so that
changes can be attributed to increasing variational capacity rather than to
a change of ansatz family. The dependence of VQE convergence on entangling
structure, circuit depth, and problem hardness has also been studied
directly~\cite{Woitzik2020Entanglement,DiezValle2021EntanglementHardness}.
Hamiltonian-informed ansatze (VHA) may reach the same target with different depth
and landscape structure and are not considered here.

We apply the depth-dependent protocol defined in Sec.~II D to
$N=8,10,12$ for $p=1,\ldots,6$. The $p=1$ results reproduce the original
Q3 reachability values, providing a direct reference for the effect of
increasing depth.

\begin{figure*}[htpb]
\centering
\begin{minipage}{0.38\textwidth}
    \centering
    \includegraphics[width=\linewidth]{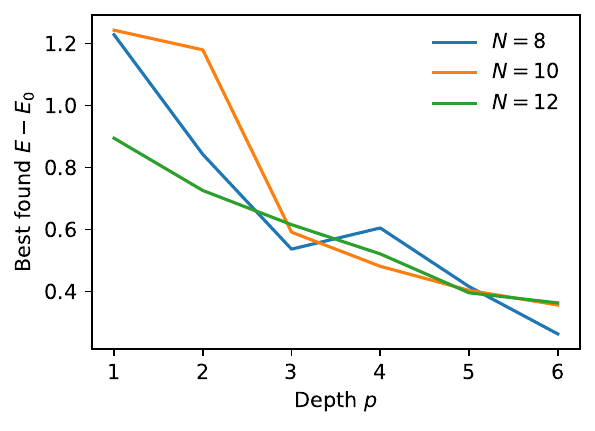}\\[-1mm]
    \small (a) Ground-state energy error
\end{minipage}%
\hspace{0.03\textwidth}%
\begin{minipage}{0.38\textwidth}
    \centering
    \includegraphics[width=\linewidth]{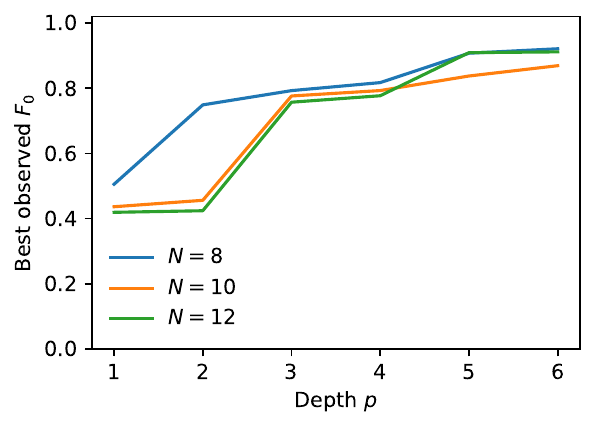}\\[-1mm]
    \small (b) Ground-state fidelity
\end{minipage}

\vspace{2mm}

\begin{minipage}{0.38\textwidth}
    \centering
    \includegraphics[width=\linewidth]{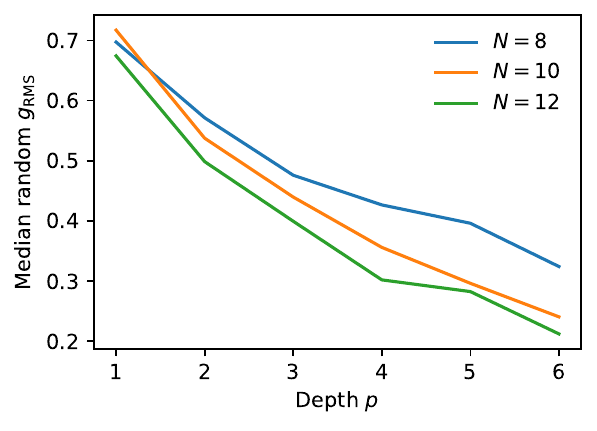}\\[-1mm]
    \small (c) Random-gradient scale
\end{minipage}%
\hspace{0.03\textwidth}%
\begin{minipage}{0.38\textwidth}
    \centering
    \includegraphics[width=\linewidth]{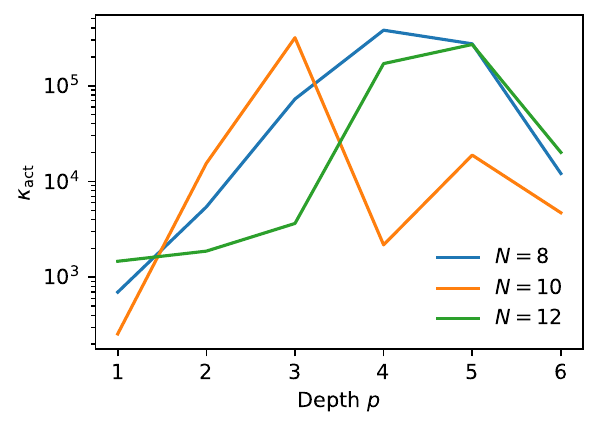}\\[-1mm]
    \small (d) Hessian conditioning
\end{minipage}

\caption{Effect of increasing the Q3 hardware-efficient ansatz depth.
(a) Best-found ground-state energy error $E_{\mathrm{best}}-E_0$.
(b) Best observed exact ground-state fidelity.
(c) Median RMS parameter-shift gradient over random parameter vectors.
(d) Active Hessian condition number at the best-found energy point.
Depth improves physical reachability, while the deeper landscapes remain
strongly anisotropic. The random-gradient scale decreases with depth but
remains finite over the studied range.}
\label{fig:q3-depth-main}
\end{figure*}

Increasing depth reduces the best-found ground-state energy error and increases the
best observed ground-state fidelity overall, as shown in
Fig.~\ref{fig:q3-depth-main}. At $p=6$, the best-found ground-state energy errors are $0.264$,
$0.357$, and $0.364$ for $N=8$, $10$, and $12$, respectively, compared with
$1.229$, $1.243$, and $0.895$ at $p=1$. The corresponding best observed
ground-state fidelities increase from $(0.505,0.436,0.419)$ to
$(0.921,0.869,0.911)$.

The exact ground-state Schmidt spectra give a simple structural lower bound on
the required depth. The exact state is ruled out by Schmidt rank for
$p<4$, $p<5$, and $p<6$ at $N=8$, $10$, and $12$, respectively.
Thus the first depths not excluded by Schmidt rank are $p=4$, $5$, and $6$.
Passing this threshold does not guarantee exact representability by the
restricted $R_y$--CNOT architecture or exact recovery by the numerical
search.

\begin{table}[t]
\centering
\caption{Compact Q3 ansatz-depth reachability summary.
$p_{\rm S}$ is the first depth not excluded by the exact ground-state Schmidt
rank. $\Delta_{\mathcal M}=E_{\rm best}-E_0$ is the best-found ground-state energy error. $F_{\rm best}$ is the largest ground-state fidelity observed among the
energy- and fidelity-targeted searches.}
\label{tab:q3-depth-main}
\begin{tabular}{cc|cc|cc}
\toprule
$N$ & $p_{\rm S}$
& \multicolumn{2}{c|}{$p=1$}
& \multicolumn{2}{c}{$p=6$} \\
& & $\Delta_{\mathcal M}$ & $F_{\rm best}$
  & $\Delta_{\mathcal M}$ & $F_{\rm best}$ \\
\midrule
8  & 4 & 1.229 & 0.505 & 0.264 & 0.921 \\
10 & 5 & 1.243 & 0.436 & 0.357 & 0.869 \\
12 & 6 & 0.895 & 0.419 & 0.364 & 0.911 \\
\bottomrule
\end{tabular}
\end{table}

The median random-gradient scale decreases from $0.67$--$0.72$ at $p=1$ to
$0.21$--$0.32$ at $p=6$, with no sampled random point below
$g_{\mathrm{RMS}}=10^{-3}$. At the same time, the Hessian condition number
reaches $10^4$--$10^5$ at several depths. The deeper circuits therefore improve
reachability while producing strongly anisotropic and increasingly coupled
local geometry, rather than a finite-size globally vanishing-gradient
landscape.

\subsection{Cross-model comparison at $N=12$}
\label{sec:results_cross_model}

\begin{figure*}[htpb]
    \centering
    \includegraphics[width=\textwidth]{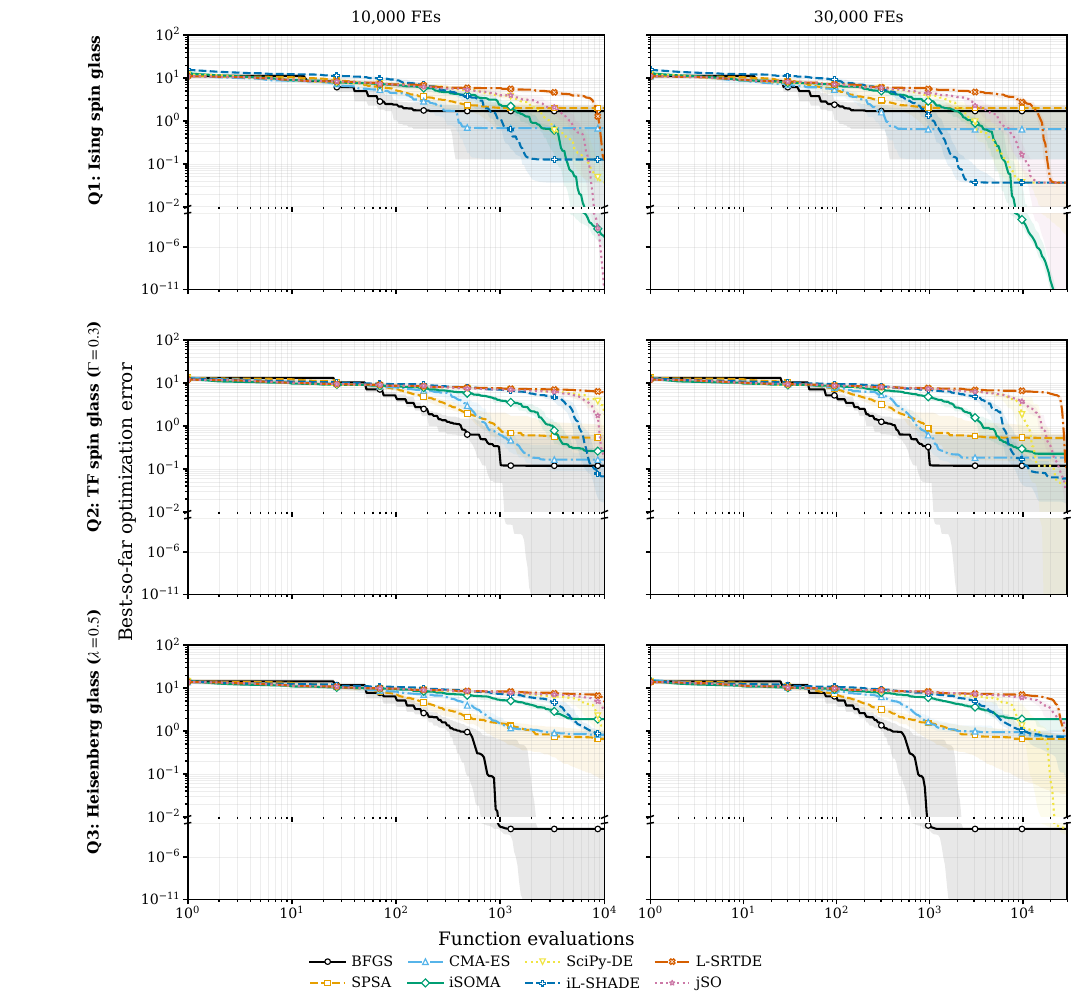}
    \caption{Median best-so-far optimization error with interquartile ranges for
    the three $N=12$ benchmarks at $10\,000$ and $30\,000$ FEs. For Q1 the
    reference is $E_0$, for Q2 the stored continuation reference, and for Q3
    $E_{\mathrm{ref}}$. The broken logarithmic axis preserves resolution
    in both the main error band and the near-zero regime.}
    \label{fig:cross-convergence-n12}
\end{figure*}

Figure~\ref{fig:cross-convergence-n12} compares the three $N=12$ convergence
profiles directly. Q1 shows continued late-budget progress by several
population methods, Q2 exhibits budget-dependent crossovers, and Q3 shows BFGS
entering the best-known variational region much earlier. The broken vertical
axis retains resolution near numerical zero.

\begin{figure*}[htpb]
    \centering
    \includegraphics[width=\textwidth]{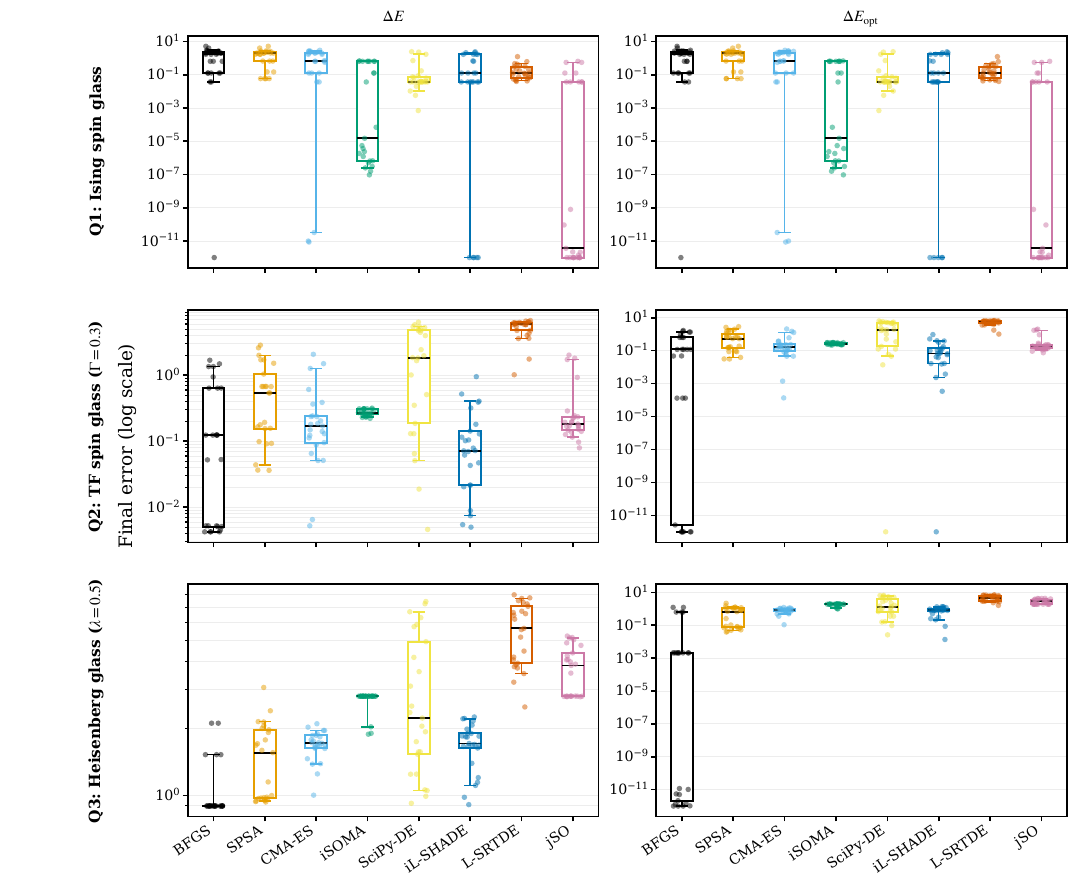}
    \caption{Final-error distributions for $N=12$ at $10\,000$ FEs. Boxes show
    the median and interquartile range, whiskers span the 5th--95th percentiles,
    and points are individual runs. Numerical zeros are shown at the plotting
    floor $10^{-12}$.}
    \label{fig:cross-dist-n12-10k}
\end{figure*}

\begin{figure*}[htpb]
    \centering
    \includegraphics[width=\textwidth]{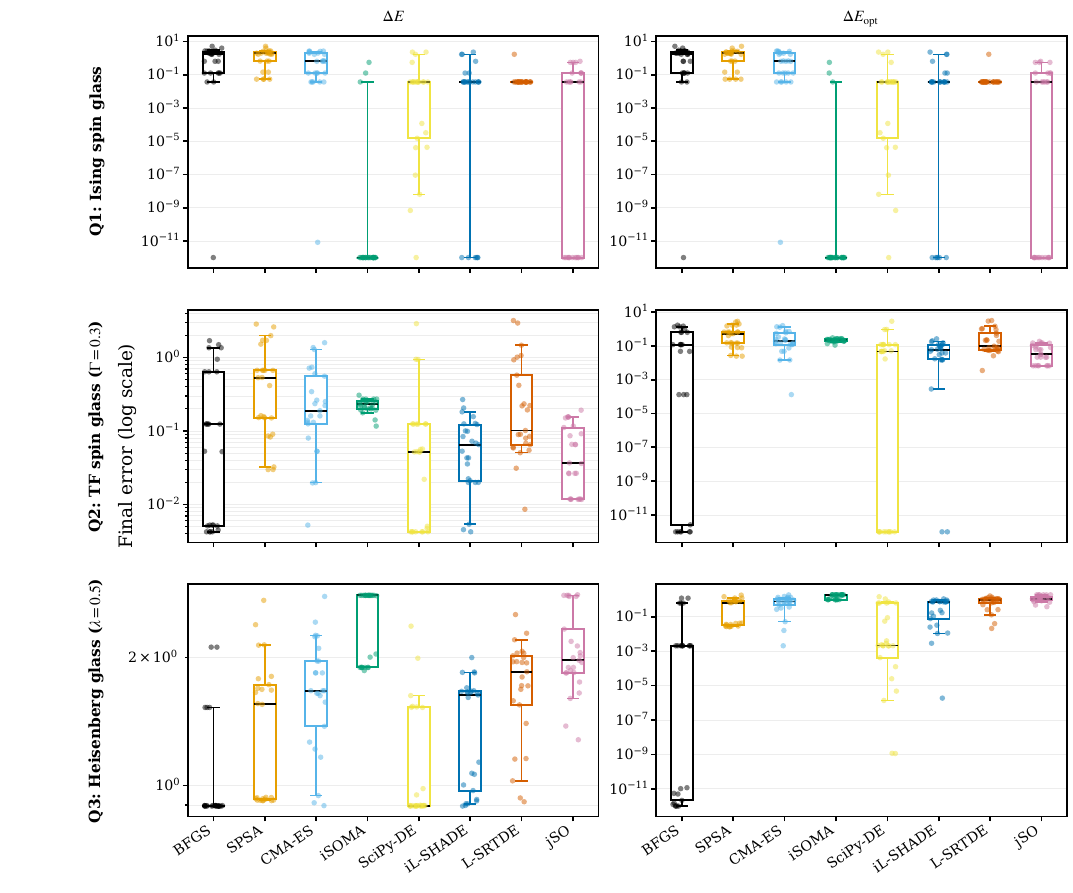}
    \caption{Final-error distributions for $N=12$ at $30\,000$ FEs, using the
    same conventions as Fig.~\ref{fig:cross-dist-n12-10k}. The larger budget
    exposes which population-based methods continue to improve after the
    smaller-budget comparison has saturated.}
    \label{fig:cross-dist-n12-30k}
\end{figure*}

Figures~\ref{fig:cross-dist-n12-10k} and~\ref{fig:cross-dist-n12-30k} resolve the
endpoint dispersion hidden by the medians. Several Q1 and Q2 methods mix
near-optimal and poor runs, whereas Q3 at $10\,000$ FEs has a concentrated
low-error BFGS population. The paired $\Delta E$ and $\Delta E_{\mathrm{opt}}$
panels separate physical ground-state error from error to the model-specific
variational reference.

\subsection{Landscape geometry and ansatz reachability}
\label{sec:results_landscape}

Table~\ref{tab:landscape-main} compares optimizer behavior with independent
properties of the same variational objectives. Q1 is exactly expressive and
strongly multimodal: standardized local descent terminates in 14--22 distinct
energy levels, with only $4.2$--$8.3\%$ of starts within $10^{-2}$ of the best
observed variational energy. Q2 remains fragmented, with 15--17 endpoint
levels. Q3 at $N=12$ has only three levels and 22 of 24 quenches in the same
near-optimal window. Q3 at $N=10$ has a comparable variational gap, yet
only 1 of 24 quenches reaches that window.

\begin{table*}[t]
\centering
\caption{Central landscape diagnostics on the fixed benchmark instances.
$\Delta_{\mathcal M}$ is zero when the physical ground state is known to lie in
the ansatz and otherwise denotes the best observed variational energy minus
$E_0$. $F_{\max}$ is the best ground-state fidelity found by direct fidelity
maximization (unity for Q1 by construction). $K_E$ is the number of distinct
endpoint energy levels from 24 standardized local quenches,
$p_{0.01}$ is the fraction ending within $10^{-2}$ of the best observed
variational energy, $\kappa_{\rm act}$ is the active Hessian condition number at
the best sampled minimum, and $g_{\rm RMS}$ is the median over 64 uniform
random parameter vectors.}
\label{tab:landscape-main}
\small
\begin{tabular}{llrrrrrr}
\toprule
Model & $N$ & $\Delta_{\mathcal M}$ & $F_{\max}$ & $K_E$ & $p_{0.01}$ & $\kappa_{\rm act}$ & med. $g_{\rm RMS}$ \\
\midrule
Q1 & 12 & 0 & 1 & 14 & 0.083 & 3.65 & 0.837 \\
Q1 & 14 & 0 & 1 & 22 & 0.042 & 3.73 & 0.881 \\
Q1 & 16 & 0 & 1 & 19 & 0.083 & 3.79 & 0.902 \\
Q2 & 10 & 0.0255 & 0.461 & 17 & 0.125 & $1.69\times10^3$ & 0.627 \\
Q2 & 12 & 0.00425 & 0.999 & 15 & 0.167 & $3.98\times10^4$ & 0.620 \\
Q3 & 8 & 1.229 & 0.505 & 5 & 0.250 & $6.96\times10^2$ & 0.716 \\
Q3 & 10 & 1.243 & 0.436 & 12 & 0.042 & $2.55\times10^2$ & 0.710 \\
Q3 & 12 & 0.895 & 0.419 & 3 & 0.917 & $1.46\times10^3$ & 0.697 \\
\bottomrule
\end{tabular}%

\end{table*}

\begin{figure*}[htpb]
    \centering
    \includegraphics[width=0.86\textwidth]{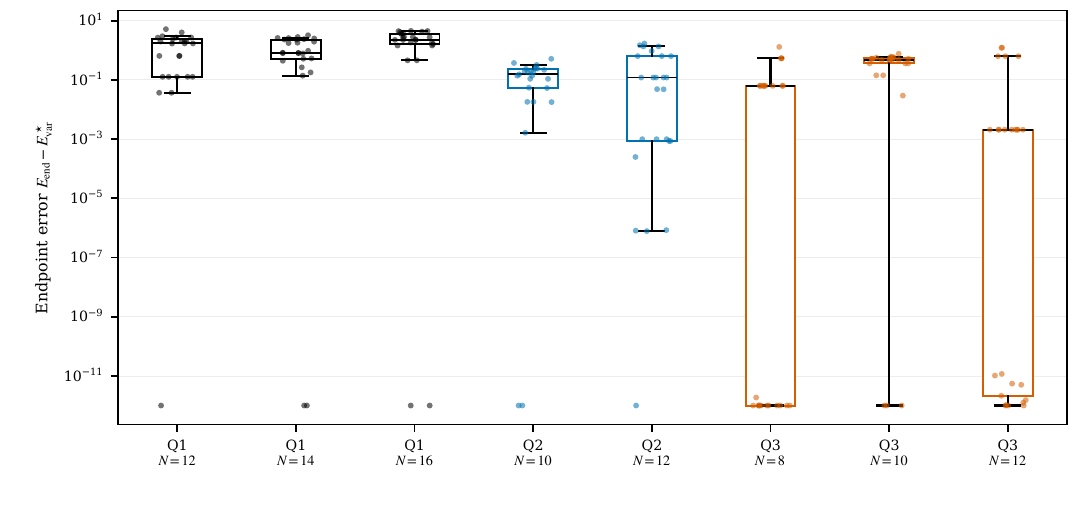}
    \caption{Endpoint-error distributions from the 25 archived maximum-budget
    BFGS runs for each model--size condition. Errors are measured relative to
    the corresponding saved variational reference. This archive-based view is
    complementary to the independent standardized-quench statistics in
    Table~\ref{tab:landscape-main}.}
    \label{fig:archived-bfgs-endpoints}
\end{figure*}

The independent standardized quenches are complemented by the actual
maximum-budget BFGS endpoint populations in
Fig.~\ref{fig:archived-bfgs-endpoints}. The archive-based distributions make
the same model dependence visible directly: Q3 at $N=12$ has
$p_{0.01}=0.80$, whereas the Q1 conditions have $p_{0.01}=0.04$--$0.08$ and
Q3 at $N=10$ has $p_{0.01}=0.12$. These archived endpoints are used only as a
complementary view and are distinct from the 24 standardized L-BFGS-B quenches
reported in Table~\ref{tab:landscape-main}.

\begin{figure*}[htpb]
    \centering
    \includegraphics[width=0.90\textwidth]{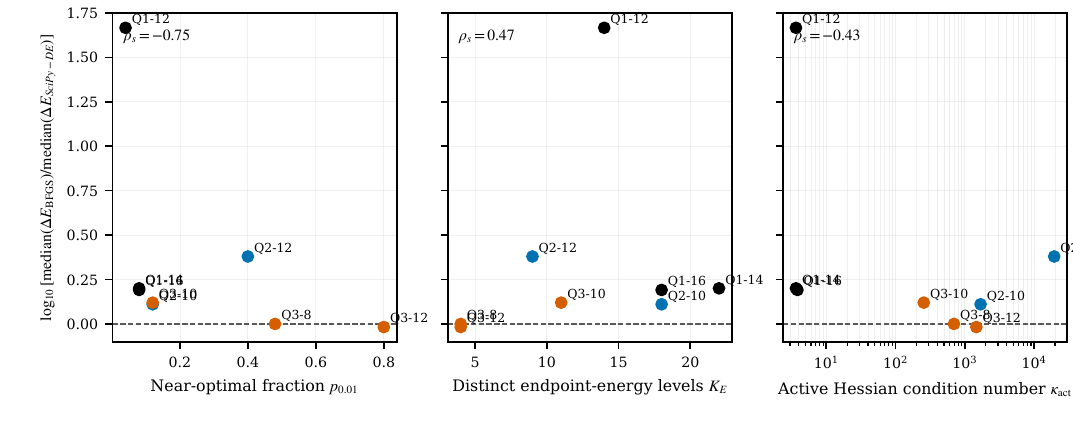}
    \caption{Direct relation between $30\,000$-FE BFGS/SciPy-DE performance
    and landscape descriptors across the eight fixed model--size conditions.
    The vertical axis is
    $\log_{10}\operatorname{median}(\Delta E_{\mathrm{BFGS}})/
    \operatorname{median}(\Delta E_{\text{SciPy-DE}})$; positive values favor
    SciPy-DE and negative values favor BFGS. The $p_{0.01}$ and $K_E$ values in
    this figure are computed from archived maximum-budget BFGS endpoints.}
    \label{fig:bfgs-descriptor-scatter}
\end{figure*}

Figure~\ref{fig:bfgs-descriptor-scatter} compares BFGS and SciPy-DE at
$30\,000$ FEs. Relative performance is most strongly associated with the
archive-derived near-optimal endpoint fraction ($\rho_s=-0.75$); the
associations with $K_E$ ($\rho_s=0.47$) and $\kappa_{\rm act}$
($\rho_s=-0.43$) are weaker. The standardized-quench rank analysis gives the
same trend. These eight-condition correlations are descriptive.

The curvature and gradient diagnostics separate multimodality from global
flatness. Q1 minima are well conditioned,
$\kappa_{\rm act}\simeq3.6$--$3.8$, while the best sampled Q2/Q3 minima reach
$\kappa_{\rm act}\simeq2.6\times10^2$--$4.0\times10^4$. The active
off-diagonal Hessian-coupling fraction rises from $0.20$--$0.27$ for Q1 to
approximately $0.69$--$0.83$ for Q2/Q3
(Appendix~\ref{app:landscape-curvature}). Median random-point $g_{\rm RMS}$
remains 0.620--0.902, and none of the 512 sampled points has
$g_{\rm RMS}<10^{-3}$. At these sizes the difficult landscapes contain
competing basins and anisotropic, coupled valleys with no evidence of a globally vanishing-gradient regime over the studied sizes \cite{AnschuetzKiani2022}.

\paragraph{Physical reachability and spectral content}

Physical reachability varies strongly across the HEA instances. Direct fidelity
maximization gives $F_{\max}=0.461$ for Q2 at $N=10$ and $0.9986$ at $N=12$.
For $N=10$, the physical spectral gap is $5.93\times10^{-3}$ and the continuation state
has weights 0.410 and 0.523 on the ground and first excited states. At $N=12$,
the smaller gap $2.06\times10^{-3}$ coexists with a continuation-state
ground-state weight of 0.9978. The gap alone therefore does not determine the
fidelity of a low-energy variational state.

For Q3, the estimated variational gaps are $0.895$--$1.243$, and direct
fidelity maximization gives only
$F_{\max}=0.419$--$0.505$. At $N=10$, the best standardized energy minimum has
ground-state fidelity $1.7\times10^{-3}$ although direct fidelity optimization
reaches 0.436. Energy minimization and state-overlap optimization can therefore
select different regions of the restricted manifold. The strong BFGS result
for Q3 at $N=12$ concerns optimization of that restricted variational
objective.

\paragraph{Controlled parameter interpolation}

\begin{figure*}[htpb]
    \centering
    \includegraphics[width=0.75\textwidth]{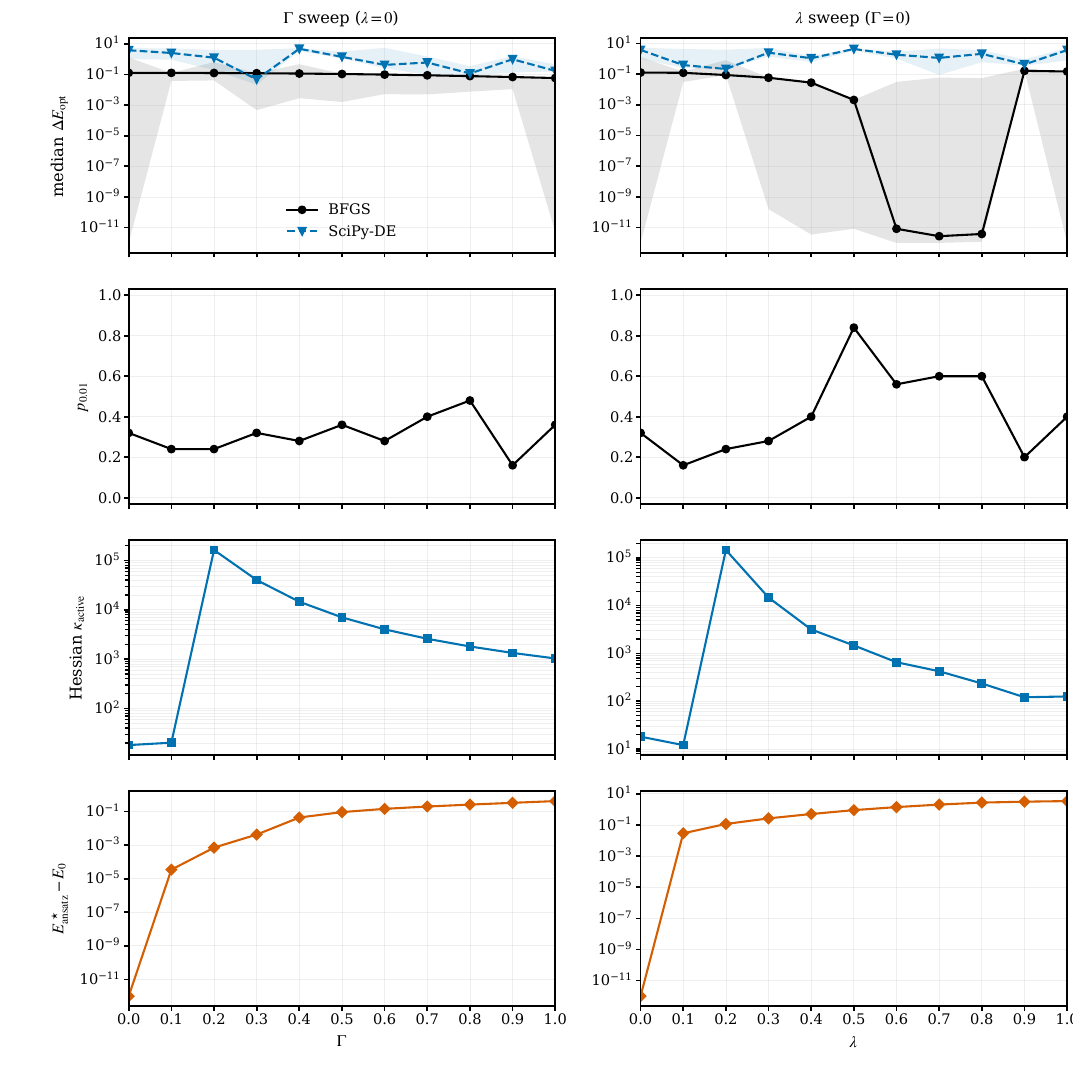}
    \caption{Controlled one-dimensional parameter sweeps for the fixed
    $N=12$ disorder realization and shallow $R_y$--CNOT--$R_y$ ansatz.
    Left: transverse-field cut $(\Gamma,0)$. Right: anisotropic-interaction
    cut $(0,\lambda)$. From top to bottom: median optimization error relative
    to the best-known variational reference, near-optimal BFGS-quench
    fraction $p_{0.01}$, active Hessian condition number, and variational
    gap $E_{\mathrm{ref}}-E_0$.}
    \label{fig:parameter-sweeps}
\end{figure*}

Figure~\ref{fig:parameter-sweeps} follows the transverse-field cut
$(\Gamma,0)$ and anisotropic-interaction cut $(0,\lambda)$ at fixed $N=12$,
disorder, and ansatz. Along the $\Gamma$ cut,
$E_{\mathrm{ref}}-E_0$ rises from zero to 0.420 while the median BFGS
optimization error decreases from $1.27\times10^{-1}$ to
$5.81\times10^{-2}$. Ground-state fidelity remains near unity through
$\Gamma=0.3$ ($F=0.9978$) and drops sharply at $\Gamma=0.4$. Physical
reachability and optimization difficulty therefore evolve differently along
this cut.

The $\lambda$ cut is nonmonotonic. The variational gap grows to 3.41 at
$\lambda=1$, while BFGS reaches median $\Delta E_{\mathrm{opt}}$ of order
$10^{-12}$ for $\lambda=0.6$--$0.8$. The near-optimal quench fraction peaks at
$p_{0.01}=0.84$ for $\lambda=0.5$, remains 0.56--0.60 for
$\lambda=0.6$--$0.8$, and then decreases. Stronger interaction therefore does
not imply harder classical optimization on this fixed variational manifold.

Both cuts show a peak in local anisotropy near interaction strength $0.2$,
where $\kappa_{\rm act}=\mathcal{O}(10^5)$, followed by lower conditioning
ratios at stronger coupling. Curvature, basin accessibility, and physical
reachability therefore provide complementary information across the sweep.

\section{Discussion}
\label{sec:discussion}

The cross-model results connect optimizer reversals to changes in the
Hamiltonian--ansatz landscape. Q1 combines exact reachability with many
separated energetic endpoints and a small near-optimal basin fraction. Q2
remains fragmented and strongly anisotropic, whereas Q3 at $N=12$ has a large near-optimal basin. The controlled $\Gamma$ and $\lambda$
sweeps further show that reachability, curvature, and basin accessibility vary
differently as the Hamiltonian is changed. Classical optimization difficulty
is therefore not determined by the variational gap alone.

The Q3 depth study extends this separation from Hamiltonian variation to
ansatz variation. Increasing the entangling-block depth substantially improves
the best-found ground-state energy and fidelity, while the Schmidt spectra give
necessary depth thresholds independent of optimizer dynamics. The auxiliary
$N=4,6$ controls show that satisfying the Schmidt-rank condition does not imply
immediate exact recovery by the restricted $R_y$--CNOT architecture. For the
benchmark instances, deeper circuits retain finite random-point gradients but
develop stronger parameter coupling and Hessian condition numbers reaching
$10^4$--$10^5$. Increased variational capacity therefore does not simply make
the optimization landscape uniformly flatter; it changes reachability and
local geometry simultaneously.

The FE-matched multistart BFGS control sharpens the optimizer interpretation.
Repeated local restarts use the same total FE budgets as the population
methods and reach the exact Q1 reference and the saved Q3 variational
reference to numerical precision, while Q2 improves to median physical
ground-state errors of $7.75\times10^{-3}$ and $4.25\times10^{-3}$ at
$10\,000$ and $30\,000$ FEs. The population-method advantage in the
single-start benchmark therefore reflects, in part, the basin coverage
obtained within the FE budget. Independent local restarts provide another
mechanism for obtaining this coverage on deterministic objectives.

This conclusion is specific to the exact-objective numerical-gradient
implementation. Independent additive objective noise produces abrupt
degradation of MS-BFGS, consistent with amplification of objective noise by
finite-difference gradient estimation. The noise control should not be
interpreted as a general robustness comparison between population methods and
quasi-Newton optimization; analytic-gradient, stochastic-quasi-Newton, and
finite-shot-aware methods define different optimization regimes~\cite{Scriva2024ShotNoise}.

The observed behavior is consistent with prior VQE studies in which DE escapes
local minima that trap local optimizers \cite{Failde2023} and with broader
comparisons showing strong problem and budget dependence \cite{Jones2025}.
It also limits direct transfer from generic continuous-optimization
benchmarks: iL-SHADE, jSO, and L-SRTDE are strong CEC solvers
\cite{Brest2016iLSHADE,Brest2017jSO,Stanovov2024LSRTDE,Novak2026CEC}, but
their ordering changes across the present VQE landscapes.

The conclusions remain finite-size and instance specific. The main optimizer
benchmark uses selected disorder realizations rather than a disorder ensemble,
and the depth study is restricted to Q3 and to the tested
hardware-efficient architecture. Exact-statevector evaluation excludes
realistic finite-shot measurement statistics and hardware errors; the
additive-Gaussian control isolates numerical sensitivity rather than a
complete measurement model. Broader disorder ensembles, structured ansatze,
and noise-aware or analytic-gradient optimization are therefore the main tests
of the generality of the reported landscape--optimizer relations.

\section{Conclusion}

Across a matched hierarchy of frustrated spin models, optimizer rankings vary with basin accessibility, local curvature, and the physical reachability of the restricted variational manifold.
 Q1 combines exact reachability with strong
multimodality; Q2 combines fragmentation with strong Hessian anisotropy and
budget-dependent optimizer crossovers; Q3 at $N=12$ has a broad near-optimal
local-descent catchment. Controlled Hamiltonian sweeps show that these
landscape descriptors evolve differently from the variational gap.

The Q3 ansatz depth study shows that increasing variational capacity improves
ground-state reachability while producing more anisotropic and coupled local
geometry rather than a finite-size collapse of the gradient scale, consistent
with the broader observation that structured variational solutions can retain
favorable local landscapes with nonvanishing gradients~\cite{Mele2022Smooth}.
Schmidt-rank constraints provide necessary depth thresholds but do not
guarantee exact recovery by the restricted hardware-efficient architecture.
FE-matched multistart BFGS further shows that basin coverage within a fixed
evaluation budget explains much of the apparent local/global optimizer
contrast on exact objectives. Its rapid degradation under additive objective
noise restricts this conclusion to the numerical-gradient deterministic
regime. Together, the results identify ansatz capacity, basin accessibility,
and local landscape geometry as distinct controls of VQE performance.

\begin{acknowledgments}
This project has received funding from the Research Council of Lithuania (LMTLT), agreement No. P-ITP-24-9. This research was also supported by research grant SGS No. SP2026/063 of VSB--Technical University of Ostrava, Czech Republic, and by the Ministry of Education, Youth and Sports of the Czech Republic through e-INFRA CZ (ID:90254).
\end{acknowledgments}

\section*{Data availability}
The numerical dataset is archived on Zenodo~\cite{novak_2026_22210896}. The source
code required to reproduce the numerical experiments is available at
\url{https://github.com/VojtechNovak/BenchmarkingFrustrated}.

\section*{Competing interests}
The authors declare no known competing financial interests or personal relationships that could have influenced the work reported here.

\appendix
\section{Additional characterization of the optimization landscapes}
\label{app:landscape-characterization}

This appendix gives the detailed diagnostics supporting
Section~\ref{sec:results_landscape}. All calculations use the same fixed
disorder realizations as the benchmark. Q1 denotes the diagonal Ising glass,
Q2 its transverse-field extension, and Q3 the anisotropic Heisenberg glass.

\subsection{Standardized local-quench and reachability protocol}
\label{app:standardized-quench}

The primary basin statistic in Section~\ref{sec:results_landscape} is independent
of the benchmark archives. For each model--size condition, 24 uniform parameter
vectors are quenched with the same L-BFGS-B local-descent rule using exact
gradients. Energies separated by at most $10^{-6}$ are merged into one endpoint
energy level. Parameter-space clustering is additionally performed using a
periodic RMS distance threshold of 0.5 radians, but energy-level multiplicity is
used for the main interpretation because ansatz symmetries can generate many
parameter-distinct representations of the same variational energy. Direct
fidelity maximization uses 32 independent starts for each Q2/Q3 condition. The 32-probe gradient statistics in this appendix are the retained
archive-based diagnostics and are distinct from the 64 independent
random-gradient probes reported in Table~\ref{tab:landscape-main}.

\begin{table}[htpb]
\centering
\caption{Independent standardized local-quench statistics. $K_E$ is the number
of endpoint energy levels among 24 starts; $p_{0.01}$ and $p_{0.1}$ are the
fractions within the stated energy distance of the best observed variational
minimum.}
\label{tab:app-standardized-quench}
\small
\begin{tabular}{llrrrr}
\toprule
Model & $N$ & $K_E$ & $p_{0.01}$ & $p_{0.1}$ & best $E-E_0$ \\
\midrule
Q1 & 12 & 14 & 0.083 & 0.208 & $\approx0$ \\
Q1 & 14 & 22 & 0.042 & 0.083 & 0.139$^\dagger$ \\
Q1 & 16 & 19 & 0.083 & 0.083 & $\approx0$ \\
Q2 & 10 & 17 & 0.125 & 0.500 & 0.0255 \\
Q2 & 12 & 15 & 0.167 & 0.333 & 0.00425 \\
Q3 & 8 & 5 & 0.250 & 0.833 & 1.229 \\
Q3 & 10 & 12 & 0.042 & 0.250 & 1.243 \\
Q3 & 12 & 3 & 0.917 & 0.917 & 0.895 \\
\bottomrule
\end{tabular}
\par\vspace{1mm}
\footnotesize $^\dagger$The Q1 $N=14$ ansatz still contains the exact physical ground state by
construction; the 24 standardized quenches simply did not sample its basin.
\end{table}

\subsection{Post-processing protocol and metric definitions}
\label{app:landscape-methods}

The original archive-based post-processing is retained below as a complementary
view. It used analysis seed 20260823, 32 random gradient probes, 64 global
structure probes, and two low-energy Hessian representatives per condition.
These archive-based quantities are distinct from the independent standardized
quench statistics above and no benchmark optimizer was rerun.

For Q1, the complete diagonal energy landscape is available by enumeration.
A bit string $s$ is a one-spin-flip minimum when its energy is no larger than
that of every Hamming-distance-one neighbor. Consequently, the Q1
multimodality count is exact. An additional exact geometric basin measure is
obtained by deterministic steepest one-spin-flip descent from every basis
configuration. The primary basin-frequency statistic nevertheless uses the
already archived maximum-budget BFGS runs: each final Q1 BFGS energy is mapped
to the corresponding exact one-spin-flip minimum.

For Q2 and Q3, multimodality is necessarily an empirical archive-based
quantity in this inexpensive post-processing analysis. All archived
maximum-budget optimizer endpoints are greedily clustered jointly in energy
and periodic parameter space. Two endpoints can belong to the same cluster
only when their energy difference is below
\[
 \varepsilon_E =
 \max\!\left(10^{-5},
 10^{-4}\max(1,|E_{\rm best}|)\right)
\]
and their periodic root-mean-square parameter distance is no larger than
$0.5$ radians. Basin frequencies are then evaluated using only the archived
maximum-budget BFGS endpoints. The Q2/Q3 cluster counts therefore describe
\emph{observed low-energy endpoint/basin structure at the stated resolution};
they are not an exhaustive count of stationary points.

For basin probabilities $p_i$, we report the ground/best-basin frequency
$p_0$, the complementary basin difficulty
\[
 D_{\rm basin}=1-p_0,
\]
the normalized basin entropy
\[
 \widetilde H =
 \frac{-\sum_i p_i\log p_i}{\log K},
\]
where $K$ is the number of occupied basins, and the effective basin count
$K_{\rm eff}=\exp(-\sum_i p_i\log p_i)$. For Q1 the degeneracy entries are
exact multiplicities of one-spin-flip minimum energy levels. For Q2/Q3 they
are empirical multiplicities of distinct parameter-space basin clusters that
share the same endpoint energy level within $\varepsilon_E$; they are
therefore not Hamiltonian spectral degeneracies.

Conditioning is characterized from the Hessian eigenvalues at low-energy
representatives. Active eigenvalues satisfy
\[
 |\lambda| >
 \max\!\left(10^{-8},
 10^{-6}\max(1,\max_j|\lambda_j|)\right),
\]
and the active condition number is
$\kappa_{\rm act}=\lambda_{\max}^+/\lambda_{\min}^+$ over active positive
eigenvalues. A representative is treated as a stationary-minimum proxy when
its RMS gradient is below $10^{-4}$ and no active negative Hessian eigenvalue
is present. For Q2 and Q3 the derivatives are evaluated by exact
parameter-shift identities; Q1 uses its analytic objective derivatives.

Separability is probed with one deterministic random Hessian per condition.
We report the off-diagonal Frobenius ratio
\[
 R_{\rm off} =
 \frac{\|H-\operatorname{diag}(H)\|_F}{\|H\|_F}
\]
and the density of active off-diagonal couplings. A coupling $H_{ij}$ is
active when
\[
 |H_{ij}| >
 \max\!\left(10^{-8},10^{-3}\max_{kl}|H_{kl}|\right).
\]
Trainability is summarized by
$g_{\rm RMS}=\|\nabla E\|_2/\sqrt D$ over random parameters. Finally, global
structure is characterized in two ways: (i) the Spearman correlation between
Hamming distance and the mean energy of each diagonal-energy Hamming shell,
and (ii) the Spearman correlation between the full variational energy and the
periodic RMS parameter distance from the saved low-energy anchor. For Q2 and
Q3 the Hamming diagnostic refers only to the shared classical diagonal
backbone; the parameter-distance diagnostic evaluates the complete quantum
objective.

\subsection{Multimodality, basin structure, and degeneracy}
\label{app:landscape-basins}

Table~\ref{tab:app-topology} summarizes the primary topology statistics.
For Q1, the exact number of one-spin-flip minima grows from 28 at $N=12$ to 72
at $N=14$ and 130 at $N=16$. The archived BFGS runs reached the exact
ground-state basin in only $4$--$8\%$ of runs, giving
$D_{\rm basin}=0.92$--$0.96$. The high normalized basin entropies
($0.943$--$0.981$) show that the BFGS endpoints were distributed across many
competing minima rather than being dominated by a single attraction basin.

For Q2 and Q3, the empirical endpoint clustering is highly fragmented:
164--188 distinct clusters are observed among 200 archived maximum-budget
endpoints. At the chosen clustering resolution, all 25 archived BFGS endpoints
occupy distinct clusters in every Q2/Q3 condition, giving
$\widetilde H=1$ and $K_{\rm eff}=25$. The best observed basin is reached by
BFGS with frequency $0$--$4\%$. These values demonstrate endpoint dispersion
at the stated energy/parameter resolution; they should not be interpreted as
an exact topological count of continuous local minima.

\begin{table*}[t]
\centering
\caption{Landscape topology and archived-BFGS basin statistics. $M_{\rm obs}$
is exact for Q1 and is the number of empirical endpoint clusters for Q2/Q3.
The degeneracy columns are exact minimum-energy-level multiplicities for Q1
but empirical same-energy basin multiplicities for Q2/Q3. $\Delta E_1$ is
the separation between the lowest and first distinct excited minimum/basin
energy levels, not a Hamiltonian spectral gap.}
\label{tab:app-topology}
\setlength{\tabcolsep}{3.0pt}
\renewcommand{\arraystretch}{0.94}
\scriptsize
\begin{tabular}{llrrrrrrrr}
\toprule
Model & $N$ & $M_{\rm obs}$ & $p_0$ & $D_{\rm basin}$ &
$\widetilde H$ & $K_{\rm eff}$ & $g_0$ & $g_{\rm exc,max}$ &
$\Delta E_1$ \\
\midrule
Q1 & 12 & 28 & 4.0\% & 96.0\% & 0.943 & 12.04 & 1 & 1 & 0.0368 \\
Q1 & 14 & 72 & 8.0\% & 92.0\% & 0.981 & 20.73 & 1 & 1 & 0.139 \\
Q1 & 16 & 130 & 8.0\% & 92.0\% & 0.980 & 17.92 & 1 & 1 & 0.458 \\
Q2 & 10 & 188 & 4.0\% & 96.0\% & 1.000 & 25.00 & 7 & 8 & $1.61\times 10^{-3}$ \\
Q2 & 12 & 185 & 4.0\% & 96.0\% & 1.000 & 25.00 & 23 & 13 & $3.99\times 10^{-3}$ \\
Q3 & 8 & 164 & 0.0\% & 100.0\% & 1.000 & 25.00 & 29 & 36 & $1.17\times 10^{-3}$ \\
Q3 & 10 & 188 & 0.0\% & 100.0\% & 1.000 & 25.00 & 6 & 13 & 0.0205 \\
Q3 & 12 & 182 & 4.0\% & 96.0\% & 1.000 & 25.00 & 19 & 19 & $2.15\times 10^{-3}$ \\
\bottomrule
\end{tabular}%
\end{table*}

The exact Q1 basin geometry, which does not depend on any optimizer endpoint,
is shown separately in Table~\ref{tab:app-q1-geometry}. The fraction of all
basis states flowing to the ground-state basin decreases monotonically from
$11.6\%$ at $N=12$ to $3.9\%$ at $N=16$, while the effective number of
geometric basins increases from 18.6 to 70.3. Hence the reduction in the
ground basin is visible directly in the discrete landscape and is not an
artifact of BFGS behavior.

\begin{table}[t]
\centering
\caption{Exact Q1 discrete steepest-one-flip basin geometry. The final column
shows the archived-BFGS ground-basin frequency for comparison.}
\label{tab:app-q1-geometry}
\small
\begin{tabular}{rrrrrr}
\toprule
$N$ & minima & $p_0^{\rm geom}$ & $\widetilde H_{\rm geom}$ &
$K_{\rm eff}^{\rm geom}$ & $p_0^{\rm BFGS}$ \\
\midrule
12 & 28 & 11.6\% & 0.877 & 18.56 & 4.0\% \\
14 & 72 & 7.3\% & 0.866 & 40.52 & 8.0\% \\
16 & 130 & 3.9\% & 0.874 & 70.30 & 8.0\% \\
\bottomrule
\end{tabular}
\end{table}

The exact Q1 minimum energies are non-degenerate for all three sizes:
$g_0=g_{\rm exc,max}=1$. This is consistent with the use of small longitudinal
fields to break trivial discrete degeneracies. By contrast, Q2/Q3 exhibit
several distinct parameter-space basin representatives at the same empirical
energy level. For example, the lowest empirical level contains 23 distinct
basin clusters for Q2 at $N=12$ and 29 for Q3 at $N=8$. These multiplicities
most naturally quantify repeated parameter-space representations and
near-equivalent low-energy basins; no claim of physical eigenstate degeneracy
is made.

\subsection{Local curvature and parameter coupling}
\label{app:landscape-curvature}

The curvature diagnostics separate Q1 sharply from the two non-commuting
models. All 16 low-energy Hessian probes satisfy the stationary-minimum proxy
criterion, and none contains an active negative eigenvalue. Q1 remains
well-conditioned at the sampled minima, with median active condition numbers
between 3.65 and 3.83. Q2 is substantially more anisotropic:
$\kappa_{\rm act}\approx1.69\times10^3$ at $N=10$ and
$1.95\times10^4$ at $N=12$. Q3 is also strongly ill-conditioned, with median
values from $2.55\times10^2$ to $1.46\times10^3$ over the studied sizes. The
large condition numbers arise from small positive-curvature directions rather
than from negative curvature at the selected low-energy representatives.

The random-probe Hessians also show a marked increase in interaction density.
For Q1, the number of active off-diagonal couplings is exactly $18/66$,
$21/91$, and $24/120$ for $N=12,14,16$, respectively, reflecting the sparse
degree-three interaction graph. Q2 and Q3 are much denser, with active-coupling
fractions of approximately $0.69$--$0.83$. Thus the quantum extensions combine
strong curvature anisotropy with substantially less separable parameter
interactions.

\begin{table*}[t]
\centering
\caption{Conditioning and Hessian-coupling summaries. The eigenvalue columns
are medians over the two low-energy probes. ``Stat.'' is the number of
stationary-minimum proxies among the two probes; ``neg.'' is the total number
of active negative eigenvalues. $R_{\rm off}$ and coupling density are measured
at one deterministic random parameter probe.}
\label{tab:app-curvature}
\setlength{\tabcolsep}{3.0pt}
\renewcommand{\arraystretch}{0.94}
\scriptsize
\begin{tabular}{llrrrrrrrr}
\toprule
Model & $N$ & med. $\lambda_{\min}$ & med. $\lambda_{\max}$ &
med. $\kappa_{\rm act}$ & Stat. & neg. & $R_{\rm off}$ &
active edges & density \\
\midrule
Q1 & 12 & 0.868 & 3.17 & 3.65 & 2/2 & 0 & 0.912 & 18/66 & 0.273 \\
Q1 & 14 & 0.86 & 3.14 & 3.65 & 2/2 & 0 & 0.662 & 21/91 & 0.231 \\
Q1 & 16 & 0.822 & 3.15 & 3.83 & 2/2 & 0 & 0.859 & 24/120 & 0.200 \\
Q2 & 10 & $3.41\times 10^{-3}$ & 5.74 & $1.69\times 10^{3}$ & 2/2 & 0 & 0.823 & 144/190 & 0.758 \\
Q2 & 12 & $8.33\times 10^{-4}$ & 5.52 & $1.95\times 10^{4}$ & 2/2 & 0 & 0.778 & 196/276 & 0.710 \\
Q3 & 8 & $8.37\times 10^{-3}$ & 5.83 & 696 & 2/2 & 0 & 0.876 & 100/120 & 0.833 \\
Q3 & 10 & 0.0204 & 5.21 & 255 & 2/2 & 0 & 0.893 & 131/190 & 0.689 \\
Q3 & 12 & $2.84\times 10^{-3}$ & 4.16 & $1.46\times 10^{3}$ & 2/2 & 0 & 0.878 & 190/276 & 0.688 \\
\bottomrule
\end{tabular}%
\end{table*}

For completeness, Table~\ref{tab:app-hessian-probes} reports the individual
low-energy probes rather than only the condition-level medians. This is
particularly important for Q2 at $N=12$, where the two sampled representatives
have active condition numbers $3.25\times10^3$ and $3.58\times10^4$, showing
substantial variation even within the low-energy region.

\begin{table*}[t]
\centering
\caption{Individual low-energy Hessian probes. $g_{\rm RMS}$ is the RMS
gradient at the probe. All listed probes satisfy $g_{\rm RMS}<10^{-4}$ and have
no active negative Hessian eigenvalues.}
\label{tab:app-hessian-probes}
\setlength{\tabcolsep}{3.0pt}
\renewcommand{\arraystretch}{0.92}
\scriptsize
\begin{tabular}{lllrrrrr}
\toprule
Model & $N$ & probe & $E$ & $g_{\rm RMS}$ & $\lambda_{\min}$ &
$\lambda_{\max}$ & $\kappa_{\rm act}$ \\
\midrule
Q1 & 12 & minimum\_0 & -12.341270 & $1.91\times 10^{-16}$ & 0.861 & 3.14 & 3.65 \\
Q1 & 12 & minimum\_1 & -12.304496 & $2.09\times 10^{-16}$ & 0.876 & 3.2 & 3.65 \\
Q1 & 14 & minimum\_0 & -13.480440 & $2.16\times 10^{-16}$ & 0.881 & 3.14 & 3.57 \\
Q1 & 14 & minimum\_1 & -13.341080 & $2.10\times 10^{-16}$ & 0.84 & 3.13 & 3.73 \\
Q1 & 16 & minimum\_0 & -16.228986 & $1.65\times 10^{-16}$ & 0.837 & 3.17 & 3.79 \\
Q1 & 16 & minimum\_1 & -15.771014 & $2.14\times 10^{-16}$ & 0.806 & 3.13 & 3.88 \\
Q2 & 10 & bfgs\_basin\_0 & -9.575173 & $2.74\times 10^{-7}$ & $3.41\times 10^{-3}$ & 5.74 & $1.69\times 10^{3}$ \\
Q2 & 10 & saved\_or\_best\_anchor & -9.575173 & $1.95\times 10^{-7}$ & $3.41\times 10^{-3}$ & 5.74 & $1.69\times 10^{3}$ \\
Q2 & 12 & bfgs\_basin\_0 & -12.698730 & $2.19\times 10^{-6}$ & $1.73\times 10^{-4}$ & 6.2 & $3.58\times 10^{4}$ \\
Q2 & 12 & saved\_or\_best\_anchor & -12.697867 & $3.28\times 10^{-7}$ & $1.49\times 10^{-3}$ & 4.84 & $3.25\times 10^{3}$ \\
Q3 & 8 & empirical\_basin\_0 & -9.160190 & $4.78\times 10^{-8}$ & $8.37\times 10^{-3}$ & 5.83 & 696 \\
Q3 & 8 & saved\_or\_best\_anchor & -9.160190 & $1.38\times 10^{-7}$ & $8.37\times 10^{-3}$ & 5.83 & 696 \\
Q3 & 10 & empirical\_basin\_0 & -10.745266 & $3.97\times 10^{-8}$ & 0.0204 & 5.21 & 255 \\
Q3 & 10 & saved\_or\_best\_anchor & -10.745266 & $1.32\times 10^{-7}$ & 0.0204 & 5.21 & 255 \\
Q3 & 12 & bfgs\_basin\_0 & -14.242884 & $1.90\times 10^{-7}$ & $2.84\times 10^{-3}$ & 4.16 & $1.46\times 10^{3}$ \\
Q3 & 12 & saved\_or\_best\_anchor & -14.242884 & $2.37\times 10^{-7}$ & $2.84\times 10^{-3}$ & 4.16 & $1.46\times 10^{3}$ \\
\bottomrule
\end{tabular}%
\end{table*}

For Q1 at \(N=14\), the analytic structure permits an exhaustive curvature check over all 72 exact one-spin-flip minima. The active Hessian condition number has median \(3.64\) and ranges from \(1.34\) to \(3.74\). The two low-energy probes are therefore representative of the complete Q1 minimum set; Q1 difficulty is dominated by global multimodality, with only mild local ill-conditioning.

\subsection{Trainability and global structure}
\label{app:landscape-global}

The random-parameter gradient distributions do not show a barren-plateau-like
suppression over the system sizes considered here. The median $g_{\rm RMS}$
lies between 0.617 and 0.893 for all eight conditions, and even the 10th
percentile is at least 0.529. None of the 256 sampled parameter vectors has
$g_{\rm RMS}<10^{-3}$. These observations do not establish an asymptotic
scaling law, but within the benchmark sizes they indicate that the difficulty
is not explained by globally vanishing gradients.

The distance diagnostics instead show weak global funnel structure. On the
diagonal backbone, a single spin flip away from the ground configuration raises
the \emph{mean} shell energy by approximately 3.65--4.10 energy units, and even
the best state at Hamming distance one remains 1.65--1.76 units above the
diagonal ground energy. Nevertheless, the Spearman correlation between Hamming
distance and the mean energy across the complete set of Hamming shells is only
0.088--0.167. The immediate ground-state neighborhood is locally uphill, but the ordering
does not extend to a monotonic funnel over the full discrete configuration space.

The continuous parameter-space picture is similarly non-funnel-like. Across
64 random parameter vectors per condition, the Spearman correlation between
full variational energy and periodic RMS distance from the low-energy anchor
ranges only from $-0.152$ to $0.238$, with no consistent sign across model
families or sizes. Consequently, distance from a known good parameter vector
is generally a poor predictor of objective value on the global torus.

\begin{table*}[htpb]
\centering
\caption{Trainability and global-structure diagnostics. Gradient statistics are
based on 32 random parameter vectors per condition.
$\Delta\bar E_{d=1}$ and $\Delta E_{d=1}^{\min}$ refer to the diagonal-energy
Hamming shell one spin flip from the classical ground configuration. $\rho_H$
is the shell-mean energy/Hamming-distance Spearman coefficient, and
$\rho_\theta$ is the full objective energy/periodic-parameter-distance
coefficient from 64 random points.}
\label{tab:app-trainability-global}
\setlength{\tabcolsep}{3.0pt}
\renewcommand{\arraystretch}{0.94}
\scriptsize
\begin{tabular}{llrrrrrrrr}
\toprule
Model & $N$ & med. $g_{\rm RMS}$ & P10 & P90 &
$P(g_{\rm RMS}<10^{-3})$ &
$\Delta\bar E_{d=1}$ & $\Delta E_{d=1}^{\min}$ &
$\rho_H$ & $\rho_\theta$ \\
\midrule
Q1 & 12 & 0.802 & 0.561 & 1.105 & 0.0\% & 4.057 & 1.722 & 0.115 & -0.040 \\
Q1 & 14 & 0.893 & 0.732 & 1.061 & 0.0\% & 3.783 & 1.761 & 0.100 & 0.238 \\
Q1 & 16 & 0.877 & 0.734 & 1.025 & 0.0\% & 4.029 & 1.674 & 0.088 & -0.119 \\
Q2 & 10 & 0.629 & 0.529 & 0.732 & 0.0\% & 3.648 & 1.729 & 0.136 & -0.008 \\
Q2 & 12 & 0.617 & 0.537 & 0.730 & 0.0\% & 4.057 & 1.722 & 0.115 & -0.080 \\
Q3 & 8 & 0.727 & 0.601 & 0.904 & 0.0\% & 4.099 & 1.648 & 0.167 & -0.152 \\
Q3 & 10 & 0.741 & 0.567 & 0.925 & 0.0\% & 3.648 & 1.729 & 0.136 & 0.201 \\
Q3 & 12 & 0.720 & 0.547 & 0.848 & 0.0\% & 4.057 & 1.722 & 0.115 & 0.066 \\
\bottomrule
\end{tabular}%
\end{table*}

\subsection{Ansatz reachability and WHRF-inspired diagnostics}
\label{app:reachability-whrf}

Direct fidelity maximization over 32 starts gives best-found physical
ground-state fidelities $(0.461,0.999)$ for Q2 at $N=(10,12)$ and
$(0.505,0.436,0.419)$ for Q3 at $N=(8,10,12)$. For Q2 $N=10$, the stored
continuation state has ground- and first-excited-state \cite{beseda2024state,illesova2025transformation} weights 0.410 and 0.523;
for Q2 $N=12$ the corresponding ground-state weight is 0.9978. For Q3
$N=10$, the best standardized energy minimum has ground-state fidelity
$1.7\times10^{-3}$, although direct fidelity optimization reaches 0.436.

Following the local random-field picture of Anschuetz and Kiani
\cite{AnschuetzKiani2022}, we also evaluate a reverse-light-cone
underparameterization proxy $\gamma_{\rm AK}=\ell/(2m)$ term by term. Median
values lie between approximately 0.007 and 0.024 for all studied conditions,
placing all of them in a strongly underparameterized regime according to this
proxy. We also evaluated the spectrum-based finite-size diagnostic
\begin{equation}
 m_{\rm spec}=\frac{\|H-E_0 I\|_*^2}{\|H-\bar E I\|_F^2},
 \qquad \gamma_{\rm spec}=\frac{D}{2m_{\rm spec}},
\end{equation}
used in recent WHRF numerical analysis \cite{Anschuetz2022CriticalPoints, Michalek2026}. Across the present
conditions $\gamma_{\rm spec}$ ranges from $1.1\times10^{-5}$ to
$5.3\times10^{-3}$ and gives the same qualitative underparameterized label.
Neither proxy quantitatively distinguishes the benign Q3 $N=12$ restricted
landscape. We do not identify the concrete landscapes with Wishart hypertoroidal
random fields: the deterministic $R_y$ and $R_y$--CNOT--$R_y$ ansatze have
not been shown to satisfy the approximate local-scrambling/$t$-design
assumptions required for WHRF convergence. The fact that Q3 $N=12$ is benign
under standardized local descent despite its small $\gamma_{\rm AK}$ further
shows that the proxy is qualitative here rather than a quantitative predictor
of trap severity.

\subsection{Implications for the optimizer benchmark}
\label{app:landscape-implications}

Q1 combines many competing discrete minima with a ground-state basin that
shrinks as $N$ increases. Q2 and Q3 add much stronger local anisotropy and a
dense parameter-interaction graph. Neither the discrete Hamming geometry nor
the continuous parameter torus shows a strong global energy--distance funnel,
while random-point gradients remain of order unity at the tested sizes.

These diagnostics associate optimization difficulty with basin structure,
parameter coupling, and ill-conditioned low-energy valleys. They characterize
the objective functions independently of the main optimizer comparison and do
not indicate a uniformly flat finite-size landscape.

The model-to-model quantities are not all equally exhaustive. Q1 multimodality
and geometric basin measures are exact discrete results, whereas Q2/Q3 basin
counts are empirical endpoint statistics. Conditioning is exhaustive for Q1 at
$N=14$ but sampled at low-energy representatives for Q2/Q3; separability and
random-parameter diagnostics are also sampled. The appendix should therefore
be read as a reproducible structural characterization, not as a complete
stationary-point census.

\section{Q3 ansatz-depth expressivity and landscape diagnostics}
\label{app:q3-depth-study}

This appendix gives the detailed depth study supporting
Section~\ref{sec:results_q3_depth}. For the benchmark instances
$N=8,10,12$, the Hamiltonian, disorder realization, $\lambda=0.5$, and
exact-statevector objective are unchanged from the main benchmark. We also
include auxiliary $N=4,6$ instances generated with the same Q3 construction
rule. These smaller systems are used only as controls and are not part of the
main optimizer benchmark.

For a contiguous cut $c=A\vert B$, local $R_y$ rotations do not change Schmidt
rank, while each entangling block contains exactly one CNOT crossing the cut.
Define the architecture-level Schmidt-rank bound
\begin{equation}
 R_{p,c}=\min\!\left(2^p,2^{\min(|A|,|B|)}\right).
\end{equation}
Every state generated by the depth-$p$ ansatz satisfies
$r_{p,c}\leq R_{p,c}$. If $s_{j,c}$ are the exact ground-state Schmidt
coefficients across cut $c$, then its ground-state fidelity satisfies
\begin{equation}
 |\langle\psi_0|\psi(\boldsymbol{\theta})\rangle|^2
 \leq F_{\mathrm{Sch},c}^{(p)}
 =\sum_{j=1}^{R_{p,c}}s_{j,c}^2 .
\end{equation}
We evaluate all $N-1$ contiguous cuts and report the strongest bound,
\begin{equation}
 F_{\mathrm{Sch}}^{(p)}=\min_c F_{\mathrm{Sch},c}^{(p)} .
\end{equation}
Exact Schmidt ranks use singular values larger than $10^{-10}$, and
$p_{\rm S}$ is the smallest depth for which no contiguous cut excludes the
exact ground state by Schmidt rank. Thus $p\geq p_{\rm S}$ removes this
particular structural obstruction but does not guarantee that the restricted
$R_y$--CNOT manifold contains the exact state.

\subsection{Numerical protocol}

The depth study uses PyADE iL-SHADE for global exploration and SciPy BFGS for
local refinement. iL-SHADE uses its implementation defaults except for bounds
$[-\pi,\pi]^D$, the FE budget, the explicit seed, and initial population
$4D$. Each iL-SHADE endpoint is refined by BFGS using the exact
parameter-shift gradient with $\mathrm{gtol}=10^{-8}$. The FE budgets in
Table~\ref{tab:app-q3-depth-protocol} apply to iL-SHADE; subsequent BFGS
refinement is counted separately and terminates by its convergence test or
maximum iteration count. Fidelity searches minimize
$1-|\langle\psi_0|\psi(\theta)\rangle|^2$ and are used only as a reachability
diagnostic. The depth study is not FE-matched across $p$. Search effort is increased
with variational dimension to obtain strong best-found reachability
estimates. Consequently, the energy- and fidelity-versus-depth curves are
interpreted as reachability diagnostics rather than as controlled
comparisons of optimization efficiency across depths.

\begin{table*}[t]
\centering
\caption{Numerical settings for the Q3 depth study. $N_E$ and $B_E$ are the
number and iL-SHADE FE budget of independent energy searches, $N_F$ and $B_F$
the corresponding fidelity searches, $N_Q$ the number of random-start BFGS
quenches, and $N_G$ the number of random parameter vectors used for gradient
statistics.}
\label{tab:app-q3-depth-protocol}
\small
\begin{tabular}{ccrrrrrrr}
\toprule
$N$ & $p$ & $N_E$ & $B_E$ & $N_F$ & $B_F$ & $N_Q$ & $N_G$ & BFGS maxiter \\
\midrule
8  & $1$--$3$ & 1  & 4500  & 1  & 3500  & 4  & 16 & 450 \\
8  & $4$--$6$ & 3  & 15000 & 2  & 9000  & 8  & 16 & 450 \\
10 & $1$--$3$ & 2  & 8000  & 2  & 7000  & 8  & 20 & 500 \\
10 & $4$--$6$ & 3  & 25000 & 3  & 18000 & 8  & 20 & 500 \\
12 & $1$--$3$ & 2  & 10000 & 2  & 8000  & 8  & 16 & 500 \\
12 & $4$--$6$ & 2  & 30000 & 2  & 20000 & 8  & 16 & 500 \\
4  & $1$--$4$ & 12 & 8000  & 8  & 6000  & 24 & 64 & 800 \\
6  & $1$--$5$ & 12 & 8000  & 8  & 6000  & 24 & 64 & 800 \\
6  & $6$--$7$ & 16 & 12000 & 12 & 10000 & 32 & 64 & 1200 \\
\bottomrule
\end{tabular}
\end{table*}

Before Hessian and QFIM evaluation, the selected best-energy point is polished
with BFGS at $\mathrm{gtol}=10^{-10}$. The final-polish iteration limit is
1200 for $N=10,12$ and for the initial auxiliary scan, 2000 for the $N=6$,
$p=6,7$ extension, and 5000 for the final $N=8$ stationarity repolish. The
benchmark-depth calculations and initial auxiliary scan use base seed
$20260830$; the $N=6$, $p=6,7$ extension uses $20260831$. Derived seeds are
\begin{equation}
 s=(s_0+1009N+9176p+7919q+104729r)\bmod(2^{32}-1),
\end{equation}
where $r$ is the run index and $q$ distinguishes the search or diagnostic
stream.

\subsection{Small-system control and Schmidt threshold}

Table~\ref{tab:app-q3-depth-threshold} summarizes the resulting thresholds.
For the auxiliary $N=4$ system, the Schmidt obstruction disappears at $p=2$,
but the exact ground state is first recovered at $p=3$. At this depth all
iL-SHADE--BFGS energy searches and all random-start BFGS quenches recover the
ground energy to the numerical target, providing a simple positive control for
the optimization procedure.

For $N=6$, the Schmidt obstruction disappears at $p=3$, while the error
continues to decrease over several additional layers. At $p=6$ the best gap is
$1.34\times10^{-7}$, and at $p=7$ the exact ground state is recovered to
numerical precision. The exact solution is nevertheless reached only by a
fraction of independent runs at $p=7$, showing that expressivity and
optimization difficulty remain distinct even for this small system.

\begin{table}[t]
\centering
\caption{Summary of the Q3 depth study. $p_{\rm S}$ is the first depth not
excluded by exact ground-state Schmidt rank, and $p_{\rm exact}$ is the first
tested depth at which the physical ground state was recovered to numerical
precision. A dash indicates that exact recovery was not observed within the
tested depths. The $N=4,6$ instances are auxiliary controls.}
\label{tab:app-q3-depth-threshold}
\begin{tabular}{ccccc}
\toprule
$N$ & Instance & $p_{\rm S}$ & $p_{\rm exact}$ &
$\Delta_{\mathcal M}$ at largest $p$ \\
\midrule
4  & auxiliary & 2 & 3 & $<10^{-14}$ \\
6  & auxiliary & 3 & 7 & $<10^{-14}$ \\
8  & benchmark & 4 & -- & $0.264$ \\
10 & benchmark & 5 & -- & $0.357$ \\
12 & benchmark & 6 & -- & $0.364$ \\
\bottomrule
\end{tabular}
\end{table}

\begin{figure*}[t]
    \centering
    \includegraphics[width=0.82\textwidth]
    {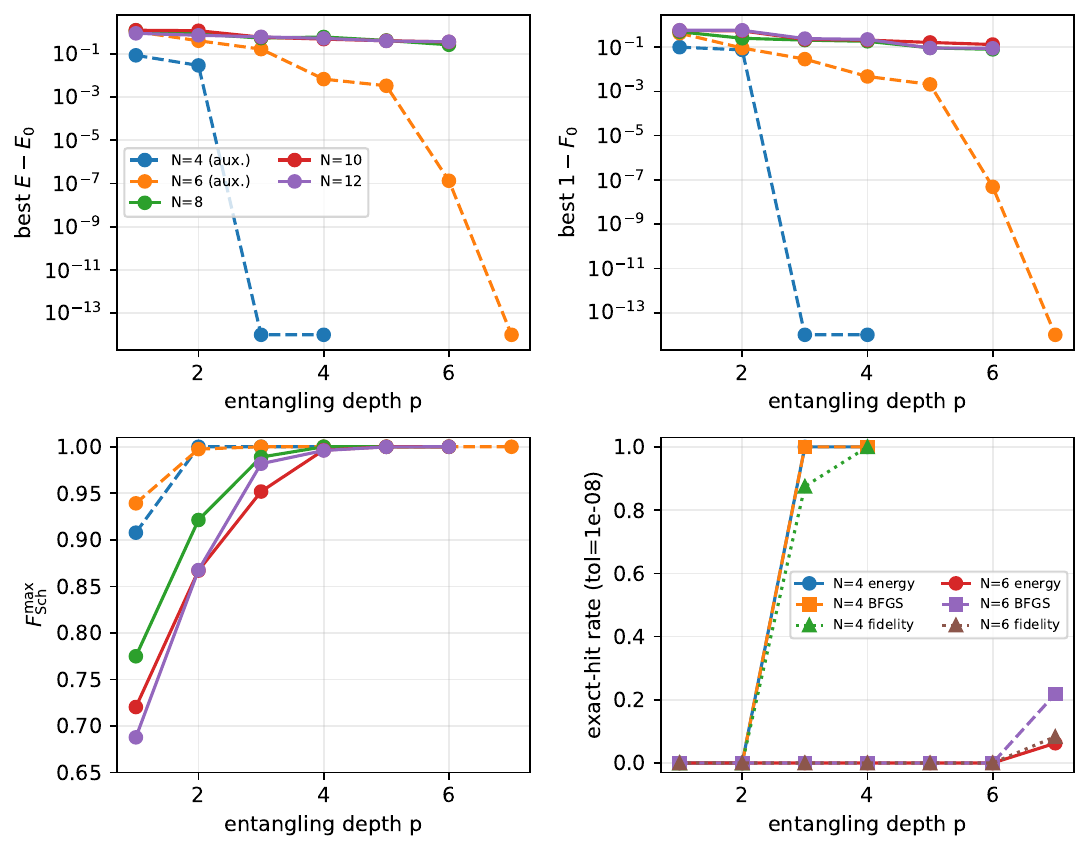}
    \caption{Q3 reachability with increasing hardware-efficient ansatz depth.
    The $N=4,6$ curves are auxiliary small-system controls, while
    $N=8,10,12$ are the benchmark instances. The panels show the best physical
    energy error, ground-state infidelity, the Schmidt-rank fidelity bound,
    and exact-solution hit rates for the auxiliary systems. Schmidt rank gives
    a necessary depth threshold but does not by itself guarantee exact
    representability by the restricted $R_y$--CNOT circuit.}
    \label{fig:app-q3-depth-allN}
\end{figure*}

The combined results make the role of the Schmidt bound explicit. The first
Schmidt-allowed depths are $p_{\rm S}=2,3,4,5,6$ for
$N=4,6,8,10,12$, respectively. Exact recovery occurs later for both auxiliary
systems: one additional layer for $N=4$ and four additional layers for $N=6$.
Thus the Schmidt criterion identifies the minimum entangling capacity required
by the target state, but not the minimum depth of this particular
hardware-efficient circuit.

For the larger benchmark instances, increasing depth substantially improves
reachability but does not produce exact recovery within $p\leq6$. At the
largest tested depth the best observed ground-state fidelities are $0.921$,
$0.869$, and $0.911$ for $N=8,10,12$, respectively. Since the small-system
controls demonstrate that the same search procedure can recover exact
solutions when they are accessible, the remaining large-$N$ gaps are
consistent with a combination of restricted-manifold expressivity and
increasing optimization difficulty rather than with the Schmidt-rank
constraint alone.

\subsection{Depth dependence of the landscape geometry}

\begin{figure*}[t]
\centering
\begin{minipage}{0.35\textwidth}
    \centering
    \includegraphics[width=\linewidth]{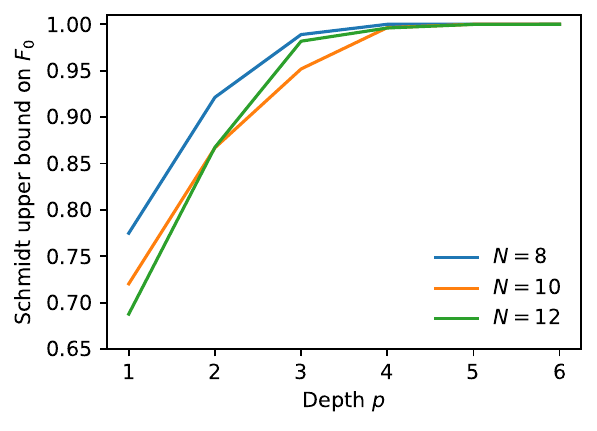}\\[-1mm]
    \small (a) Schmidt fidelity bound
\end{minipage}%
\hspace{0.04\textwidth}%
\begin{minipage}{0.35\textwidth}
    \centering
    \includegraphics[width=\linewidth]{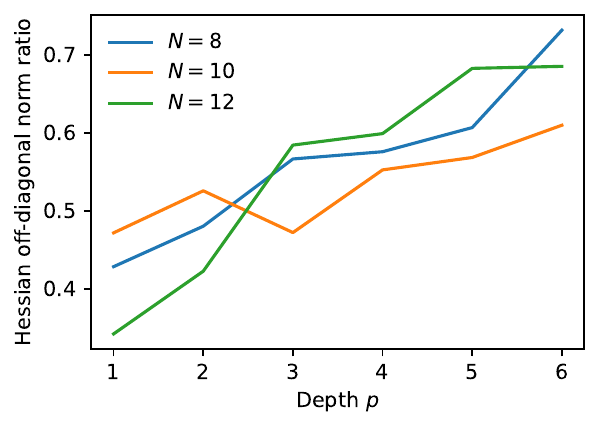}\\[-1mm]
    \small (b) Hessian parameter coupling
\end{minipage}

\vspace{2mm}

\begin{minipage}{0.35\textwidth}
    \centering
    \includegraphics[width=\linewidth]{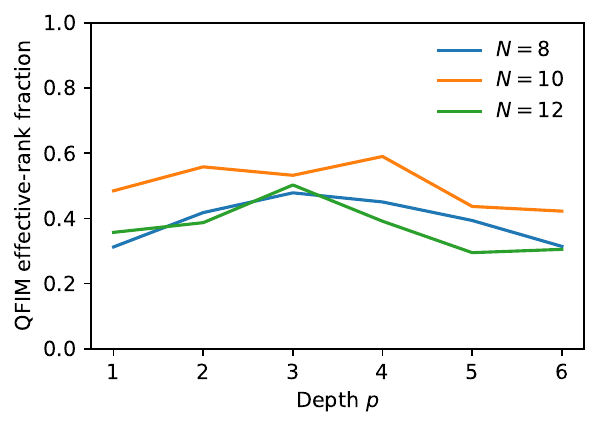}\\[-1mm]
    \small (c) QFIM effective rank
\end{minipage}%
\hspace{0.04\textwidth}%
\begin{minipage}{0.35\textwidth}
    \centering
    \includegraphics[width=\linewidth]{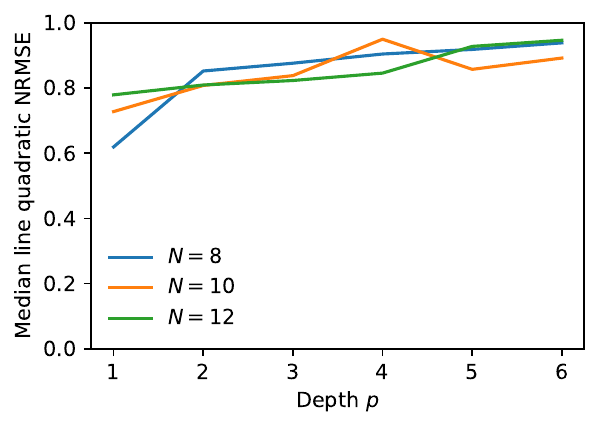}\\[-1mm]
    \small (d) Random-line nonquadraticity
\end{minipage}

\caption{Additional Q3 depth diagnostics for the $N=8,10,12$ benchmark
instances. (a) Schmidt-truncation upper bound on ground-state fidelity.
(b) Relative off-diagonal Hessian norm.
(c) QFIM effective-rank fraction at the best-energy point.
(d) Median normalized quadratic-fit error on random one-dimensional parameter
scans.}
\label{fig:app-q3-depth-geometry}
\end{figure*}

The landscape changes systematically as the circuit is deepened.
Figure~\ref{fig:app-q3-depth-geometry} collects the main diagnostics for the
benchmark instances. The median random-point gradient decreases with depth,
from $0.67$--$0.72$ at $p=1$ to $0.21$--$0.32$ at $p=6$, but no sampled
random point has $g_{\rm RMS}<10^{-3}$. The observed behavior therefore does
not correspond to a finite-size globally vanishing-gradient plateau.

Hessian and QFIM diagnostics are evaluated at the BFGS-polished best-energy
point for each benchmark $(N,p)$ condition. Stationarity is checked using the
exact parameter-shift gradient; after the final repolish all 18 reported
benchmark points satisfy $g_{\rm RMS}\leq4.3\times10^{-8}$. The pure-state
QFIM is
\begin{equation}
 \mathcal F_{ij}=4\,\mathrm{Re}\!\left[
 \langle\partial_i\psi|\partial_j\psi\rangle-
 \langle\partial_i\psi|\psi\rangle
 \langle\psi|\partial_j\psi\rangle\right].
\end{equation}
For QFIM eigenvalues $\nu_i$, active positive values satisfy
\begin{equation}
 \nu_i>\tau_Q,\qquad
 \tau_Q=\max\!\left(10^{-10},10^{-8}\max(1,\max_j|\nu_j|)\right).
\end{equation}
We report the participation-ratio effective rank
\begin{equation}
 r_{\rm eff}=\frac{(\sum_i\nu_i)^2}{\sum_i\nu_i^2}
\end{equation}
over the active positive spectrum, normalized by $D$.

The random-line diagnostic uses eight independent directions per benchmark
condition. Anchors are uniform on $[-\pi,\pi]^D$; Gaussian directions are
rescaled to unit RMS, and the energy is evaluated at 41 equally spaced points
for $t\in[-\pi,\pi]$ with periodic parameter wrapping. After fitting a
quadratic $\widehat E(t)$, we report the median normalized residual
\begin{equation}
 e_{\rm quad}=
 \frac{\sqrt{M^{-1}\sum_k[E(t_k)-\widehat E(t_k)]^2}}
 {\operatorname{std}_k E(t_k)} .
\end{equation}
It increases overall from $(0.619,0.728,0.779)$ at $p=1$ to
$(0.939,0.892,0.947)$ at $p=6$ for $N=(8,10,12)$, so the deeper random-line
restrictions are less well described by a quadratic model, although the trend
is not monotonic at every intermediate depth.

The local geometry becomes more strongly coupled. The relative off-diagonal
Hessian norm rises overall from $0.34$--$0.47$ at $p=1$ to $0.61$--$0.69$ at
$p=6$. Hessian conditioning is strongly nonmonotonic and reaches
$10^4$--$10^5$ at several depths. The QFIM effective rank varies with depth
but does not show a monotonic collapse. Increasing expressivity therefore
comes with more anisotropic and coupled optimization geometry rather than a
simple loss of all gradient signal.

\section{Optimizer configurations}
\label{app:optimizer-configurations}

All optimizers were evaluated on the same exact objective within a given
benchmark condition and were subject to the common FE
budget $B\in\{10\,000,30\,000\}$. Parameters were initialized or constrained
on $[-\pi,\pi]^D$, with $D=N$ for Q1 and $D=2N$ for Q2 and Q3. The optimizer
settings used by the benchmark wrappers are summarized in
Table~\ref{tab:optimizer-configurations}. Hyperparameters were not retuned
between the benchmark models. When an implementation exposed additional
internal controls that were not overridden by the benchmark wrapper, its
implementation defaults were retained.

\begin{table*}[htpb]
\centering
\caption{Optimizer configurations used in the benchmark. Here $B$ denotes the
FE budget and $D$ the variational dimension. The objective-level FE guard
provides the common hard evaluation cap; an optimizer may terminate earlier
according to its own convergence criterion.}
\label{tab:optimizer-configurations}
\setlength{\tabcolsep}{3.0pt}
\renewcommand{\arraystretch}{0.94}
\scriptsize
\resizebox{\textwidth}{!}{%
\begin{tabular}{p{0.11\textwidth}p{0.19\textwidth}p{0.48\textwidth}p{0.15\textwidth}}
\toprule
Optimizer & Initialization / population & Main settings & Termination / budget handling \\
\midrule
BFGS &
$x_0\sim\mathcal U[-\pi,\pi]^D$ &
SciPy BFGS; numerical finite-difference gradient; $\mathrm{gtol}=10^{-8}$.
Q1 used $\mathrm{maxiter}=1000$; Q2/Q3 used
$\mathrm{maxiter}=\max(1,\lfloor B/(D+1)\rfloor)$. &
BFGS convergence or common FE cap. \\

SPSA &
$x_0\sim\mathcal U[-\pi,\pi]^D$ &
$a=0.30$, $c=0.10$, $A=100$, $\alpha=0.602$, $\gamma=0.101$;
one Rademacher perturbation per iteration (two FEs). &
Iterated until fewer than two FEs remained. \\

CMA-ES &
$x_0\sim\mathcal U[-\pi,\pi]^D$;
$\lambda=4+\lfloor3\ln D\rfloor$ &
$\sigma_0=1.0$, $\mathrm{tolx}=10^{-10}$,
$\mathrm{tolfun}=10^{-10}$, bounds $[-\pi,\pi]^D$. &
Internal convergence or common FE cap; nominal generation count derived from $B/\lambda$. \\

iSOMA &
$\mathrm{PopSize}=50$ (or default $150$); uniform in $[-\pi,\pi]^D$ &
$N_{\mathrm{jump}}=10$, $\mathrm{Step}=0.3$, $m=10$, $n=5$, $k=15$; adaptive $\mathrm{PRT}=0.1+0.9(\mathrm{FEs}/B)$; boundary clipping; stagnation reinitialization ($10\%$ pop) after $50\times\mathrm{PopSize}$ stagnant steps. &
$B$ evaluations ($\mathrm{Max\_FEs}=B$) or $\mathrm{Max\_Migration}=10000$. \\

SciPy-DE &
Population size $2D$ (\texttt{popsize=2}) &
\texttt{currenttobest1bin}; mutation factor dithered in $(0.4,0.9)$;
recombination $0.8$; \texttt{tol=0}; no polishing; deferred population updating. &
$\max(1,\lfloor(B-2D)/(2D)\rfloor)$ generations, with the common FE cap. \\

iL-SHADE &
Initial population $4D$ &
PyADE iL-SHADE defaults for dimension $D$, except
\texttt{population\_size}$=4D$, \texttt{max\_evals}$=B$,
bounds $[-\pi,\pi]^D$, and the replicate seed supplied explicitly. &
\texttt{max\_evals}$=B$ / implementation stopping rule. \\

L-SRTDE &
$N_{\mathrm{init}}=20D$, linearly reduced (LPSR) to $N_{\min}=4$; uniform in $[-\pi,\pi]^D$ &
Historical memory $H=5$ for $\mathrm{Cr}$ (init $1.0$); adaptive mean $\mu_F=0.4+0.25\tanh(5\cdot\mathrm{success\_rate})$, $\sigma_F=0.02$; adaptive $p$-best size $p_{\mathrm{size}}=\max(2,\lfloor N\cdot 0.7 e^{-7\cdot\mathrm{success\_rate}}\rfloor)$; rank-weighted base vector from sorted elite front; random bound reinitialization on violation; Lehmer mean $\mathrm{Cr}$ update. &
Budget $B$ ($\mathrm{max\_evals}=B$) / common FE cap. \\

jSO &
$N_{\mathrm{init}}=\max(30,\lfloor 25\sqrt{D}\log_{10} D\rfloor)$, LPSR to $N_{\min}=4$; uniform in $[-\pi,\pi]^D$ &
Historical memory $H=5$ ($M_F=0.5, M_{\mathrm{CR}}=0.8$); Cauchy $F$, Gaussian $\mathrm{CR}$; current-to-$p\text{best}$-w/1 mutation with external archive; dynamic $p=0.25 - 0.125(\mathrm{evals}/B)$; weighted Lehmer mean update for $M_F$ and arithmetic mean for $M_{\mathrm{CR}}$; boundary clipping. &
Budget $B$ ($\mathrm{max\_evals}=B$) / common FE cap. \\
\bottomrule
\end{tabular}%
}
\end{table*}

\section{FE-matched multistart BFGS and noise sensitivity}
\label{app:multistart-bfgs}

The main benchmark uses one random initialization per optimizer and FE budget.
For the $N=12$ control below, BFGS is instead restarted from independent random
parameters until a shared budget $B\in\{10\,000,30\,000\}$ is exhausted.
Twenty-five independent restart sequences are used for each model and budget,
so MS-BFGS receives the same total FE allowance as the population methods.

\begin{figure*}[htpb]
    \centering
    \includegraphics[width=0.65\textwidth]{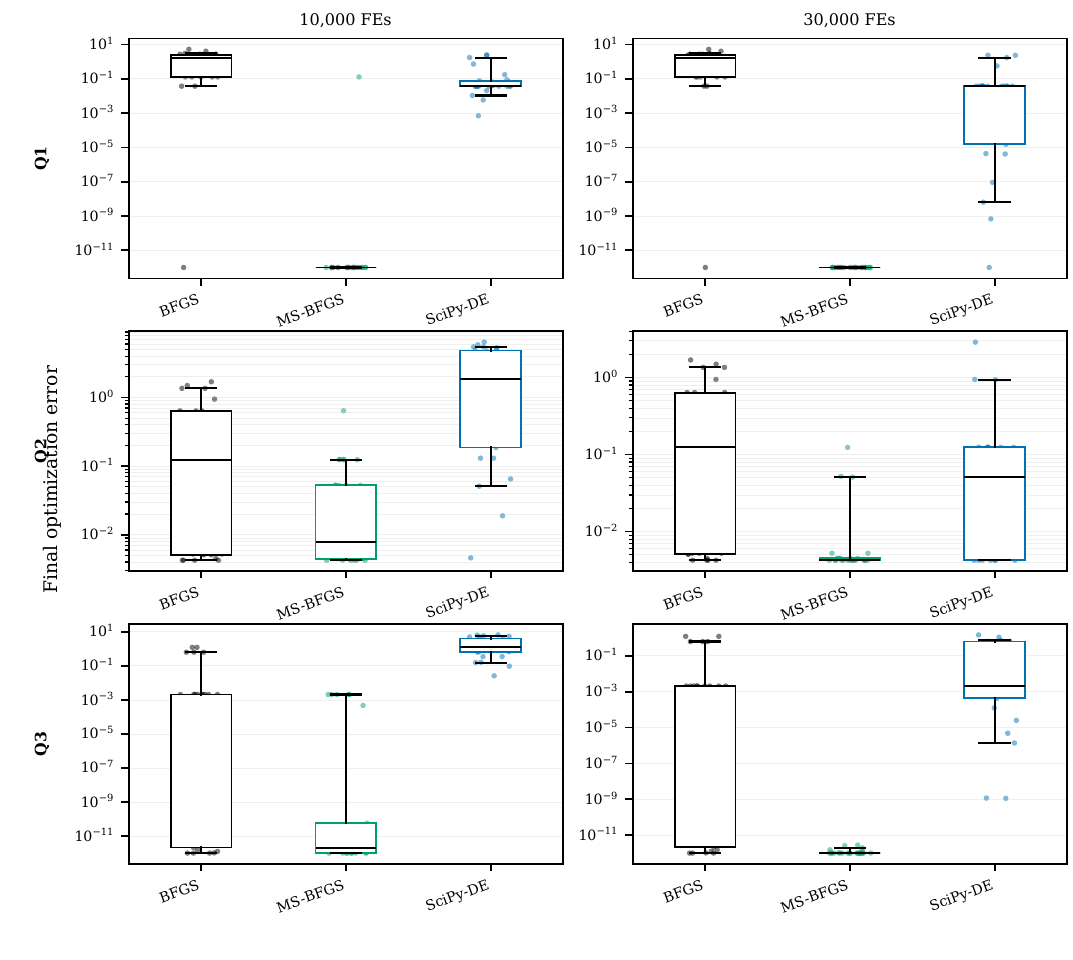}
    \caption{FE-matched multistart BFGS control at $N=12$. The original
    single-start BFGS, FE-matched multistart BFGS (MS-BFGS), and SciPy-DE
    endpoint distributions are compared at $10\,000$ and $30\,000$ FEs.
    For Q1 and Q2, errors are measured relative to the exact physical
    ground-state energy $E_0$; for Q3,
    $\Delta E_{\mathrm{opt}}=E_{\mathrm{best}}-E_{\mathrm{ref}}$.}
    \label{fig:multistart-bfgs}
\end{figure*}

MS-BFGS reaches the exact Q1 ground-state reference and the saved Q3
variational reference $E_{\mathrm{ref}}$ to numerical precision at both
budgets; for Q3 this does not imply recovery of the physical ground state
$E_0$. For Q2 the median errors are
$7.75\times10^{-3}$ and $4.25\times10^{-3}$. The median numbers of completed or
attempted restarts are approximately $24/65$ for Q1, $3/8$ for Q2, and $4/11$
for Q3 at $10\,000/30\,000$ FEs. The result identifies basin coverage within the
FE budget as an important control variable for the deterministic benchmark.
The main tables retain the original single-start protocols; this appendix
isolates the effect of restart-based exploration.

\paragraph{Sensitivity to stochastic objective evaluations.}
MS-BFGS uses finite-difference gradients and is highly sensitive to objective
noise. We repeat the control with independent additive Gaussian noise at every
FE,
\begin{equation}
    \widetilde E(\boldsymbol{\theta})
    =
    E(\boldsymbol{\theta})+\epsilon,
    \qquad
    \epsilon\sim\mathcal{N}(0,\sigma^2),
    \label{eq:appendix-bfgs-gaussian-noise}
\end{equation}

The final score is always the exact statevector energy of the parameter vector
selected using the noisy objective.

\begin{figure*}[htpb]
    \centering
    \includegraphics[width=0.75\textwidth]{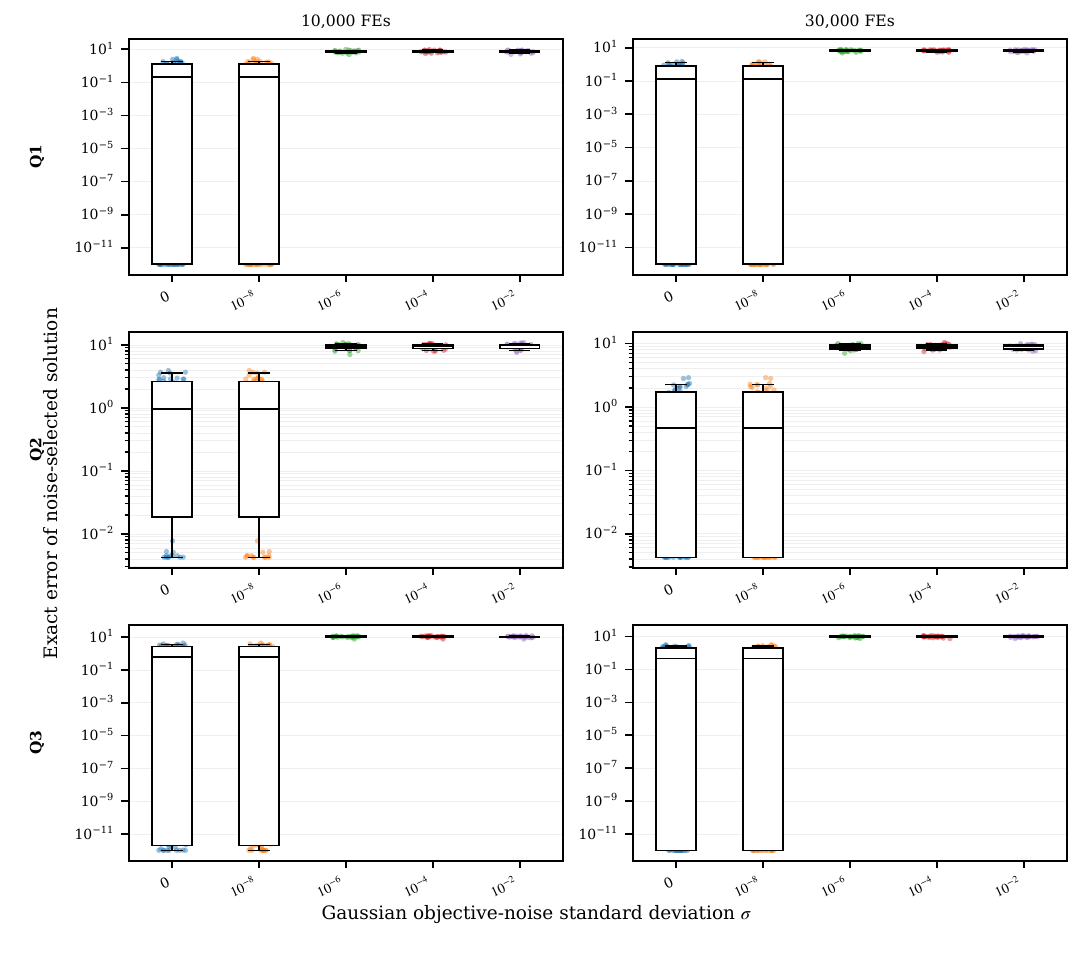}
    \caption{Sensitivity of FE-matched MS-BFGS to independent additive Gaussian
    noise in every objective evaluation.  Each panel reports the exact
    statevector error of the parameter vector selected using the noisy
    objective.  The column labels give the common total FE budget.  The
    $\sigma=0$ distributions are the noiseless MS-BFGS control from
    Fig.~\ref{fig:multistart-bfgs}.}
    \label{fig:multistart-bfgs-noise}
\end{figure*}

Figure~\ref{fig:multistart-bfgs-noise} shows abrupt degradation: at
$\sigma=10^{-8}$ the median exact errors are already of order unity for Q1--Q3,
and at $\sigma=10^{-6}$ they are of order $6$--$11$. A shot-count interpretation
is estimator dependent; with effective single-shot variance $V_1$ and $M$
shots, the usual scaling is
\begin{equation}
    \sigma_E^2 \simeq \frac{V_1}{M},
    \qquad
    M \simeq \frac{V_1}{\sigma_E^2}.
    \label{eq:appendix-shot-equivalent}
\end{equation}

For $V_1\simeq1$, the noise levels at which numerical MS-BFGS remains
effective imply many orders of magnitude more shots per objective evaluation
than typical finite-shot VQE. Combined with the $10\,000$--$30\,000$ FE
budgets, the resulting effective sampling cost would be impractically large.
This estimate is only illustrative because
finite-shot VQE noise is state-, Hamiltonian-, grouping-, and
allocation-dependent.

The near-machine-precision MS-BFGS result is therefore specific to the
exact-objective regime; finite-shot optimizer comparisons are treated in
related work \cite{Illesova2025VHA,Novak2025NoisyLandscapes,Novak2025Reliable}.

\section{Statistical comparison of optimizer performance}
\label{app:statistical-comparison}

The inferential analysis is reported here rather than in the main Results
section so that the latter remains focused on the observed energy, fidelity,
and success-probability outcomes. Although the same integer seed labels were
used across optimizers for reproducibility, heterogeneous optimizers do not
consume randomness in a common way and do not form a genuine paired
common-random-number design. The 25 runs for each optimizer are therefore
treated as independent samples. The saved data contain 25 successful finite
observations for every optimizer in every model--size--budget condition.

For each condition, an omnibus Kruskal--Wallis test compares the eight
optimizer final-error distributions. Pairwise differences are assessed with
all 28 two-sided Mann--Whitney $U$ tests, with Holm correction applied within
each condition. We additionally report the independent-sample rank-biserial
effect size
\[
 r_{\rm rb} = \frac{2U_A}{n_A n_B}-1 ,
\]
where $U_A$ is the Mann--Whitney statistic for optimizer A. Because lower
error is better, $r_{\rm rb}<0$ indicates that optimizer A tends to produce
lower errors than optimizer B, while $r_{\rm rb}>0$ favors B. The family-wise
significance level is $\alpha=0.05$.

All 16 omnibus tests reject equality of the optimizer distributions.
Table~\ref{tab:app-kruskal} gives the complete omnibus results and the number
of Holm-significant pairwise comparisons in each condition. The weakest
pairwise separation occurs for Q2 at $N=10$ and $30\,000$ FEs, where only
2 of 28 pairwise comparisons remain significant after correction despite a
significant omnibus test. The complete Holm-adjusted pairwise results and
effect sizes are reported in Tables~\ref{tab:app-mw-q1}--\ref{tab:app-mw-q3}.

\begin{table}[t]
\centering
\caption{Independent-sample Kruskal--Wallis tests across the eight optimizers.
The final column gives the number of the 28 pairwise Mann--Whitney comparisons
that remain significant after Holm correction at $\alpha=0.05$.}
\label{tab:app-kruskal}
\setlength{\tabcolsep}{1.8pt}
\renewcommand{\arraystretch}{0.92}
\scriptsize
\begin{tabular}{lrrrcc}
\toprule
Model & $N$ & \begin{tabular}[c]{@{}c@{}}FE\\budget\end{tabular} & $H(7)$ & $p$ & \begin{tabular}[c]{@{}c@{}}Holm-significant\\pairs\end{tabular} \\
\midrule
Q1 & 12 & 10k & 71.74 & $6.57\times10^{-13}$ & 17/28 \\
Q1 & 12 & 30k & 113.92 & $1.41\times10^{-21}$ & 21/28 \\
Q1 & 14 & 10k & 108.49 & $1.89\times10^{-20}$ & 14/28 \\
Q1 & 14 & 30k & 119.80 & $8.44\times10^{-23}$ & 13/28 \\
Q1 & 16 & 10k & 42.94 & $3.42\times10^{-7}$ & 9/28 \\
Q1 & 16 & 30k & 57.16 & $5.56\times10^{-10}$ & 15/28 \\
Q2 & 10 & 10k & 77.11 & $5.33\times10^{-14}$ & 13/28 \\
Q2 & 10 & 30k & 21.72 & $0.003$ & 2/28 \\
Q2 & 12 & 10k & 91.77 & $5.35\times10^{-17}$ & 14/28 \\
Q2 & 12 & 30k & 55.32 & $1.29\times10^{-9}$ & 10/28 \\
Q3 & 8 & 10k & 105.37 & $8.35\times10^{-20}$ & 20/28 \\
Q3 & 8 & 30k & 115.28 & $7.37\times10^{-22}$ & 20/28 \\
Q3 & 10 & 10k & 127.57 & $2.02\times10^{-24}$ & 19/28 \\
Q3 & 10 & 30k & 77.15 & $5.24\times10^{-14}$ & 8/28 \\
Q3 & 12 & 10k & 139.69 & $5.91\times10^{-27}$ & 21/28 \\
Q3 & 12 & 30k & 108.52 & $1.86\times10^{-20}$ & 20/28 \\
\bottomrule
\end{tabular}
\end{table}

The pairwise tables use the compact cell format
$p_{\rm Holm}\,[r_{\rm rb}]$; statistically significant adjusted $p$-values
are shown in bold. The sign of $r_{\rm rb}$ is defined for the first optimizer
in each row relative to the second.

\begin{table*}[p]
\centering
\caption{Q1 pairwise Mann--Whitney comparisons. Cells report
$p_{\rm Holm}\,[r_{\rm rb}]$. Bold adjusted $p$-values are significant at
$\alpha=0.05$.}
\label{tab:app-mw-q1}
\setlength{\tabcolsep}{2.2pt}
\renewcommand{\arraystretch}{0.92}
\scriptsize
\begin{tabular}{lcccccc}
\toprule
Pair & $N=12$, 10k & $N=12$, 30k & $N=14$, 10k & $N=14$, 30k & $N=16$, 10k & $N=16$, 30k \\
\midrule
BFGS--SPSA & $1.000\;[-0.12]$ & $1.000\;[-0.11]$ & $1.000\;[-0.17]$ & $1.000\;[-0.17]$ & $1.000\;[+0.02]$ & $1.000\;[+0.03]$ \\
BFGS--CMA-ES & $1.000\;[+0.04]$ & $1.000\;[+0.15]$ & $1.000\;[-0.04]$ & $1.000\;[+0.07]$ & $1.000\;[+0.28]$ & $1.000\;[-0.03]$ \\
BFGS--iSOMA & $\mathbf{0.004}\;[+0.61]$ & $\mathbf{4.02\times10^{-8}}\;[+0.96]$ & $\mathbf{1.16\times10^{-5}}\;[+0.82]$ & $\mathbf{6.37\times10^{-9}}\;[+0.99]$ & $\mathbf{0.002}\;[+0.65]$ & $\mathbf{3.85\times10^{-5}}\;[+0.80]$ \\
BFGS--SciPy-DE & $\mathbf{0.004}\;[+0.62]$ & $\mathbf{1.90\times10^{-4}}\;[+0.72]$ & $0.825\;[+0.27]$ & $0.245\;[+0.39]$ & $0.208\;[+0.41]$ & $\mathbf{0.022}\;[+0.53]$ \\
BFGS--iL-SHADE & $\mathbf{0.036}\;[+0.49]$ & $\mathbf{3.68\times10^{-5}}\;[+0.78]$ & $1.000\;[+0.13]$ & $1.000\;[+0.22]$ & $\mathbf{0.032}\;[+0.52]$ & $\mathbf{0.001}\;[+0.66]$ \\
BFGS--L-SRTDE & $\mathbf{0.009}\;[+0.57]$ & $\mathbf{1.85\times10^{-4}}\;[+0.72]$ & $0.312\;[+0.36]$ & $0.110\;[+0.44]$ & $0.650\;[+0.32]$ & $\mathbf{0.011}\;[+0.57]$ \\
BFGS--jSO & $\mathbf{4.97\times10^{-5}}\;[+0.79]$ & $\mathbf{6.98\times10^{-6}}\;[+0.84]$ & $\mathbf{7.38\times10^{-6}}\;[+0.84]$ & $\mathbf{8.96\times10^{-9}}\;[+0.98]$ & $\mathbf{0.007}\;[+0.60]$ & $\mathbf{1.97\times10^{-4}}\;[+0.74]$ \\
SPSA--CMA-ES & $1.000\;[+0.20]$ & $0.207\;[+0.35]$ & $1.000\;[+0.15]$ & $1.000\;[+0.26]$ & $1.000\;[+0.20]$ & $1.000\;[-0.02]$ \\
SPSA--iSOMA & $\mathbf{2.47\times10^{-5}}\;[+0.81]$ & $\mathbf{2.52\times10^{-8}}\;[+0.97]$ & $\mathbf{7.70\times10^{-8}}\;[+0.97]$ & $\mathbf{4.92\times10^{-9}}\;[+1.00]$ & $\mathbf{0.002}\;[+0.65]$ & $\mathbf{3.23\times10^{-4}}\;[+0.72]$ \\
SPSA--SciPy-DE & $\mathbf{1.62\times10^{-4}}\;[+0.74]$ & $\mathbf{1.06\times10^{-5}}\;[+0.83]$ & $0.259\;[+0.39]$ & $0.054\;[+0.48]$ & $0.610\;[+0.33]$ & $\mathbf{0.031}\;[+0.51]$ \\
SPSA--iL-SHADE & $\mathbf{0.005}\;[+0.61]$ & $\mathbf{8.39\times10^{-6}}\;[+0.83]$ & $0.911\;[+0.25]$ & $0.979\;[+0.27]$ & $\mathbf{0.032}\;[+0.53]$ & $\mathbf{0.010}\;[+0.57]$ \\
SPSA--L-SRTDE & $\mathbf{0.001}\;[+0.67]$ & $\mathbf{1.16\times10^{-7}}\;[+0.97]$ & $0.061\;[+0.47]$ & $\mathbf{0.027}\;[+0.52]$ & $1.000\;[+0.19]$ & $\mathbf{0.020}\;[+0.54]$ \\
SPSA--jSO & $\mathbf{8.00\times10^{-7}}\;[+0.92]$ & $\mathbf{2.56\times10^{-6}}\;[+0.87]$ & $\mathbf{4.46\times10^{-8}}\;[+0.99]$ & $\mathbf{8.96\times10^{-9}}\;[+0.98]$ & $\mathbf{0.019}\;[+0.56]$ & $\mathbf{0.002}\;[+0.65]$ \\
CMA-ES--iSOMA & $\mathbf{0.028}\;[+0.51]$ & $\mathbf{7.26\times10^{-8}}\;[+0.94]$ & $\mathbf{2.01\times10^{-6}}\;[+0.88]$ & $\mathbf{5.14\times10^{-9}}\;[+0.99]$ & $\mathbf{0.032}\;[+0.53]$ & $\mathbf{2.57\times10^{-4}}\;[+0.73]$ \\
CMA-ES--SciPy-DE & $0.057\;[+0.46]$ & $\mathbf{5.99\times10^{-4}}\;[+0.67]$ & $1.000\;[+0.22]$ & $0.783\;[+0.30]$ & $1.000\;[+0.20]$ & $\mathbf{0.045}\;[+0.49]$ \\
CMA-ES--iL-SHADE & $0.074\;[+0.44]$ & $\mathbf{2.18\times10^{-4}}\;[+0.71]$ & $1.000\;[+0.09]$ & $1.000\;[+0.09]$ & $0.447\;[+0.36]$ & $\mathbf{0.003}\;[+0.63]$ \\
CMA-ES--L-SRTDE & $\mathbf{0.049}\;[+0.48]$ & $\mathbf{0.015}\;[+0.52]$ & $0.823\;[+0.28]$ & $0.357\;[+0.36]$ & $1.000\;[-0.04]$ & $\mathbf{0.045}\;[+0.49]$ \\
CMA-ES--jSO & $\mathbf{4.27\times10^{-5}}\;[+0.79]$ & $\mathbf{5.81\times10^{-5}}\;[+0.76]$ & $\mathbf{3.28\times10^{-7}}\;[+0.93]$ & $\mathbf{9.13\times10^{-9}}\;[+0.97]$ & $0.126\;[+0.45]$ & $\mathbf{6.59\times10^{-4}}\;[+0.69]$ \\
iSOMA--SciPy-DE & $0.500\;[-0.31]$ & $\mathbf{6.98\times10^{-6}}\;[-0.81]$ & $\mathbf{3.96\times10^{-8}}\;[-1.00]$ & $\mathbf{4.87\times10^{-9}}\;[-1.00]$ & $0.187\;[-0.42]$ & $1.000\;[-0.25]$ \\
iSOMA--iL-SHADE & $1.000\;[-0.20]$ & $\mathbf{8.87\times10^{-4}}\;[-0.59]$ & $\mathbf{3.96\times10^{-8}}\;[-1.00]$ & $\mathbf{4.46\times10^{-9}}\;[-1.00]$ & $1.000\;[-0.08]$ & $1.000\;[-0.08]$ \\
iSOMA--L-SRTDE & $0.520\;[-0.30]$ & $\mathbf{2.05\times10^{-6}}\;[-0.85]$ & $\mathbf{3.96\times10^{-8}}\;[-1.00]$ & $\mathbf{4.87\times10^{-9}}\;[-1.00]$ & $\mathbf{7.78\times10^{-4}}\;[-0.69]$ & $1.000\;[-0.00]$ \\
iSOMA--jSO & $\mathbf{0.027}\;[+0.52]$ & $\mathbf{0.024}\;[-0.41]$ & $\mathbf{5.65\times10^{-5}}\;[+0.77]$ & $1.000\;[-0.00]$ & $1.000\;[-0.16]$ & $1.000\;[+0.12]$ \\
SciPy-DE--iL-SHADE & $1.000\;[-0.01]$ & $0.956\;[+0.20]$ & $0.293\;[+0.37]$ & $1.000\;[+0.23]$ & $1.000\;[+0.23]$ & $1.000\;[+0.25]$ \\
SciPy-DE--L-SRTDE & $\mathbf{0.019}\;[-0.54]$ & $\mathbf{0.005}\;[-0.58]$ & $\mathbf{0.015}\;[-0.55]$ & $1.000\;[+0.02]$ & $0.657\;[-0.32]$ & $1.000\;[-0.02]$ \\
SciPy-DE--jSO & $\mathbf{0.006}\;[+0.59]$ & $0.430\;[+0.28]$ & $\mathbf{3.96\times10^{-8}}\;[+1.00]$ & $\mathbf{4.92\times10^{-9}}\;[+1.00]$ & $1.000\;[+0.25]$ & $1.000\;[+0.22]$ \\
iL-SHADE--L-SRTDE & $1.000\;[-0.11]$ & $\mathbf{0.012}\;[-0.54]$ & $0.313\;[-0.36]$ & $1.000\;[-0.12]$ & $0.064\;[-0.48]$ & $0.277\;[-0.38]$ \\
iL-SHADE--jSO & $0.203\;[+0.38]$ & $1.000\;[+0.14]$ & $\mathbf{3.96\times10^{-8}}\;[+1.00]$ & $\mathbf{4.46\times10^{-9}}\;[+1.00]$ & $1.000\;[-0.01]$ & $1.000\;[+0.06]$ \\
L-SRTDE--jSO & $\mathbf{7.56\times10^{-4}}\;[+0.69]$ & $0.134\;[+0.39]$ & $\mathbf{3.96\times10^{-8}}\;[+1.00]$ & $\mathbf{4.87\times10^{-9}}\;[+1.00]$ & $\mathbf{0.019}\;[+0.55]$ & $1.000\;[+0.04]$ \\
\bottomrule
\end{tabular}%
\end{table*}

\begin{table*}[p]
\centering
\caption{Q2 pairwise Mann--Whitney comparisons. Cells report
$p_{\rm Holm}\,[r_{\rm rb}]$. Bold adjusted $p$-values are significant at
$\alpha=0.05$.}
\label{tab:app-mw-q2}
\setlength{\tabcolsep}{2.2pt}
\renewcommand{\arraystretch}{0.92}
\scriptsize
\begin{tabular}{lcccc}
\toprule
Pair & $N=10$, 10k & $N=10$, 30k & $N=12$, 10k & $N=12$, 30k \\
\midrule
BFGS--SPSA & $0.600\;[-0.32]$ & $1.000\;[-0.23]$ & $0.076\;[-0.46]$ & $0.119\;[-0.44]$ \\
BFGS--CMA-ES & $0.541\;[-0.34]$ & $1.000\;[-0.11]$ & $0.651\;[-0.25]$ & $0.838\;[-0.31]$ \\
BFGS--iSOMA & $\mathbf{0.042}\;[-0.50]$ & $1.000\;[-0.22]$ & $0.243\;[-0.36]$ & $0.568\;[-0.34]$ \\
BFGS--SciPy-DE & $0.438\;[-0.36]$ & $1.000\;[+0.14]$ & $\mathbf{0.002}\;[-0.65]$ & $1.000\;[+0.14]$ \\
BFGS--iL-SHADE & $0.600\;[-0.32]$ & $1.000\;[-0.24]$ & $1.000\;[-0.03]$ & $1.000\;[+0.05]$ \\
BFGS--L-SRTDE & $\mathbf{5.48\times10^{-8}}\;[-0.99]$ & $1.000\;[-0.17]$ & $\mathbf{5.94\times10^{-8}}\;[-0.99]$ & $1.000\;[-0.24]$ \\
BFGS--jSO & $\mathbf{8.12\times10^{-5}}\;[-0.77]$ & $1.000\;[+0.27]$ & $0.161\;[-0.40]$ & $1.000\;[+0.04]$ \\
SPSA--CMA-ES & $1.000\;[+0.05]$ & $1.000\;[+0.14]$ & $0.375\;[+0.31]$ & $1.000\;[+0.22]$ \\
SPSA--iSOMA & $1.000\;[-0.13]$ & $1.000\;[+0.00]$ & $1.000\;[+0.12]$ & $1.000\;[+0.15]$ \\
SPSA--SciPy-DE & $1.000\;[-0.06]$ & $0.561\;[+0.37]$ & $0.243\;[-0.37]$ & $\mathbf{0.004}\;[+0.63]$ \\
SPSA--iL-SHADE & $1.000\;[+0.04]$ & $1.000\;[+0.01]$ & $\mathbf{0.004}\;[+0.61]$ & $\mathbf{7.90\times10^{-4}}\;[+0.69]$ \\
SPSA--L-SRTDE & $\mathbf{1.58\times10^{-7}}\;[-0.96]$ & $1.000\;[+0.05]$ & $\mathbf{1.07\times10^{-7}}\;[-0.97]$ & $1.000\;[+0.24]$ \\
SPSA--jSO & $\mathbf{0.018}\;[-0.55]$ & $\mathbf{0.017}\;[+0.56]$ & $1.000\;[+0.16]$ & $\mathbf{2.02\times10^{-4}}\;[+0.74]$ \\
CMA-ES--iSOMA & $1.000\;[-0.20]$ & $1.000\;[-0.10]$ & $0.054\;[-0.48]$ & $1.000\;[-0.12]$ \\
CMA-ES--SciPy-DE & $1.000\;[-0.08]$ & $1.000\;[+0.29]$ & $\mathbf{0.023}\;[-0.52]$ & $\mathbf{0.009}\;[+0.58]$ \\
CMA-ES--iL-SHADE & $1.000\;[+0.03]$ & $1.000\;[-0.11]$ & $0.243\;[+0.37]$ & $\mathbf{0.012}\;[+0.57]$ \\
CMA-ES--L-SRTDE & $\mathbf{5.48\times10^{-8}}\;[-0.99]$ & $1.000\;[-0.06]$ & $\mathbf{5.94\times10^{-8}}\;[-0.99]$ & $1.000\;[+0.07]$ \\
CMA-ES--jSO & $\mathbf{0.006}\;[-0.60]$ & $0.224\;[+0.43]$ & $1.000\;[-0.16]$ & $\mathbf{0.004}\;[+0.63]$ \\
iSOMA--SciPy-DE & $1.000\;[+0.11]$ & $1.000\;[+0.32]$ & $0.086\;[-0.44]$ & $\mathbf{1.92\times10^{-4}}\;[+0.74]$ \\
iSOMA--iL-SHADE & $1.000\;[+0.10]$ & $1.000\;[-0.13]$ & $\mathbf{0.005}\;[+0.60]$ & $\mathbf{2.07\times10^{-6}}\;[+0.89]$ \\
iSOMA--L-SRTDE & $\mathbf{4.47\times10^{-8}}\;[-1.00]$ & $1.000\;[-0.09]$ & $\mathbf{3.96\times10^{-8}}\;[-1.00]$ & $1.000\;[+0.22]$ \\
iSOMA--jSO & $\mathbf{0.042}\;[-0.50]$ & $\mathbf{2.27\times10^{-4}}\;[+0.74]$ & $\mathbf{0.006}\;[+0.59]$ & $\mathbf{1.84\times10^{-7}}\;[+0.96]$ \\
SciPy-DE--iL-SHADE & $1.000\;[+0.04]$ & $0.688\;[-0.36]$ & $\mathbf{9.68\times10^{-4}}\;[+0.67]$ & $1.000\;[-0.12]$ \\
SciPy-DE--L-SRTDE & $\mathbf{3.33\times10^{-4}}\;[-0.72]$ & $1.000\;[-0.29]$ & $\mathbf{6.11\times10^{-4}}\;[-0.69]$ & $0.065\;[-0.48]$ \\
SciPy-DE--jSO & $0.165\;[-0.42]$ & $1.000\;[+0.02]$ & $0.076\;[+0.45]$ & $1.000\;[-0.10]$ \\
iL-SHADE--L-SRTDE & $\mathbf{5.71\times10^{-8}}\;[-0.99]$ & $1.000\;[+0.09]$ & $\mathbf{3.96\times10^{-8}}\;[-1.00]$ & $0.074\;[-0.47]$ \\
iL-SHADE--jSO & $\mathbf{0.008}\;[-0.59]$ & $0.102\;[+0.48]$ & $\mathbf{0.009}\;[-0.57]$ & $1.000\;[+0.09]$ \\
L-SRTDE--jSO & $\mathbf{0.034}\;[+0.51]$ & $0.064\;[+0.50]$ & $\mathbf{6.18\times10^{-8}}\;[+0.98]$ & $\mathbf{0.028}\;[+0.53]$ \\
\bottomrule
\end{tabular}%
\end{table*}

\begin{table*}[p]
\centering
\caption{Q3 pairwise Mann--Whitney comparisons. Cells report
$p_{\rm Holm}\,[r_{\rm rb}]$. Bold adjusted $p$-values are significant at
$\alpha=0.05$.}
\label{tab:app-mw-q3}
\setlength{\tabcolsep}{2.2pt}
\renewcommand{\arraystretch}{0.92}
\scriptsize
\resizebox{\textwidth}{!}{%
\begin{tabular}{lcccccc}
\toprule
Pair & $N=8$, 10k & $N=8$, 30k & $N=10$, 10k & $N=10$, 30k & $N=12$, 10k & $N=12$, 30k \\
\midrule
BFGS--SPSA & $\mathbf{0.003}\;[-0.62]$ & $\mathbf{0.007}\;[-0.56]$ & $0.241\;[-0.35]$ & $1.000\;[-0.12]$ & $\mathbf{6.74\times10^{-5}}\;[-0.75]$ & $\mathbf{9.55\times10^{-5}}\;[-0.75]$ \\
BFGS--CMA-ES & $\mathbf{4.92\times10^{-5}}\;[-0.78]$ & $\mathbf{0.003}\;[-0.61]$ & $\mathbf{0.029}\;[-0.49]$ & $0.104\;[-0.46]$ & $\mathbf{2.14\times10^{-5}}\;[-0.79]$ & $\mathbf{5.59\times10^{-5}}\;[-0.78]$ \\
BFGS--iSOMA & $\mathbf{9.17\times10^{-7}}\;[-0.91]$ & $\mathbf{5.89\times10^{-7}}\;[-0.92]$ & $\mathbf{3.96\times10^{-8}}\;[-1.00]$ & $\mathbf{3.96\times10^{-8}}\;[-1.00]$ & $\mathbf{6.09\times10^{-8}}\;[-0.98]$ & $\mathbf{3.56\times10^{-7}}\;[-0.94]$ \\
BFGS--SciPy-DE & $0.059\;[-0.44]$ & $0.503\;[+0.20]$ & $1.000\;[+0.01]$ & $1.000\;[+0.24]$ & $\mathbf{2.59\times10^{-6}}\;[-0.87]$ & $0.059\;[-0.44]$ \\
BFGS--iL-SHADE & $\mathbf{1.63\times10^{-4}}\;[-0.73]$ & $\mathbf{0.007}\;[-0.56]$ & $0.067\;[-0.44]$ & $1.000\;[-0.05]$ & $\mathbf{1.42\times10^{-5}}\;[-0.81]$ & $\mathbf{3.15\times10^{-4}}\;[-0.71]$ \\
BFGS--L-SRTDE & $\mathbf{6.33\times10^{-7}}\;[-0.92]$ & $\mathbf{0.011}\;[-0.54]$ & $\mathbf{2.36\times10^{-6}}\;[-0.87]$ & $1.000\;[-0.10]$ & $\mathbf{3.96\times10^{-8}}\;[-1.00]$ & $\mathbf{2.71\times10^{-5}}\;[-0.80]$ \\
BFGS--jSO & $\mathbf{9.45\times10^{-7}}\;[-0.91]$ & $\mathbf{1.02\times10^{-6}}\;[-0.90]$ & $\mathbf{3.96\times10^{-8}}\;[-1.00]$ & $1.000\;[-0.13]$ & $\mathbf{3.96\times10^{-8}}\;[-1.00]$ & $\mathbf{2.65\times10^{-6}}\;[-0.88]$ \\
SPSA--CMA-ES & $\mathbf{0.010}\;[-0.56]$ & $0.156\;[-0.36]$ & $0.968\;[-0.17]$ & $0.405\;[-0.37]$ & $0.501\;[-0.23]$ & $0.223\;[-0.30]$ \\
SPSA--iSOMA & $\mathbf{1.12\times10^{-4}}\;[-0.75]$ & $\mathbf{8.24\times10^{-5}}\;[-0.76]$ & $\mathbf{1.58\times10^{-7}}\;[-0.96]$ & $\mathbf{1.20\times10^{-7}}\;[-0.96]$ & $\mathbf{2.69\times10^{-6}}\;[-0.86]$ & $\mathbf{2.96\times10^{-6}}\;[-0.87]$ \\
SPSA--SciPy-DE & $\mathbf{0.018}\;[+0.52]$ & $\mathbf{4.68\times10^{-4}}\;[+0.69]$ & $0.274\;[+0.33]$ & $0.427\;[+0.35]$ & $\mathbf{0.021}\;[-0.50]$ & $\mathbf{0.005}\;[+0.58]$ \\
SPSA--iL-SHADE & $0.916\;[-0.20]$ & $0.522\;[+0.11]$ & $1.000\;[+0.01]$ & $1.000\;[+0.12]$ & $0.501\;[-0.25]$ & $1.000\;[+0.04]$ \\
SPSA--L-SRTDE & $\mathbf{8.15\times10^{-6}}\;[-0.84]$ & $0.107\;[+0.40]$ & $\mathbf{0.001}\;[-0.64]$ & $1.000\;[+0.01]$ & $\mathbf{3.96\times10^{-8}}\;[-1.00]$ & $0.102\;[-0.40]$ \\
SPSA--jSO & $\mathbf{3.15\times10^{-4}}\;[-0.71]$ & $\mathbf{0.001}\;[-0.65]$ & $\mathbf{1.58\times10^{-7}}\;[-0.96]$ & $1.000\;[-0.09]$ & $\mathbf{9.90\times10^{-8}}\;[-0.96]$ & $\mathbf{8.98\times10^{-4}}\;[-0.66]$ \\
CMA-ES--iSOMA & $1.000\;[-0.13]$ & $0.153\;[-0.37]$ & $\mathbf{4.82\times10^{-7}}\;[-0.92]$ & $\mathbf{1.03\times10^{-4}}\;[-0.76]$ & $\mathbf{9.28\times10^{-8}}\;[-0.97]$ & $\mathbf{8.48\times10^{-5}}\;[-0.76]$ \\
CMA-ES--SciPy-DE & $\mathbf{1.59\times10^{-4}}\;[+0.73]$ & $\mathbf{1.01\times10^{-5}}\;[+0.83]$ & $\mathbf{0.029}\;[+0.50]$ & $\mathbf{0.004}\;[+0.62]$ & $0.387\;[-0.29]$ & $\mathbf{3.22\times10^{-4}}\;[+0.70]$ \\
CMA-ES--iL-SHADE & $0.097\;[+0.40]$ & $0.104\;[+0.41]$ & $0.968\;[+0.21]$ & $0.141\;[+0.44]$ & $0.703\;[-0.07]$ & $0.128\;[+0.37]$ \\
CMA-ES--L-SRTDE & $0.097\;[-0.40]$ & $\mathbf{0.002}\;[+0.63]$ & $\mathbf{0.012}\;[-0.55]$ & $0.178\;[+0.43]$ & $\mathbf{3.96\times10^{-8}}\;[-1.00]$ & $1.000\;[-0.08]$ \\
CMA-ES--jSO & $1.000\;[-0.08]$ & $0.503\;[-0.23]$ & $\mathbf{3.61\times10^{-7}}\;[-0.93]$ & $0.477\;[+0.34]$ & $\mathbf{3.96\times10^{-8}}\;[-1.00]$ & $0.062\;[-0.43]$ \\
iSOMA--SciPy-DE & $\mathbf{7.75\times10^{-7}}\;[+0.92]$ & $\mathbf{3.30\times10^{-8}}\;[+1.00]$ & $\mathbf{5.14\times10^{-7}}\;[+0.92]$ & $\mathbf{3.96\times10^{-8}}\;[+1.00]$ & $0.703\;[+0.16]$ & $\mathbf{2.94\times10^{-7}}\;[+0.94]$ \\
iSOMA--iL-SHADE & $\mathbf{0.010}\;[+0.56]$ & $\mathbf{8.93\times10^{-7}}\;[+0.91]$ & $\mathbf{3.96\times10^{-8}}\;[+1.00]$ & $\mathbf{3.96\times10^{-8}}\;[+1.00]$ & $\mathbf{2.35\times10^{-7}}\;[+0.94]$ & $\mathbf{9.54\times10^{-8}}\;[+0.97]$ \\
iSOMA--L-SRTDE & $0.059\;[-0.44]$ & $\mathbf{3.56\times10^{-8}}\;[+1.00]$ & $\mathbf{1.67\times10^{-5}}\;[+0.80]$ & $\mathbf{3.96\times10^{-8}}\;[+1.00]$ & $\mathbf{3.12\times10^{-7}}\;[-0.93]$ & $\mathbf{0.001}\;[+0.65]$ \\
iSOMA--jSO & $1.000\;[+0.15]$ & $\mathbf{5.84\times10^{-5}}\;[+0.77]$ & $0.968\;[-0.22]$ & $\mathbf{9.90\times10^{-8}}\;[+0.97]$ & $\mathbf{0.013}\;[-0.53]$ & $\mathbf{0.047}\;[+0.46]$ \\
SciPy-DE--iL-SHADE & $\mathbf{0.002}\;[-0.63]$ & $\mathbf{1.08\times10^{-4}}\;[-0.74]$ & $0.099\;[-0.41]$ & $0.417\;[-0.36]$ & $0.350\;[+0.32]$ & $\mathbf{0.005}\;[-0.58]$ \\
SciPy-DE--L-SRTDE & $\mathbf{9.17\times10^{-7}}\;[-0.91]$ & $\mathbf{2.09\times10^{-4}}\;[-0.72]$ & $\mathbf{4.94\times10^{-5}}\;[-0.77]$ & $0.178\;[-0.42]$ & $\mathbf{0.002}\;[-0.61]$ & $\mathbf{9.22\times10^{-5}}\;[-0.75]$ \\
SciPy-DE--jSO & $\mathbf{1.03\times10^{-6}}\;[-0.90]$ & $\mathbf{3.30\times10^{-8}}\;[-1.00]$ & $\mathbf{5.45\times10^{-7}}\;[-0.92]$ & $0.178\;[-0.42]$ & $0.350\;[-0.32]$ & $\mathbf{4.13\times10^{-6}}\;[-0.86]$ \\
iL-SHADE--L-SRTDE & $\mathbf{1.40\times10^{-4}}\;[-0.74]$ & $0.200\;[+0.32]$ & $\mathbf{7.72\times10^{-6}}\;[-0.83]$ & $1.000\;[-0.00]$ & $\mathbf{3.96\times10^{-8}}\;[-1.00]$ & $\mathbf{0.016}\;[-0.52]$ \\
iL-SHADE--jSO & $\mathbf{0.036}\;[-0.48]$ & $\mathbf{3.15\times10^{-5}}\;[-0.79]$ & $\mathbf{3.96\times10^{-8}}\;[-1.00]$ & $1.000\;[-0.07]$ & $\mathbf{3.96\times10^{-8}}\;[-1.00]$ & $\mathbf{7.75\times10^{-5}}\;[-0.76]$ \\
L-SRTDE--jSO & $\mathbf{0.037}\;[+0.48]$ & $\mathbf{4.19\times10^{-8}}\;[-0.99]$ & $\mathbf{1.47\times10^{-5}}\;[-0.81]$ & $1.000\;[-0.06]$ & $\mathbf{0.003}\;[+0.60]$ & $0.166\;[-0.34]$ \\
\bottomrule
\end{tabular}%
}
\end{table*}

\clearpage
\bibliography{bibfile}

\begin{thebibliography}{47}%
\makeatletter
\providecommand \@ifxundefined [1]{%
 \@ifx{#1\undefined}
}%
\providecommand \@ifnum [1]{%
 \ifnum #1\expandafter \@firstoftwo
 \else \expandafter \@secondoftwo
 \fi
}%
\providecommand \@ifx [1]{%
 \ifx #1\expandafter \@firstoftwo
 \else \expandafter \@secondoftwo
 \fi
}%
\providecommand \natexlab [1]{#1}%
\providecommand \enquote  [1]{``#1''}%
\providecommand \bibnamefont  [1]{#1}%
\providecommand \bibfnamefont [1]{#1}%
\providecommand \citenamefont [1]{#1}%
\providecommand \href@noop [0]{\@secondoftwo}%
\providecommand \href [0]{\begingroup \@sanitize@url \@href}%
\providecommand \@href[1]{\@@startlink{#1}\@@href}%
\providecommand \@@href[1]{\endgroup#1\@@endlink}%
\providecommand \@sanitize@url [0]{\catcode `\\12\catcode `\$12\catcode `\&12\catcode `\#12\catcode `\^12\catcode `\_12\catcode `\%12\relax}%
\providecommand \@@startlink[1]{}%
\providecommand \@@endlink[0]{}%
\providecommand \url  [0]{\begingroup\@sanitize@url \@url }%
\providecommand \@url [1]{\endgroup\@href {#1}{\urlprefix }}%
\providecommand \urlprefix  [0]{URL }%
\providecommand \Eprint [0]{\href }%
\providecommand \doibase [0]{https://doi.org/}%
\providecommand \selectlanguage [0]{\@gobble}%
\providecommand \bibinfo  [0]{\@secondoftwo}%
\providecommand \bibfield  [0]{\@secondoftwo}%
\providecommand \translation [1]{[#1]}%
\providecommand \BibitemOpen [0]{}%
\providecommand \bibitemStop [0]{}%
\providecommand \bibitemNoStop [0]{.\EOS\space}%
\providecommand \EOS [0]{\spacefactor3000\relax}%
\providecommand \BibitemShut  [1]{\csname bibitem#1\endcsname}%
\let\auto@bib@innerbib\@empty
\bibitem [{\citenamefont {Cerezo}\ \emph {et~al.}(2021)\citenamefont {Cerezo}, \citenamefont {Arrasmith}, \citenamefont {Babbush}, \citenamefont {Benjamin}, \citenamefont {Endo}, \citenamefont {Fujii}, \citenamefont {McClean}, \citenamefont {Mitarai}, \citenamefont {Yuan}, \citenamefont {Cincio},\ and\ \citenamefont {Coles}}]{Cerezo2021}%
  \BibitemOpen
  \bibfield  {author} {\bibinfo {author} {\bibfnamefont {M.}~\bibnamefont {Cerezo}}, \bibinfo {author} {\bibfnamefont {A.}~\bibnamefont {Arrasmith}}, \bibinfo {author} {\bibfnamefont {R.}~\bibnamefont {Babbush}}, \bibinfo {author} {\bibfnamefont {S.~C.}\ \bibnamefont {Benjamin}}, \bibinfo {author} {\bibfnamefont {S.}~\bibnamefont {Endo}}, \bibinfo {author} {\bibfnamefont {K.}~\bibnamefont {Fujii}}, \bibinfo {author} {\bibfnamefont {J.~R.}\ \bibnamefont {McClean}}, \bibinfo {author} {\bibfnamefont {K.}~\bibnamefont {Mitarai}}, \bibinfo {author} {\bibfnamefont {X.}~\bibnamefont {Yuan}}, \bibinfo {author} {\bibfnamefont {L.}~\bibnamefont {Cincio}},\ and\ \bibinfo {author} {\bibfnamefont {P.~J.}\ \bibnamefont {Coles}},\ }\bibfield  {title} {\bibinfo {title} {Variational quantum algorithms},\ }\href {https://doi.org/10.1038/s42254-021-00348-9} {\bibfield  {journal} {\bibinfo  {journal} {Nature Reviews Physics}\ }\textbf {\bibinfo {volume} {3}},\ \bibinfo {pages} {625} (\bibinfo {year} {2021})}\BibitemShut
  {NoStop}%
\bibitem [{\citenamefont {Fedorov}\ \emph {et~al.}(2022)\citenamefont {Fedorov}, \citenamefont {Peng}, \citenamefont {Govind},\ and\ \citenamefont {Alexeev}}]{fedorov2022vqe}%
  \BibitemOpen
  \bibfield  {author} {\bibinfo {author} {\bibfnamefont {D.~A.}\ \bibnamefont {Fedorov}}, \bibinfo {author} {\bibfnamefont {B.}~\bibnamefont {Peng}}, \bibinfo {author} {\bibfnamefont {N.}~\bibnamefont {Govind}},\ and\ \bibinfo {author} {\bibfnamefont {Y.}~\bibnamefont {Alexeev}},\ }\bibfield  {title} {\bibinfo {title} {Vqe method: a short survey and recent developments},\ }\href@noop {} {\bibfield  {journal} {\bibinfo  {journal} {Materials Theory}\ }\textbf {\bibinfo {volume} {6}},\ \bibinfo {pages} {2} (\bibinfo {year} {2022})}\BibitemShut {NoStop}%
\bibitem [{\citenamefont {McClean}\ \emph {et~al.}(2018)\citenamefont {McClean}, \citenamefont {Boixo}, \citenamefont {Smelyanskiy}, \citenamefont {Babbush},\ and\ \citenamefont {Neven}}]{McClean2018}%
  \BibitemOpen
  \bibfield  {author} {\bibinfo {author} {\bibfnamefont {J.~R.}\ \bibnamefont {McClean}}, \bibinfo {author} {\bibfnamefont {S.}~\bibnamefont {Boixo}}, \bibinfo {author} {\bibfnamefont {V.~N.}\ \bibnamefont {Smelyanskiy}}, \bibinfo {author} {\bibfnamefont {R.}~\bibnamefont {Babbush}},\ and\ \bibinfo {author} {\bibfnamefont {H.}~\bibnamefont {Neven}},\ }\bibfield  {title} {\bibinfo {title} {Barren plateaus in quantum neural network training landscapes},\ }\href {https://doi.org/10.1038/s41467-018-07090-4} {\bibfield  {journal} {\bibinfo  {journal} {Nature Communications}\ }\textbf {\bibinfo {volume} {9}},\ \bibinfo {pages} {4812} (\bibinfo {year} {2018})}\BibitemShut {NoStop}%
\bibitem [{\citenamefont {Anschuetz}\ and\ \citenamefont {Kiani}(2022)}]{AnschuetzKiani2022}%
  \BibitemOpen
  \bibfield  {author} {\bibinfo {author} {\bibfnamefont {E.~R.}\ \bibnamefont {Anschuetz}}\ and\ \bibinfo {author} {\bibfnamefont {B.~T.}\ \bibnamefont {Kiani}},\ }\bibfield  {title} {\bibinfo {title} {Quantum variational algorithms are swamped with traps},\ }\href {https://doi.org/10.1038/s41467-022-35364-5} {\bibfield  {journal} {\bibinfo  {journal} {Nature Communications}\ }\textbf {\bibinfo {volume} {13}},\ \bibinfo {pages} {7760} (\bibinfo {year} {2022})}\BibitemShut {NoStop}%
\bibitem [{\citenamefont {Nemkov}\ \emph {et~al.}(2025)\citenamefont {Nemkov}, \citenamefont {Kiktenko},\ and\ \citenamefont {Fedorov}}]{Nemkov2025Traps}%
  \BibitemOpen
  \bibfield  {author} {\bibinfo {author} {\bibfnamefont {N.~A.}\ \bibnamefont {Nemkov}}, \bibinfo {author} {\bibfnamefont {E.~O.}\ \bibnamefont {Kiktenko}},\ and\ \bibinfo {author} {\bibfnamefont {A.~K.}\ \bibnamefont {Fedorov}},\ }\bibfield  {title} {\bibinfo {title} {Barren plateaus swamped with traps},\ }\href {https://doi.org/10.1103/PhysRevA.111.012441} {\bibfield  {journal} {\bibinfo  {journal} {Phys. Rev. A}\ }\textbf {\bibinfo {volume} {111}},\ \bibinfo {pages} {012441} (\bibinfo {year} {2025})}\BibitemShut {NoStop}%
\bibitem [{\citenamefont {Jones}\ \emph {et~al.}(2025)\citenamefont {Jones}, \citenamefont {Mineh},\ and\ \citenamefont {Montanaro}}]{Jones2025}%
  \BibitemOpen
  \bibfield  {author} {\bibinfo {author} {\bibfnamefont {B.~D.~M.}\ \bibnamefont {Jones}}, \bibinfo {author} {\bibfnamefont {L.}~\bibnamefont {Mineh}},\ and\ \bibinfo {author} {\bibfnamefont {A.}~\bibnamefont {Montanaro}},\ }\bibfield  {title} {\bibinfo {title} {Benchmarking a wide range of optimisers for solving the fermi--hubbard model using the variational quantum eigensolver},\ }\href {https://doi.org/10.1088/2058-9565/adfe15} {\bibfield  {journal} {\bibinfo  {journal} {Quantum Science and Technology}\ }\textbf {\bibinfo {volume} {10}},\ \bibinfo {pages} {045032} (\bibinfo {year} {2025})},\ \Eprint {https://arxiv.org/abs/2411.13742} {arXiv:2411.13742} \BibitemShut {NoStop}%
\bibitem [{\citenamefont {Nov{\'a}k}\ \emph {et~al.}(2025{\natexlab{a}})\citenamefont {Nov{\'a}k}, \citenamefont {Zelinka},\ and\ \citenamefont {Sn{\'a}{\v s}el}}]{Novak2025NoisyLandscapes}%
  \BibitemOpen
  \bibfield  {author} {\bibinfo {author} {\bibfnamefont {V.}~\bibnamefont {Nov{\'a}k}}, \bibinfo {author} {\bibfnamefont {I.}~\bibnamefont {Zelinka}},\ and\ \bibinfo {author} {\bibfnamefont {V.}~\bibnamefont {Sn{\'a}{\v s}el}},\ }\bibfield  {title} {\bibinfo {title} {Optimization strategies for variational quantum algorithms in noisy landscapes},\ }\bibfield  {journal} {\bibinfo  {journal} {arXiv preprint arXiv:2506.01715}\ }\href {https://doi.org/10.48550/arXiv.2506.01715} {10.48550/arXiv.2506.01715} (\bibinfo {year} {2025}{\natexlab{a}}),\ \Eprint {https://arxiv.org/abs/2506.01715} {arXiv:2506.01715 [quant-ph]} \BibitemShut {NoStop}%
\bibitem [{\citenamefont {Ill{\'e}sov{\'a}}\ \emph {et~al.}(2025{\natexlab{a}})\citenamefont {Ill{\'e}sov{\'a}}, \citenamefont {Nov{\'a}k}, \citenamefont {Bezd{\v{e}}k}, \citenamefont {Possel},\ and\ \citenamefont {Beseda}}]{Illesova2025VHA}%
  \BibitemOpen
  \bibfield  {author} {\bibinfo {author} {\bibfnamefont {S.}~\bibnamefont {Ill{\'e}sov{\'a}}}, \bibinfo {author} {\bibfnamefont {V.}~\bibnamefont {Nov{\'a}k}}, \bibinfo {author} {\bibfnamefont {T.}~\bibnamefont {Bezd{\v{e}}k}}, \bibinfo {author} {\bibfnamefont {C.}~\bibnamefont {Possel}},\ and\ \bibinfo {author} {\bibfnamefont {M.}~\bibnamefont {Beseda}},\ }\bibfield  {title} {\bibinfo {title} {Numerical optimization strategies for the variational hamiltonian ansatz in noisy quantum environments},\ }\bibfield  {journal} {\bibinfo  {journal} {arXiv preprint arXiv:2505.22398}\ }\href {https://doi.org/10.48550/arXiv.2505.22398} {10.48550/arXiv.2505.22398} (\bibinfo {year} {2025}{\natexlab{a}}),\ \Eprint {https://arxiv.org/abs/2505.22398} {arXiv:2505.22398 [quant-ph]} \BibitemShut {NoStop}%
\bibitem [{\citenamefont {Ill{\'e}sov{\'a}}\ \emph {et~al.}(2025{\natexlab{b}})\citenamefont {Ill{\'e}sov{\'a}}, \citenamefont {Bezd{\v{e}}k}, \citenamefont {Nov{\'a}k}, \citenamefont {Senjean},\ and\ \citenamefont {Beseda}}]{Illesova2025Statistical}%
  \BibitemOpen
  \bibfield  {author} {\bibinfo {author} {\bibfnamefont {S.}~\bibnamefont {Ill{\'e}sov{\'a}}}, \bibinfo {author} {\bibfnamefont {T.}~\bibnamefont {Bezd{\v{e}}k}}, \bibinfo {author} {\bibfnamefont {V.}~\bibnamefont {Nov{\'a}k}}, \bibinfo {author} {\bibfnamefont {B.}~\bibnamefont {Senjean}},\ and\ \bibinfo {author} {\bibfnamefont {M.}~\bibnamefont {Beseda}},\ }\bibfield  {title} {\bibinfo {title} {Statistical benchmarking of optimization methods for variational quantum eigensolver under quantum noise},\ }\bibfield  {journal} {\bibinfo  {journal} {arXiv preprint arXiv:2510.08727}\ }\href {https://doi.org/10.48550/arXiv.2510.08727} {10.48550/arXiv.2510.08727} (\bibinfo {year} {2025}{\natexlab{b}}),\ \Eprint {https://arxiv.org/abs/2510.08727} {arXiv:2510.08727 [quant-ph]} \BibitemShut {NoStop}%
\bibitem [{\citenamefont {Bezd{\v{e}}k}\ \emph {et~al.}(2025)\citenamefont {Bezd{\v{e}}k}, \citenamefont {Yuan}, \citenamefont {Nov{\'a}k}, \citenamefont {Ill{\'e}sov{\'a}},\ and\ \citenamefont {Beseda}}]{Bezdek2025ClassicalOptimization}%
  \BibitemOpen
  \bibfield  {author} {\bibinfo {author} {\bibfnamefont {T.}~\bibnamefont {Bezd{\v{e}}k}}, \bibinfo {author} {\bibfnamefont {H.}~\bibnamefont {Yuan}}, \bibinfo {author} {\bibfnamefont {V.}~\bibnamefont {Nov{\'a}k}}, \bibinfo {author} {\bibfnamefont {S.}~\bibnamefont {Ill{\'e}sov{\'a}}},\ and\ \bibinfo {author} {\bibfnamefont {M.}~\bibnamefont {Beseda}},\ }\bibfield  {title} {\bibinfo {title} {Classical optimization strategies for variational quantum algorithms: A systematic study of noise effects and parameter efficiency},\ }\bibfield  {journal} {\bibinfo  {journal} {arXiv preprint arXiv:2511.09314}\ }\href {https://doi.org/10.48550/arXiv.2511.09314} {10.48550/arXiv.2511.09314} (\bibinfo {year} {2025}),\ \Eprint {https://arxiv.org/abs/2511.09314} {arXiv:2511.09314 [quant-ph]} \BibitemShut {NoStop}%
\bibitem [{\citenamefont {Nov{\'a}k}\ \emph {et~al.}(2025{\natexlab{b}})\citenamefont {Nov{\'a}k}, \citenamefont {Ill{\'e}sov{\'a}}, \citenamefont {Bezd{\v{e}}k}, \citenamefont {Zelinka},\ and\ \citenamefont {Beseda}}]{Novak2025Reliable}%
  \BibitemOpen
  \bibfield  {author} {\bibinfo {author} {\bibfnamefont {V.}~\bibnamefont {Nov{\'a}k}}, \bibinfo {author} {\bibfnamefont {S.}~\bibnamefont {Ill{\'e}sov{\'a}}}, \bibinfo {author} {\bibfnamefont {T.}~\bibnamefont {Bezd{\v{e}}k}}, \bibinfo {author} {\bibfnamefont {I.}~\bibnamefont {Zelinka}},\ and\ \bibinfo {author} {\bibfnamefont {M.}~\bibnamefont {Beseda}},\ }\bibfield  {title} {\bibinfo {title} {Reliable optimization under noise in quantum variational algorithms},\ }\bibfield  {journal} {\bibinfo  {journal} {arXiv preprint arXiv:2511.08289}\ }\href {https://doi.org/10.48550/arXiv.2511.08289} {10.48550/arXiv.2511.08289} (\bibinfo {year} {2025}{\natexlab{b}}),\ \Eprint {https://arxiv.org/abs/2511.08289} {arXiv:2511.08289 [quant-ph]} \BibitemShut {NoStop}%
\bibitem [{\citenamefont {Fa{\'i}lde}\ \emph {et~al.}(2023)\citenamefont {Fa{\'i}lde}, \citenamefont {Viqueira}, \citenamefont {Mussa~Juane},\ and\ \citenamefont {G{\'o}mez}}]{Failde2023}%
  \BibitemOpen
  \bibfield  {author} {\bibinfo {author} {\bibfnamefont {D.}~\bibnamefont {Fa{\'i}lde}}, \bibinfo {author} {\bibfnamefont {J.~D.}\ \bibnamefont {Viqueira}}, \bibinfo {author} {\bibfnamefont {M.}~\bibnamefont {Mussa~Juane}},\ and\ \bibinfo {author} {\bibfnamefont {A.}~\bibnamefont {G{\'o}mez}},\ }\bibfield  {title} {\bibinfo {title} {Using differential evolution to avoid local minima in variational quantum algorithms},\ }\href {https://doi.org/10.1038/s41598-023-43404-3} {\bibfield  {journal} {\bibinfo  {journal} {Scientific Reports}\ }\textbf {\bibinfo {volume} {13}},\ \bibinfo {pages} {16230} (\bibinfo {year} {2023})}\BibitemShut {NoStop}%
\bibitem [{\citenamefont {Mei}\ \emph {et~al.}(2024)\citenamefont {Mei}, \citenamefont {Zhao}, \citenamefont {Li}, \citenamefont {Chen}, \citenamefont {Zhang}, \citenamefont {Wang}, \citenamefont {Wu},\ and\ \citenamefont {Guo}}]{Mei2024}%
  \BibitemOpen
  \bibfield  {author} {\bibinfo {author} {\bibfnamefont {H.}~\bibnamefont {Mei}}, \bibinfo {author} {\bibfnamefont {J.}~\bibnamefont {Zhao}}, \bibinfo {author} {\bibfnamefont {Q.-S.}\ \bibnamefont {Li}}, \bibinfo {author} {\bibfnamefont {Z.-Y.}\ \bibnamefont {Chen}}, \bibinfo {author} {\bibfnamefont {J.-J.}\ \bibnamefont {Zhang}}, \bibinfo {author} {\bibfnamefont {Q.}~\bibnamefont {Wang}}, \bibinfo {author} {\bibfnamefont {Y.-C.}\ \bibnamefont {Wu}},\ and\ \bibinfo {author} {\bibfnamefont {G.-P.}\ \bibnamefont {Guo}},\ }\bibfield  {title} {\bibinfo {title} {Particle swarm optimization for a variational quantum eigensolver},\ }\href {https://doi.org/10.1039/D4CP02021A} {\bibfield  {journal} {\bibinfo  {journal} {Physical Chemistry Chemical Physics}\ }\textbf {\bibinfo {volume} {26}},\ \bibinfo {pages} {29070} (\bibinfo {year} {2024})}\BibitemShut {NoStop}%
\bibitem [{\citenamefont {Lotshaw}\ \emph {et~al.}(2023)\citenamefont {Lotshaw}, \citenamefont {Xu}, \citenamefont {Khalid}, \citenamefont {Buchs}, \citenamefont {Humble},\ and\ \citenamefont {Banerjee}}]{Lotshaw2023}%
  \BibitemOpen
  \bibfield  {author} {\bibinfo {author} {\bibfnamefont {P.~C.}\ \bibnamefont {Lotshaw}}, \bibinfo {author} {\bibfnamefont {H.}~\bibnamefont {Xu}}, \bibinfo {author} {\bibfnamefont {B.}~\bibnamefont {Khalid}}, \bibinfo {author} {\bibfnamefont {G.}~\bibnamefont {Buchs}}, \bibinfo {author} {\bibfnamefont {T.~S.}\ \bibnamefont {Humble}},\ and\ \bibinfo {author} {\bibfnamefont {A.}~\bibnamefont {Banerjee}},\ }\bibfield  {title} {\bibinfo {title} {Simulations of frustrated ising hamiltonians using quantum approximate optimization},\ }\href {https://doi.org/10.1098/rsta.2021.0414} {\bibfield  {journal} {\bibinfo  {journal} {Philosophical Transactions of the Royal Society A: Mathematical, Physical and Engineering Sciences}\ }\textbf {\bibinfo {volume} {381}},\ \bibinfo {pages} {20210414} (\bibinfo {year} {2023})}\BibitemShut {NoStop}%
\bibitem [{\citenamefont {Uvarov}\ \emph {et~al.}(2020)\citenamefont {Uvarov}, \citenamefont {Biamonte},\ and\ \citenamefont {Yudin}}]{Uvarov2020Frustrated}%
  \BibitemOpen
  \bibfield  {author} {\bibinfo {author} {\bibfnamefont {A.}~\bibnamefont {Uvarov}}, \bibinfo {author} {\bibfnamefont {J.~D.}\ \bibnamefont {Biamonte}},\ and\ \bibinfo {author} {\bibfnamefont {D.}~\bibnamefont {Yudin}},\ }\bibfield  {title} {\bibinfo {title} {Variational quantum eigensolver for frustrated quantum systems},\ }\href {https://doi.org/10.1103/PhysRevB.102.075104} {\bibfield  {journal} {\bibinfo  {journal} {Phys. Rev. B}\ }\textbf {\bibinfo {volume} {102}},\ \bibinfo {pages} {075104} (\bibinfo {year} {2020})}\BibitemShut {NoStop}%
\bibitem [{\citenamefont {Jattana}\ \emph {et~al.}(2022)\citenamefont {Jattana}, \citenamefont {Jin}, \citenamefont {De~Raedt},\ and\ \citenamefont {Michielsen}}]{Jattana2022}%
  \BibitemOpen
  \bibfield  {author} {\bibinfo {author} {\bibfnamefont {M.~S.}\ \bibnamefont {Jattana}}, \bibinfo {author} {\bibfnamefont {F.}~\bibnamefont {Jin}}, \bibinfo {author} {\bibfnamefont {H.}~\bibnamefont {De~Raedt}},\ and\ \bibinfo {author} {\bibfnamefont {K.}~\bibnamefont {Michielsen}},\ }\bibfield  {title} {\bibinfo {title} {Assessment of the variational quantum eigensolver: Application to the heisenberg model},\ }\href {https://doi.org/10.3389/fphy.2022.907160} {\bibfield  {journal} {\bibinfo  {journal} {Frontiers in Physics}\ }\textbf {\bibinfo {volume} {10}},\ \bibinfo {pages} {907160} (\bibinfo {year} {2022})}\BibitemShut {NoStop}%
\bibitem [{\citenamefont {Ill{\'e}sov{\'a}}\ \emph {et~al.}(2025{\natexlab{c}})\citenamefont {Ill{\'e}sov{\'a}}, \citenamefont {Rybotycki},\ and\ \citenamefont {Beseda}}]{illesova2025qmetric}%
  \BibitemOpen
  \bibfield  {author} {\bibinfo {author} {\bibfnamefont {S.}~\bibnamefont {Ill{\'e}sov{\'a}}}, \bibinfo {author} {\bibfnamefont {T.}~\bibnamefont {Rybotycki}},\ and\ \bibinfo {author} {\bibfnamefont {M.}~\bibnamefont {Beseda}},\ }\bibfield  {title} {\bibinfo {title} {Qmetric: Benchmarking quantum neural networks across circuits, features, and training dimensions},\ }in\ \href {https://ceur-ws.org/Vol-4080/paper3.pdf} {\emph {\bibinfo {booktitle} {Proceedings of QualITA 2025: The Fourth Conference on System and Service Quality}}},\ \bibinfo {series} {CEUR Workshop Proceedings}, Vol.\ \bibinfo {volume} {4080}\ (\bibinfo  {publisher} {CEUR-WS.org},\ \bibinfo {address} {Catania, Italy},\ \bibinfo {year} {2025})\BibitemShut {NoStop}%
\bibitem [{\citenamefont {Kirmani}\ \emph {et~al.}(2025)\citenamefont {Kirmani}, \citenamefont {Pelofske}, \citenamefont {B{\"a}rtschi}, \citenamefont {Eidenbenz},\ and\ \citenamefont {Zhu}}]{Kirmani2025}%
  \BibitemOpen
  \bibfield  {author} {\bibinfo {author} {\bibfnamefont {A.}~\bibnamefont {Kirmani}}, \bibinfo {author} {\bibfnamefont {E.}~\bibnamefont {Pelofske}}, \bibinfo {author} {\bibfnamefont {A.}~\bibnamefont {B{\"a}rtschi}}, \bibinfo {author} {\bibfnamefont {S.}~\bibnamefont {Eidenbenz}},\ and\ \bibinfo {author} {\bibfnamefont {J.-X.}\ \bibnamefont {Zhu}},\ }\bibfield  {title} {\bibinfo {title} {Variational quantum simulations of a two-dimensional frustrated transverse-field ising model on a trapped-ion quantum computer},\ }\href@noop {} {\bibfield  {journal} {\bibinfo  {journal} {arXiv preprint arXiv:2505.22932}\ } (\bibinfo {year} {2025})},\ \Eprint {https://arxiv.org/abs/2505.22932} {arXiv:2505.22932 [quant-ph]} \BibitemShut {NoStop}%
\bibitem [{\citenamefont {Pelofske}\ and\ \citenamefont {Eidenbenz}(2026)}]{Pelofske2026}%
  \BibitemOpen
  \bibfield  {author} {\bibinfo {author} {\bibfnamefont {E.}~\bibnamefont {Pelofske}}\ and\ \bibinfo {author} {\bibfnamefont {S.}~\bibnamefont {Eidenbenz}},\ }\bibfield  {title} {\bibinfo {title} {{VQE} as initial state preparation for {QPE} on heisenberg spin-glass hamiltonians},\ }\href@noop {} {\bibfield  {journal} {\bibinfo  {journal} {arXiv preprint arXiv:2606.15061}\ } (\bibinfo {year} {2026})},\ \Eprint {https://arxiv.org/abs/2606.15061} {arXiv:2606.15061 [quant-ph]} \BibitemShut {NoStop}%
\bibitem [{\citenamefont {B{\u{a}}utu}\ and\ \citenamefont {B{\u{a}}utu}(2008)}]{Bautu2008}%
  \BibitemOpen
  \bibfield  {author} {\bibinfo {author} {\bibfnamefont {A.}~\bibnamefont {B{\u{a}}utu}}\ and\ \bibinfo {author} {\bibfnamefont {E.}~\bibnamefont {B{\u{a}}utu}},\ }\bibfield  {title} {\bibinfo {title} {Searching ground states of ising spin glasses with genetic algorithms and binary particle swarm optimization},\ }in\ \href {https://doi.org/10.1007/978-3-540-78987-1_8} {\emph {\bibinfo {booktitle} {Nature Inspired Cooperative Strategies for Optimization (NICSO 2007)}}},\ \bibinfo {series} {Studies in Computational Intelligence}, Vol.\ \bibinfo {volume} {129}\ (\bibinfo  {publisher} {Springer},\ \bibinfo {year} {2008})\ pp.\ \bibinfo {pages} {85--94}\BibitemShut {NoStop}%
\bibitem [{\citenamefont {Brest}\ \emph {et~al.}(2016)\citenamefont {Brest}, \citenamefont {Sepesy~Mau{\v{c}}ec},\ and\ \citenamefont {Bo{\v{s}}kovi{\'c}}}]{Brest2016iLSHADE}%
  \BibitemOpen
  \bibfield  {author} {\bibinfo {author} {\bibfnamefont {J.}~\bibnamefont {Brest}}, \bibinfo {author} {\bibfnamefont {M.}~\bibnamefont {Sepesy~Mau{\v{c}}ec}},\ and\ \bibinfo {author} {\bibfnamefont {B.}~\bibnamefont {Bo{\v{s}}kovi{\'c}}},\ }\bibfield  {title} {\bibinfo {title} {il-shade: Improved l-shade algorithm for single objective real-parameter optimization},\ }in\ \href {https://doi.org/10.1109/CEC.2016.7743922} {\emph {\bibinfo {booktitle} {2016 IEEE Congress on Evolutionary Computation (CEC)}}}\ (\bibinfo {organization} {IEEE},\ \bibinfo {year} {2016})\ pp.\ \bibinfo {pages} {1188--1195}\BibitemShut {NoStop}%
\bibitem [{\citenamefont {Brest}\ \emph {et~al.}(2017)\citenamefont {Brest}, \citenamefont {Sepesy~Mau{\v{c}}ec},\ and\ \citenamefont {Bo{\v{s}}kovi{\'c}}}]{Brest2017jSO}%
  \BibitemOpen
  \bibfield  {author} {\bibinfo {author} {\bibfnamefont {J.}~\bibnamefont {Brest}}, \bibinfo {author} {\bibfnamefont {M.}~\bibnamefont {Sepesy~Mau{\v{c}}ec}},\ and\ \bibinfo {author} {\bibfnamefont {B.}~\bibnamefont {Bo{\v{s}}kovi{\'c}}},\ }\bibfield  {title} {\bibinfo {title} {Single objective real-parameter optimization: Algorithm jso},\ }in\ \href {https://doi.org/10.1109/CEC.2017.7969456} {\emph {\bibinfo {booktitle} {2017 IEEE Congress on Evolutionary Computation (CEC)}}}\ (\bibinfo {organization} {IEEE},\ \bibinfo {year} {2017})\ pp.\ \bibinfo {pages} {1311--1318}\BibitemShut {NoStop}%
\bibitem [{\citenamefont {Stanovov}\ and\ \citenamefont {Semenkin}(2024)}]{Stanovov2024LSRTDE}%
  \BibitemOpen
  \bibfield  {author} {\bibinfo {author} {\bibfnamefont {V.}~\bibnamefont {Stanovov}}\ and\ \bibinfo {author} {\bibfnamefont {E.}~\bibnamefont {Semenkin}},\ }\bibfield  {title} {\bibinfo {title} {Success rate-based adaptive differential evolution l-srtde for cec 2024 competition},\ }in\ \href {https://doi.org/10.1109/CEC60901.2024.10611907} {\emph {\bibinfo {booktitle} {2024 IEEE Congress on Evolutionary Computation (CEC)}}}\ (\bibinfo {organization} {IEEE},\ \bibinfo {year} {2024})\ pp.\ \bibinfo {pages} {1--8}\BibitemShut {NoStop}%
\bibitem [{\citenamefont {Nov{\'a}k}\ \emph {et~al.}(2026)\citenamefont {Nov{\'a}k}, \citenamefont {Bezd{\v{e}}k}, \citenamefont {Zelinka}, \citenamefont {Das},\ and\ \citenamefont {Beseda}}]{Novak2026CEC}%
  \BibitemOpen
  \bibfield  {author} {\bibinfo {author} {\bibfnamefont {V.}~\bibnamefont {Nov{\'a}k}}, \bibinfo {author} {\bibfnamefont {T.}~\bibnamefont {Bezd{\v{e}}k}}, \bibinfo {author} {\bibfnamefont {I.}~\bibnamefont {Zelinka}}, \bibinfo {author} {\bibfnamefont {S.}~\bibnamefont {Das}},\ and\ \bibinfo {author} {\bibfnamefont {M.}~\bibnamefont {Beseda}},\ }\bibfield  {title} {\bibinfo {title} {A longitudinal analysis of the cec single-objective competitions (2010--2024) and implications for variational quantum optimization},\ }\href {https://doi.org/https://doi.org/10.1016/j.swevo.2026.102469} {\bibfield  {journal} {\bibinfo  {journal} {Swarm and Evolutionary Computation}\ }\textbf {\bibinfo {volume} {107}},\ \bibinfo {pages} {102469} (\bibinfo {year} {2026})}\BibitemShut {NoStop}%
\bibitem [{\citenamefont {Spall}(1998)}]{spall1998implementation}%
  \BibitemOpen
  \bibfield  {author} {\bibinfo {author} {\bibfnamefont {J.~C.}\ \bibnamefont {Spall}},\ }\bibfield  {title} {\bibinfo {title} {Implementation of the simultaneous perturbation algorithm for stochastic optimization},\ }\href {https://doi.org/10.1109/7.705889} {\bibfield  {journal} {\bibinfo  {journal} {IEEE Transactions on Aerospace and Electronic Systems}\ }\textbf {\bibinfo {volume} {34}},\ \bibinfo {pages} {817} (\bibinfo {year} {1998})}\BibitemShut {NoStop}%
\bibitem [{\citenamefont {Hansen}\ \emph {et~al.}(2003)\citenamefont {Hansen}, \citenamefont {Müller},\ and\ \citenamefont {Koumoutsakos}}]{cma}%
  \BibitemOpen
  \bibfield  {author} {\bibinfo {author} {\bibfnamefont {N.}~\bibnamefont {Hansen}}, \bibinfo {author} {\bibfnamefont {S.~D.}\ \bibnamefont {Müller}},\ and\ \bibinfo {author} {\bibfnamefont {P.}~\bibnamefont {Koumoutsakos}},\ }\bibfield  {title} {\bibinfo {title} {Reducing the time complexity of the derandomized evolution strategy with covariance matrix adaptation (cma-es)},\ }\href {https://doi.org/10.1162/106365603321828970} {\bibfield  {journal} {\bibinfo  {journal} {Evolutionary Computation}\ }\textbf {\bibinfo {volume} {11}},\ \bibinfo {pages} {1} (\bibinfo {year} {2003})}\BibitemShut {NoStop}%
\bibitem [{\citenamefont {Zelinka}\ \emph {et~al.}(2023)\citenamefont {Zelinka}, \citenamefont {Kojeck{\`y}}, \citenamefont {Lampart}, \citenamefont {Nowakov{\'a}},\ and\ \citenamefont {Plucar}}]{zelinka2023isoma}%
  \BibitemOpen
  \bibfield  {author} {\bibinfo {author} {\bibfnamefont {I.}~\bibnamefont {Zelinka}}, \bibinfo {author} {\bibfnamefont {L.}~\bibnamefont {Kojeck{\`y}}}, \bibinfo {author} {\bibfnamefont {M.}~\bibnamefont {Lampart}}, \bibinfo {author} {\bibfnamefont {J.}~\bibnamefont {Nowakov{\'a}}},\ and\ \bibinfo {author} {\bibfnamefont {J.}~\bibnamefont {Plucar}},\ }\bibfield  {title} {\bibinfo {title} {isoma swarm intelligence algorithm in synthesis of quantum computing circuits},\ }\href@noop {} {\bibfield  {journal} {\bibinfo  {journal} {Applied Soft Computing}\ }\textbf {\bibinfo {volume} {142}},\ \bibinfo {pages} {110350} (\bibinfo {year} {2023})}\BibitemShut {NoStop}%
\bibitem [{\citenamefont {van Luijk}\ \emph {et~al.}(2024)\citenamefont {van Luijk}, \citenamefont {Schwonnek}, \citenamefont {Stottmeister},\ and\ \citenamefont {Werner}}]{van2024schmidt}%
  \BibitemOpen
  \bibfield  {author} {\bibinfo {author} {\bibfnamefont {L.}~\bibnamefont {van Luijk}}, \bibinfo {author} {\bibfnamefont {R.}~\bibnamefont {Schwonnek}}, \bibinfo {author} {\bibfnamefont {A.}~\bibnamefont {Stottmeister}},\ and\ \bibinfo {author} {\bibfnamefont {R.~F.}\ \bibnamefont {Werner}},\ }\bibfield  {title} {\bibinfo {title} {The schmidt rank for the commuting operator framework},\ }\href@noop {} {\bibfield  {journal} {\bibinfo  {journal} {Communications in Mathematical Physics}\ }\textbf {\bibinfo {volume} {405}},\ \bibinfo {pages} {152} (\bibinfo {year} {2024})}\BibitemShut {NoStop}%
\bibitem [{\citenamefont {Stokes}\ \emph {et~al.}(2020)\citenamefont {Stokes}, \citenamefont {Izaac}, \citenamefont {Killoran},\ and\ \citenamefont {Carleo}}]{stokes2020quantum}%
  \BibitemOpen
  \bibfield  {author} {\bibinfo {author} {\bibfnamefont {J.}~\bibnamefont {Stokes}}, \bibinfo {author} {\bibfnamefont {J.}~\bibnamefont {Izaac}}, \bibinfo {author} {\bibfnamefont {N.}~\bibnamefont {Killoran}},\ and\ \bibinfo {author} {\bibfnamefont {G.}~\bibnamefont {Carleo}},\ }\bibfield  {title} {\bibinfo {title} {Quantum natural gradient},\ }\href@noop {} {\bibfield  {journal} {\bibinfo  {journal} {Quantum}\ }\textbf {\bibinfo {volume} {4}},\ \bibinfo {pages} {269} (\bibinfo {year} {2020})}\BibitemShut {NoStop}%
\bibitem [{\citenamefont {Peruzzo}\ \emph {et~al.}(2014)\citenamefont {Peruzzo}, \citenamefont {McClean}, \citenamefont {Shadbolt}, \citenamefont {Yung}, \citenamefont {Zhou}, \citenamefont {Love}, \citenamefont {Aspuru-Guzik},\ and\ \citenamefont {O'Brien}}]{Peruzzo2014}%
  \BibitemOpen
  \bibfield  {author} {\bibinfo {author} {\bibfnamefont {A.}~\bibnamefont {Peruzzo}}, \bibinfo {author} {\bibfnamefont {J.}~\bibnamefont {McClean}}, \bibinfo {author} {\bibfnamefont {P.}~\bibnamefont {Shadbolt}}, \bibinfo {author} {\bibfnamefont {M.-H.}\ \bibnamefont {Yung}}, \bibinfo {author} {\bibfnamefont {X.-Q.}\ \bibnamefont {Zhou}}, \bibinfo {author} {\bibfnamefont {P.~J.}\ \bibnamefont {Love}}, \bibinfo {author} {\bibfnamefont {A.}~\bibnamefont {Aspuru-Guzik}},\ and\ \bibinfo {author} {\bibfnamefont {J.~L.}\ \bibnamefont {O'Brien}},\ }\bibfield  {title} {\bibinfo {title} {A variational eigenvalue solver on a photonic quantum processor},\ }\href {https://doi.org/10.1038/ncomms5213} {\bibfield  {journal} {\bibinfo  {journal} {Nature Communications}\ }\textbf {\bibinfo {volume} {5}},\ \bibinfo {pages} {4213} (\bibinfo {year} {2014})}\BibitemShut {NoStop}%
\bibitem [{\citenamefont {Nocedal}\ and\ \citenamefont {Wright}(2006)}]{nocedal2006numerical}%
  \BibitemOpen
  \bibfield  {author} {\bibinfo {author} {\bibfnamefont {J.}~\bibnamefont {Nocedal}}\ and\ \bibinfo {author} {\bibfnamefont {S.~J.}\ \bibnamefont {Wright}},\ }\href@noop {} {\emph {\bibinfo {title} {Numerical optimization}}}\ (\bibinfo  {publisher} {Springer},\ \bibinfo {year} {2006})\BibitemShut {NoStop}%
\bibitem [{\citenamefont {Spall}(1992)}]{Spall1992}%
  \BibitemOpen
  \bibfield  {author} {\bibinfo {author} {\bibfnamefont {J.~C.}\ \bibnamefont {Spall}},\ }\bibfield  {title} {\bibinfo {title} {Multivariate stochastic approximation using a simultaneous perturbation gradient approximation},\ }\href {https://doi.org/10.1109/9.119632} {\bibfield  {journal} {\bibinfo  {journal} {IEEE Transactions on Automatic Control}\ }\textbf {\bibinfo {volume} {37}},\ \bibinfo {pages} {332} (\bibinfo {year} {1992})}\BibitemShut {NoStop}%
\bibitem [{\citenamefont {Hansen}\ and\ \citenamefont {Ostermeier}(2001)}]{HansenOstermeier2001}%
  \BibitemOpen
  \bibfield  {author} {\bibinfo {author} {\bibfnamefont {N.}~\bibnamefont {Hansen}}\ and\ \bibinfo {author} {\bibfnamefont {A.}~\bibnamefont {Ostermeier}},\ }\bibfield  {title} {\bibinfo {title} {Completely derandomized self-adaptation in evolution strategies},\ }\href {https://doi.org/10.1162/106365601750190398} {\bibfield  {journal} {\bibinfo  {journal} {Evolutionary Computation}\ }\textbf {\bibinfo {volume} {9}},\ \bibinfo {pages} {159} (\bibinfo {year} {2001})}\BibitemShut {NoStop}%
\bibitem [{\citenamefont {Storn}\ and\ \citenamefont {Price}(1997)}]{StornPrice1997}%
  \BibitemOpen
  \bibfield  {author} {\bibinfo {author} {\bibfnamefont {R.}~\bibnamefont {Storn}}\ and\ \bibinfo {author} {\bibfnamefont {K.}~\bibnamefont {Price}},\ }\bibfield  {title} {\bibinfo {title} {Differential evolution---a simple and efficient heuristic for global optimization over continuous spaces},\ }\href {https://doi.org/10.1023/A:1008202821328} {\bibfield  {journal} {\bibinfo  {journal} {Journal of Global Optimization}\ }\textbf {\bibinfo {volume} {11}},\ \bibinfo {pages} {341} (\bibinfo {year} {1997})}\BibitemShut {NoStop}%
\bibitem [{\citenamefont {Diep}\ \emph {et~al.}(2022)\citenamefont {Diep}, \citenamefont {Truong}, \citenamefont {Das},\ and\ \citenamefont {Zelinka}}]{Diep2022iSOMA}%
  \BibitemOpen
  \bibfield  {author} {\bibinfo {author} {\bibfnamefont {Q.~B.}\ \bibnamefont {Diep}}, \bibinfo {author} {\bibfnamefont {T.~C.}\ \bibnamefont {Truong}}, \bibinfo {author} {\bibfnamefont {S.}~\bibnamefont {Das}},\ and\ \bibinfo {author} {\bibfnamefont {I.}~\bibnamefont {Zelinka}},\ }\bibfield  {title} {\bibinfo {title} {Self-organizing migrating algorithm with narrowing search space strategy for robot path planning},\ }\href {https://doi.org/10.1016/j.asoc.2021.108270} {\bibfield  {journal} {\bibinfo  {journal} {Applied Soft Computing}\ }\textbf {\bibinfo {volume} {116}},\ \bibinfo {pages} {108270} (\bibinfo {year} {2022})}\BibitemShut {NoStop}%
\bibitem [{\citenamefont {Akshay}\ \emph {et~al.}(2021)\citenamefont {Akshay}, \citenamefont {Philathong}, \citenamefont {Morales},\ and\ \citenamefont {Biamonte}}]{Akshay2021reachability}%
  \BibitemOpen
  \bibfield  {author} {\bibinfo {author} {\bibfnamefont {V.}~\bibnamefont {Akshay}}, \bibinfo {author} {\bibfnamefont {H.}~\bibnamefont {Philathong}}, \bibinfo {author} {\bibfnamefont {M.~E.~S.}\ \bibnamefont {Morales}},\ and\ \bibinfo {author} {\bibfnamefont {J.}~\bibnamefont {Biamonte}},\ }\bibfield  {title} {\bibinfo {title} {Reachability deficits in quantum approximate optimization of graph problems},\ }\href {https://doi.org/10.22331/q-2021-08-30-532} {\bibfield  {journal} {\bibinfo  {journal} {Quantum}\ }\textbf {\bibinfo {volume} {5}},\ \bibinfo {pages} {532} (\bibinfo {year} {2021})}\BibitemShut {NoStop}%
\bibitem [{\citenamefont {Bharti}\ \emph {et~al.}(2022)\citenamefont {Bharti}, \citenamefont {Cervera-Lierta}, \citenamefont {Kyaw}, \citenamefont {Haug}, \citenamefont {Alperin-Lea}, \citenamefont {Anand}, \citenamefont {Degroote}, \citenamefont {Heimonen}, \citenamefont {Kottmann}, \citenamefont {Menke}, \citenamefont {Mok}, \citenamefont {Sim}, \citenamefont {Kwek},\ and\ \citenamefont {Aspuru-Guzik}}]{Bharti2022NISQ}%
  \BibitemOpen
  \bibfield  {author} {\bibinfo {author} {\bibfnamefont {K.}~\bibnamefont {Bharti}}, \bibinfo {author} {\bibfnamefont {A.}~\bibnamefont {Cervera-Lierta}}, \bibinfo {author} {\bibfnamefont {T.~H.}\ \bibnamefont {Kyaw}}, \bibinfo {author} {\bibfnamefont {T.}~\bibnamefont {Haug}}, \bibinfo {author} {\bibfnamefont {S.}~\bibnamefont {Alperin-Lea}}, \bibinfo {author} {\bibfnamefont {A.}~\bibnamefont {Anand}}, \bibinfo {author} {\bibfnamefont {M.}~\bibnamefont {Degroote}}, \bibinfo {author} {\bibfnamefont {H.}~\bibnamefont {Heimonen}}, \bibinfo {author} {\bibfnamefont {J.~S.}\ \bibnamefont {Kottmann}}, \bibinfo {author} {\bibfnamefont {T.}~\bibnamefont {Menke}}, \bibinfo {author} {\bibfnamefont {W.-K.}\ \bibnamefont {Mok}}, \bibinfo {author} {\bibfnamefont {S.}~\bibnamefont {Sim}}, \bibinfo {author} {\bibfnamefont {L.-C.}\ \bibnamefont {Kwek}},\ and\ \bibinfo {author} {\bibfnamefont {A.}~\bibnamefont {Aspuru-Guzik}},\ }\bibfield  {title} {\bibinfo {title} {Noisy intermediate-scale quantum algorithms},\ }\href
  {https://doi.org/10.1103/RevModPhys.94.015004} {\bibfield  {journal} {\bibinfo  {journal} {Reviews of Modern Physics}\ }\textbf {\bibinfo {volume} {94}},\ \bibinfo {pages} {015004} (\bibinfo {year} {2022})}\BibitemShut {NoStop}%
\bibitem [{\citenamefont {Grimsley}\ \emph {et~al.}(2019)\citenamefont {Grimsley}, \citenamefont {Economou}, \citenamefont {Barnes},\ and\ \citenamefont {Mayhall}}]{Grimsley2019adaptvqe}%
  \BibitemOpen
  \bibfield  {author} {\bibinfo {author} {\bibfnamefont {H.~R.}\ \bibnamefont {Grimsley}}, \bibinfo {author} {\bibfnamefont {S.~E.}\ \bibnamefont {Economou}}, \bibinfo {author} {\bibfnamefont {E.}~\bibnamefont {Barnes}},\ and\ \bibinfo {author} {\bibfnamefont {N.~J.}\ \bibnamefont {Mayhall}},\ }\bibfield  {title} {\bibinfo {title} {An adaptive variational algorithm for exact molecular simulations on a quantum computer},\ }\href {https://doi.org/10.1038/s41467-019-10988-2} {\bibfield  {journal} {\bibinfo  {journal} {Nature Communications}\ }\textbf {\bibinfo {volume} {10}},\ \bibinfo {pages} {3007} (\bibinfo {year} {2019})}\BibitemShut {NoStop}%
\bibitem [{\citenamefont {Woitzik}\ \emph {et~al.}(2020)\citenamefont {Woitzik}, \citenamefont {Barkoutsos}, \citenamefont {Wudarski}, \citenamefont {Buchleitner},\ and\ \citenamefont {Tavernelli}}]{Woitzik2020Entanglement}%
  \BibitemOpen
  \bibfield  {author} {\bibinfo {author} {\bibfnamefont {A.~J.~C.}\ \bibnamefont {Woitzik}}, \bibinfo {author} {\bibfnamefont {P.~K.}\ \bibnamefont {Barkoutsos}}, \bibinfo {author} {\bibfnamefont {F.}~\bibnamefont {Wudarski}}, \bibinfo {author} {\bibfnamefont {A.}~\bibnamefont {Buchleitner}},\ and\ \bibinfo {author} {\bibfnamefont {I.}~\bibnamefont {Tavernelli}},\ }\bibfield  {title} {\bibinfo {title} {Entanglement production and convergence properties of the variational quantum eigensolver},\ }\href {https://doi.org/10.1103/PhysRevA.102.042402} {\bibfield  {journal} {\bibinfo  {journal} {Phys. Rev. A}\ }\textbf {\bibinfo {volume} {102}},\ \bibinfo {pages} {042402} (\bibinfo {year} {2020})}\BibitemShut {NoStop}%
\bibitem [{\citenamefont {D{\'i}ez-Valle}\ \emph {et~al.}(2021)\citenamefont {D{\'i}ez-Valle}, \citenamefont {Porras},\ and\ \citenamefont {Garc{\'i}a-Ripoll}}]{DiezValle2021EntanglementHardness}%
  \BibitemOpen
  \bibfield  {author} {\bibinfo {author} {\bibfnamefont {P.}~\bibnamefont {D{\'i}ez-Valle}}, \bibinfo {author} {\bibfnamefont {D.}~\bibnamefont {Porras}},\ and\ \bibinfo {author} {\bibfnamefont {J.~J.}\ \bibnamefont {Garc{\'i}a-Ripoll}},\ }\bibfield  {title} {\bibinfo {title} {Quantum variational optimization: The role of entanglement and problem hardness},\ }\href {https://doi.org/10.1103/PhysRevA.104.062426} {\bibfield  {journal} {\bibinfo  {journal} {Phys. Rev. A}\ }\textbf {\bibinfo {volume} {104}},\ \bibinfo {pages} {062426} (\bibinfo {year} {2021})}\BibitemShut {NoStop}%
\bibitem [{\citenamefont {Scriva}\ \emph {et~al.}(2024)\citenamefont {Scriva}, \citenamefont {Astrakhantsev}, \citenamefont {Pilati},\ and\ \citenamefont {Mazzola}}]{Scriva2024ShotNoise}%
  \BibitemOpen
  \bibfield  {author} {\bibinfo {author} {\bibfnamefont {G.}~\bibnamefont {Scriva}}, \bibinfo {author} {\bibfnamefont {N.}~\bibnamefont {Astrakhantsev}}, \bibinfo {author} {\bibfnamefont {S.}~\bibnamefont {Pilati}},\ and\ \bibinfo {author} {\bibfnamefont {G.}~\bibnamefont {Mazzola}},\ }\bibfield  {title} {\bibinfo {title} {Challenges of variational quantum optimization with measurement shot noise},\ }\href {https://doi.org/10.1103/PhysRevA.109.032408} {\bibfield  {journal} {\bibinfo  {journal} {Phys. Rev. A}\ }\textbf {\bibinfo {volume} {109}},\ \bibinfo {pages} {032408} (\bibinfo {year} {2024})}\BibitemShut {NoStop}%
\bibitem [{\citenamefont {Mele}\ \emph {et~al.}(2022)\citenamefont {Mele}, \citenamefont {Mbeng}, \citenamefont {Santoro}, \citenamefont {Collura},\ and\ \citenamefont {Torta}}]{Mele2022Smooth}%
  \BibitemOpen
  \bibfield  {author} {\bibinfo {author} {\bibfnamefont {A.~A.}\ \bibnamefont {Mele}}, \bibinfo {author} {\bibfnamefont {G.~B.}\ \bibnamefont {Mbeng}}, \bibinfo {author} {\bibfnamefont {G.~E.}\ \bibnamefont {Santoro}}, \bibinfo {author} {\bibfnamefont {M.}~\bibnamefont {Collura}},\ and\ \bibinfo {author} {\bibfnamefont {P.}~\bibnamefont {Torta}},\ }\bibfield  {title} {\bibinfo {title} {Avoiding barren plateaus via transferability of smooth solutions in a hamiltonian variational ansatz},\ }\href {https://doi.org/10.1103/PhysRevA.106.L060401} {\bibfield  {journal} {\bibinfo  {journal} {Phys. Rev. A}\ }\textbf {\bibinfo {volume} {106}},\ \bibinfo {pages} {L060401} (\bibinfo {year} {2022})}\BibitemShut {NoStop}%
\bibitem [{\citenamefont {Novák}\ \emph {et~al.}(2026)\citenamefont {Novák}, \citenamefont {Zelinka}, \citenamefont {Das},\ and\ \citenamefont {Beseda}}]{novak_2026_22210896}%
  \BibitemOpen
  \bibfield  {author} {\bibinfo {author} {\bibfnamefont {V.}~\bibnamefont {Novák}}, \bibinfo {author} {\bibfnamefont {I.}~\bibnamefont {Zelinka}}, \bibinfo {author} {\bibfnamefont {S.}~\bibnamefont {Das}},\ and\ \bibinfo {author} {\bibfnamefont {M.}~\bibnamefont {Beseda}},\ }\bibfield  {title} {\bibinfo {title} {Optimization landscape geometry in vqe for frustrated quantum spin models},\ }\href {https://doi.org/10.5281/zenodo.22210896} {10.5281/zenodo.22210896} (\bibinfo {year} {2026})\BibitemShut {NoStop}%
\bibitem [{\citenamefont {Beseda}\ \emph {et~al.}(2024)\citenamefont {Beseda}, \citenamefont {Ill{\'e}sov{\'a}}, \citenamefont {Yalouz},\ and\ \citenamefont {Senjean}}]{beseda2024state}%
  \BibitemOpen
  \bibfield  {author} {\bibinfo {author} {\bibfnamefont {M.}~\bibnamefont {Beseda}}, \bibinfo {author} {\bibfnamefont {S.}~\bibnamefont {Ill{\'e}sov{\'a}}}, \bibinfo {author} {\bibfnamefont {S.}~\bibnamefont {Yalouz}},\ and\ \bibinfo {author} {\bibfnamefont {B.}~\bibnamefont {Senjean}},\ }\bibfield  {title} {\bibinfo {title} {State-averaged orbital-optimized vqe: A quantum algorithm for the democratic description of ground and excited electronic states},\ }\href@noop {} {\bibfield  {journal} {\bibinfo  {journal} {Journal of Open Source Software}\ }\textbf {\bibinfo {volume} {9}},\ \bibinfo {pages} {6036} (\bibinfo {year} {2024})}\BibitemShut {NoStop}%
\bibitem [{\citenamefont {Ill{\'e}sová}\ \emph {et~al.}(2025)\citenamefont {Ill{\'e}sová}, \citenamefont {Beseda}, \citenamefont {Yalouz}, \citenamefont {Lasorne},\ and\ \citenamefont {Senjean}}]{illesova2025transformation}%
  \BibitemOpen
  \bibfield  {author} {\bibinfo {author} {\bibfnamefont {S.}~\bibnamefont {Ill{\'e}sová}}, \bibinfo {author} {\bibfnamefont {M.}~\bibnamefont {Beseda}}, \bibinfo {author} {\bibfnamefont {S.}~\bibnamefont {Yalouz}}, \bibinfo {author} {\bibfnamefont {B.}~\bibnamefont {Lasorne}},\ and\ \bibinfo {author} {\bibfnamefont {B.}~\bibnamefont {Senjean}},\ }\bibfield  {title} {\bibinfo {title} {Transformation-free generation of a quasi-diabatic representation from the state-average orbital-optimized variational quantum eigensolver},\ }\href {https://doi.org/10.1021/acs.jctc.5c00327} {\bibfield  {journal} {\bibinfo  {journal} {Journal of Chemical Theory and Computation}\ }\textbf {\bibinfo {volume} {21}},\ \bibinfo {pages} {5457} (\bibinfo {year} {2025})},\ \bibinfo {note} {pMID: 40408769},\ \Eprint {https://arxiv.org/abs/https://doi.org/10.1021/acs.jctc.5c00327} {https://doi.org/10.1021/acs.jctc.5c00327} \BibitemShut {NoStop}%
\bibitem [{\citenamefont {Anschuetz}(2022)}]{Anschuetz2022CriticalPoints}%
  \BibitemOpen
  \bibfield  {author} {\bibinfo {author} {\bibfnamefont {E.~R.}\ \bibnamefont {Anschuetz}},\ }\bibfield  {title} {\bibinfo {title} {Critical points in quantum generative models},\ }in\ \href {https://doi.org/10.48550/arXiv.2109.06957} {\emph {\bibinfo {booktitle} {International Conference on Learning Representations (ICLR)}}}\ (\bibinfo {year} {2022})\BibitemShut {NoStop}%
\bibitem [{\citenamefont {Mich{\'a}lek}(2026)}]{Michalek2026}%
  \BibitemOpen
  \bibfield  {author} {\bibinfo {author} {\bibfnamefont {J.}~\bibnamefont {Mich{\'a}lek}},\ }\emph {\bibinfo {title} {Application of Random Fields in Variational Quantum Algorithms}},\ \href@noop {} {\bibinfo {type} {Master's thesis}},\ \bibinfo  {school} {Brno University of Technology, Faculty of Mechanical Engineering} (\bibinfo {year} {2026})\BibitemShut {NoStop}%
\end{thebibliography}%

\end{document}